\documentclass[letterpaper,11pt,fleqn]{article}
\usepackage{jheppub}
\usepackage{graphbox}
\usepackage[utf8]{inputenc}
\usepackage[T1]{fontenc}
\usepackage{extarrows}
\usepackage{bm,amsmath,amssymb}
\usepackage[mathscr]{eucal}
\usepackage{graphicx}
\usepackage{cancel}
\usepackage{subcaption}
\usepackage{dsfont}
\usepackage{enumerate}
\usepackage[normalem]{ulem}

\long\def\comment#1{ }

\newcommand{\eqn}[1]{Eq.~\eqref{#1}}
\newcommand{\beq}{\begin{equation}}
\newcommand{\eeq}{\end{equation}}
\newcommand{\bal}{\begin{align}}
\newcommand{\eal}{\end{align}}

\newcommand{\tF}{\tilde{\mcal{F}}_g}
\newcommand{\bt}{b_\perp}

\newcommand{\rmS}{{\rm S}}
\newcommand{\del}{\partial}
\newcommand{\xP}{x_{\mathbb P}}

\newcommand{\order}[1]{\mathcal{O}{\left(#1\right)}}
\newcommand{\abar}{\bar{\alpha}_s}
\newcommand{\nn}{\nonumber\\}
\newcommand{\rmd}{{\rm d}}
\newcommand{\dif}{{\rm d}}
\newcommand{\rme}{{\rm e}}
\newcommand{\bk}{\bm{k}}

\newcommand{\bb}{\bm{b}}

\newcommand{\mcal}{\mathcal}
\newcommand{\bK}{\bm{K}}

\newcommand{\KT}{K_\perp}
\newcommand{\PT}{P_\perp}
\newcommand{\kg}{k_{g\perp}}
\newcommand{\lt}{\ell_\perp}
\newcommand{\bell}{\bm{\ell}}

\newcommand{\YP}{Y_{\mathbb P}}

\title{The quantum evolutions of the diffractive transverse-momentum dependent gluon distribution}

\author[a]{E.~Iancu,}
\author[b]{D.N.~Triantafyllopoulos,}
\author[c]{S.Y.~Wei,}
\author[d,a]{and F.~Yuan}

\affiliation[a]{Universit\'{e} Paris-Saclay, CNRS, CEA, Institut de physique th\'{e}orique, F-91191, Gif-sur-Yvette, France}
\affiliation[b]{European Centre for Theoretical Studies in Nuclear Physics and Related Areas (ECT*)\\and Fondazione Bruno Kessler, Strada delle Tabarelle 286, I-38123 Villazzano (TN), Italy}
\affiliation[c]{Key Laboratory of Particle Physics and Particle Irradiation (MOE), Institute of frontier and interdisciplinary science, Shandong University, Qingdao, Shandong 266237, China}
\affiliation[d]{Nuclear Science Division, Lawrence Berkeley National Laboratory, Berkeley, CA 94720, USA}

\emailAdd{edmond.iancu@ipht.fr}
\emailAdd{trianta@ectstar.eu}
\emailAdd{shuyi@sdu.edu.cn}
\emailAdd{fyuan@lbl.gov}

\abstract{Using the Colour Glass Condensate description of electron-nucleus collisions at high energy, we study the diffractive production of a pair of jets with transverse momenta much larger than the nuclear saturation momentum $Q_s$. At leading order in the QCD coupling, the di-jet cross-section exhibits transverse-momentum dependent (TMD) factorisation, with a gluon diffractive TMD distribution (DTMD) which is  controlled by gluon saturation and describes the transverse-momentum imbalance between the produced jets. The next-to-leading corrections generate the various quantum evolutions of the diffractive gluon distribution. We focus on the Collins-Soper-Sterman (CSS) evolution which describes the change in the gluon DTMD when increasing the ``hard scale'' (the typical transverse momentum of the di-jets). We consider two different representations for this equation, one in transverse-momentum space, the other one in transverse-coordinate space. They are not fully equivalent with each other (despite being related by a Fourier transform) because of the respective boundary conditions. These conditions encode the essential physics of gluon saturation together with the effects of two other types of quantum evolution: the BK/JIMWLK evolution over the rapidity gap (``inside the Pomeron'') and the DGLAP evolution outside the rapidity gap (``within the diffractive system''). We demonstrate that, due to gluon saturation, one can compute both the boundary conditions and the CSS solutions fully from first principles, without reference to non-perturbative physics. We numerically find a good agreement between the CSS solutions in the two aforementioned representations.
}

\begin{document}
\maketitle

\section{Introduction}
\label{sec:intro}

One of the main physics goals of the future experiments at the Electron-Ion Collider (EIC) currently under construction in USA~\cite{Boer:2011fh,Accardi:2012qut,Aschenauer:2017jsk,AbdulKhalek:2022hcn} will be the exploration of new forms of hadronic matter characterised by high partonic density and the important, but still elusive, phenomenon of {\it gluon saturation}. A consequence of strong non-linear effects in dense gluonic systems, this phenomenon limits the growth of the gluon occupation numbers when decreasing $x$ (the longitudinal momentum fraction of a gluon with respect to its parent hadron) and/or increasing the nuclear mass number $A$ (when the parent hadron is a large nucleus with $A\gg 1$). Importantly, this phenomenon introduces an intrinsic transverse momentum scale, the hadron (nucleus) saturation momentum $Q_s$, which is proportional to the gluon density per unit transverse area and increases as a power of $1/x$ and also as a power of $A$: one roughly has $Q_s^2(x, A)\sim A^\delta(1/x)^\lambda$, with $\delta\simeq 1/3$ and $\lambda\simeq 0.2$. This scale controls the running of the coupling for phenomena like multiple scattering and gets transmitted to the particles produced in the final state. For sufficiently small $x$/large $A$, the saturation scale is much larger than the QCD confinement scale $\Lambda_{\rm QCD}\sim 0.2$~GeV, thus allowing for perturbative calculations from first principles.
The proper framework for performing such calculations is the  colour glass condensate (CGC) effective theory~\cite{Iancu:2002xk,Iancu:2003xm,Gelis:2010nm,Kovchegov:2012mbw}, which allows for the resummation of the high-gluon density effects via background field techniques.

Yet, for the phenomenologically interesting conditions (in particular at the EIC), the saturation scale is {\it semi-hard}, in the ballpark of 1~GeV for a Pb nucleus ($A\simeq 200$) at $x=10^{-2}$. Hence, the associated QCD coupling is not that small, $\alpha_s\sim 0.3$, and hadron production with transverse momenta of order $Q_s$ is strongly influenced by non-perturbative phenomena like hadronisation. This explains why (theoretical and experimental) studies of gluon saturation are rather complicated and this phenomenon is still elusive in the data: so far, it has not been {\it directly} observed, despite compelling and abundant indirect evidences in its favour. But the situation may change at the EIC, notably due to measurements of {\it diffractive} particle production, which may provide a smoking gun for gluon saturation.

Indeed, it has since long been appreciated (in the related context of electron-proton deep inelastic scattering at HERA) that the diffractive structure function at small\footnote{In the context of diffraction, the relevant value of $x$ is the longitudinal momentum fraction $\xP$ carried by the Pomeron; see  Sect.~\ref{sec:tree} for details.} $x$ is strongly sensitive to gluon saturation~\cite{Wusthoff:1997fz,GolecBiernat:1999qd,Hebecker:1997gp,Buchmuller:1998jv,Hautmann:1998xn,Hautmann:1999ui,
Hautmann:2000pw,Golec-Biernat:2001gyl},  including in the hard regime where the photon virtuality $Q^2$ is much larger than the hadron saturation momentum $Q_s^2(x)$. This is so since (coherent) diffraction --- a process in which the hadronic target emerges intact in the final state --- proceeds via elastic scattering and the respective cross-section is controlled by the strong scattering off the dense gluons at saturation. Recently, this conclusion has been extended to the diffractive production of hard jets, or hadrons: the dominant (``leading power'') contributions at large transverse momenta come from special partonic configurations which undergo strong scattering~\cite{Iancu:2021rup}. This opens the way to measurements of gluon saturation via {\it hard} collisions, which allow for controlled theoretical studies and clean experimental measurements.

On the theory side, a remarkable simplification which occurs for hard processes at small $x$ is the emergence of a special form of collinear factorisation, known as {\it transverse-momentum dependent (TMD) factorisation}, from the CGC effective theory. This applies to processes characterised by two widely separated transverse momentum scales, like the production of a pair of jets with relatively large transverse momenta $k_{1\perp},\,k_{2\perp}\gg Q_s$ and which propagate nearly back-to-back in the transverse plane; that is, their relative transverse momentum $P_\perp\equiv|\bk_{1\perp}-\bk_{2\perp}|/2$ is much larger than their momentum imbalance $K_\perp\equiv|\bk_{1\perp}+\bk_{2\perp}|$. For such processes and to leading power\footnote{Throughout this work, we shall assume that $P_\perp^2$ is comparable to, or larger than, the photon virtuality $Q^2$, hence $P_\perp$ is the hardest transverse momentum scale in the problem. In particular, all our results can be evaluated in the photo-production limit $Q^2\to 0$ --- this merely amounts to a simplification of the hard factor.}
 in $1/P_\perp$, the cross-section computed within the CGC approach to leading order in $\alpha_s$ can be written as the product of a ``hard factor'' (a partonic cross-section which encodes the dependence upon $P_\perp$) and a transverse-momentum dependent parton distribution function (succinctly referred as a ``TMD''), which describes the three-dimensional distribution --- in $x$ and $\bK_\perp$ --- of the partons from the target which are available for the collision. Such a factorisation was already familiar for hard processes  at {\it moderate} values of $x$ (say, $x\gg 10^{-2}$), where the hadronic target is dilute and the collision naturally involves a single parton from the target~\cite{Boer:2011fh,Collins:2011zzd,Boussarie:2023izj}. But in the recent years, it has been shown to also emerge from CGC calculations of particle production at small $x$, with the additional, remarkable, feature that, when $K_\perp$ is not much larger than $Q_s$, the TMDs also encode the effects of gluon saturation --- so, they include multi-parton correlations. Specifically, TMD factorisation has been demonstrated (at least, to leading order) for a variety of multi-scale processes at small $x$, which refer to both inclusive~\cite{Marquet:2009ca,Dominguez:2010xd,Dominguez:2011wm,Metz:2011wb,Dominguez:2011br,Xiao:2017yya,Marquet:2017xwy,Altinoluk:2024tyx,Caucal:2025xxh}
and diffractive~\cite{Iancu:2021rup,Hatta:2022lzj,Iancu:2022lcw,Hauksson:2024bvv} particle production. As a matter of fact, the validity of TMD factorisation for diffractive jet production in $eA$ collisions has been first recognised within the CGC approach. 

In this paper, we will focus on the diffractive production of a pair of nearly back-to-back jets which are initiated by a quark and an antiquark, respectively.  Within the CGC effective theory, this process is naturally computed using the colour dipole picture, valid in a frame in which the photon exchanged between the electron and the nucleus is ultrarelativistic and develops long-lived partonic fluctuations which subsequently scatter off the gluon distribution of the nuclear target. One may think that in order to produce a pair of jets at leading order it suffices to consider the quark-antiquark ($q\bar q$) fluctuation of the virtual photon, but for hard jets ($k_{1\perp}\simeq k_{2\perp}\simeq P_\perp \gg K_\perp,\, Q_s$) this is actually not the case~\cite{Iancu:2021rup}. A hard $q\bar q$ pair in a colour singlet state (a ``colour dipole'') has a very small transverse size $r\sim 1/P_\perp\ll 1/Q_s$ and hence it scatters only weakly. For diffraction, this is further penalised by the fact that the cross-section is proportional to the elastic amplitude squared. Under these circumstances, the leading-power contribution at large $P_\perp$ rather comes from ``(2+1)-\,jet'' configurations where the $q\bar q$ pair is accompanied by a semi-hard gluon ($g$), with transverse momentum $\kg\sim Q_s$: this $q\bar qg$ system has a relatively large transverse extent $R\sim 1/Q_s$ and thus suffers strong elastic scattering, with an amplitude of order one. The recoil of the gluon emission also fixes the $q\bar q$ di-jet imbalance: $K_\perp\simeq\kg$.

As previously mentioned, this process admits TMD factorisation~\cite{Iancu:2021rup}, with a {\it diffractive} gluon TMD expressing the gluon distribution inside the ``Pomeron'' --- the colourless exchange between the  $q\bar qg$ system and the nuclear target. While natural from a physics standpoint (given the hierarchy of transverse scales in the problem), the emergence of TMD factorisation from the CGC framework is still non-trivial, as shown by the following argument:  TMD factorisation  refers to a  {\it target} picture --- the gluon described by the DTMD is supposed to be emitted inside the wavefunction of the nuclear target (and more precisely by the Pomeron) ---, whereas the CGC calculations follow the evolution of the projectile --- the gluon is now emitted by the quark-antiquark pair. Hence, in order to demonstrate TMD factorisation in this framework, one must transfer the gluon emission from  the wavefunction of the projectile to that of the target. As shown e.g. in~\cite{Iancu:2022lcw,Caucal:2024bae} this is indeed possible so long as the gluon is soft with respect to (w.r.t.) the projectile --- i.e. it carries a small fraction $z_g\ll 1$ of the longitudinal momentum of the incoming photon.

The prominence of gluon saturation is visible in the fact that the (tree-level) gluon DTMD is roughly flat at $K_\perp \lesssim Q_s$ --- corresponding to gluon occupation numbers inside the Pomeron of order one ---, but it rapidly decreases at larger momenta  $K_\perp \gg Q_s$, like $1/\KT^4$~\cite{Iancu:2021rup,Hatta:2022lzj,Iancu:2022lcw}. Hence the bulk of the gluon distribution lies at saturation. The fact that the Pomeron exchange between the projectile and the target is colourless implies the existence of a rapidity gap $\YP=\ln(1/\xP)$ --- an angular region adjacent to the direction of propagation of the final nucleus which is void of particles in the final state. Here, $\xP$ represents the fraction of the target longitudinal momentum which is taken away by the Pomeron (and transferred to the $q\bar qg$ system). At tree-level, the gluon DTMD depends upon 3 variables: $\xP$, the gluon splitting fraction $x$ w.r.t. the Pomeron\footnote{In the literature on diffraction, this splitting fraction is often denoted as $\beta$; here however we prefer to use the notation $x$ since this is also the longitudinal variable concerned by the DGLAP evolution, as we shall see.}, and its transverse momentum $K_\perp$.

The fact that the tree-level tail $1/\KT^4$ decays much faster than the standard pQCD tail $1/\KT^2$ that would be produced by bremsstrahlung suggests that the large-$\KT$ behaviour of the diffractive TMD should be strongly sensitive to higher-order radiative corrections. This is confirmed by NLO calculations for related processes within the CGC effective theory:  when computed to leading power in $1/\PT$, the NLO corrections preserve TMD factorisation, but with additional contributions to the relevant TMD, which decay like $1/\KT^2$ at large $\KT$~\cite{Taels:2022tza,Caucal:2022ulg,Caucal:2024bae,Caucal:2024vbv,Hauksson:2024bvv}. Moreover, the NLO corrections include pieces enhanced by large kinematical logarithms that can be recognised as the beginning of several types of quantum evolutions. Specifically,  one finds \texttt{(i)} corrections enhanced by the rapidity logarithm $\ln(1/x)$ which signal the high-energy, BK/JIMWLK, evolution~\cite{Balitsky:1995ub,Kovchegov:1999yj,JalilianMarian:1997jx,JalilianMarian:1997gr,Kovner:2000pt,Weigert:2000gi,Iancu:2000hn,Iancu:2001ad,Ferreiro:2001qy}, \texttt{(ii)} corrections involving the double, $\ln^2(\PT^2/\KT^2)$, and single, $\ln(\PT^2/\KT^2)$, Sudakov logarithms~\cite{Mueller:2012uf,Mueller:2013wwa,Altinoluk:2024vgg,Caucal:2024vbv}, which are consistent with the Collins-Soper-Sterman (CSS) equation~\cite{Collins:1981uk,Collins:1981uw,Collins:1984kg,Collins:2011zzd}, and \texttt{(iii)}  corrections enhanced by the collinear logarithm $\ln(\KT^2/Q_s^2)$, which express one step in the  Dokshitzer-Gribov-Lipatov-Altarelli-Parisi (DGLAP) evolution~\cite{Gribov:1972ri,Altarelli:1977zs,Dokshitzer:1977sg}. As generally with the CGC approach, the NLO corrections have been computed as gluon emissions in the wavefunction of the projectile, but the resulting evolutions refer to the wavefunction of the nuclear target. 

For the particular problem of diffractive di-jet production, the phase-spaces for the various evolutions are controlled by the four variables $\xP$, $x$, $\KT$ and $\PT$. The high-energy evolution refers to the evolution of the Pomeron with decreasing $\xP$, or increasing the rapidity gap $\YP$. In practice, this amounts to solving the non-linear BK equation for a dipole scattering amplitude which enters the structure of the gluon DTMD at tree-level. The non-linear effects encoded in this equation reflect the high gluon occupancy within the Pomeron. The two collinear evolution equations, CSS and DGLAP, refer to the gluon exchanged between the Pomeron and the $q\bar q$ dipole. They are naturally linear since there is only one such a gluon: as already mentioned, the $q\bar q$ dipole has a small size $r\sim 1/P_\perp\ll 1/Q_s$, hence it scatters only once. For the same reason these equations are {\it universal} --- they do not depend upon the source of the exchanged gluon (here, the Pomeron) and would be the same if that gluon was emitted by a (dilute or dense) hadron. This explains why the standard  CSS and DGLAP equations, as familiar within the collinear factorisation at moderate $x$, are recovered in this small-$x$ context, where saturation effects are important.

Although the CSS and DGLAP equations take their standard form in the literature, the respective solutions ``know'' about saturation via their boundary or initial conditions, which are controlled by the tree-level gluon DTMD together with its BK evolution with decreasing $\xP$. The DGLAP evolution becomes important when the gluon transverse momentum $\KT$ is much larger than $Q_s$, in which case it provides the standard (``integrated'') diffractive parton distribution function (DPDF) for gluons with a given $x$ as measured with a resolution scale $\KT^2$. Importantly, the 
initial condition for this evolution (as formulated at some scale $\mu_0$ of order $Q_s$) 
is known from first principles, due to saturation. The CSS equation refers to the gluon DTMD. It is local in $x$ but non-local in $\KT$ and it resums the effects of gluon emissions with transverse momenta comprised between $\KT$ and $\PT$ --- so its solutions acquire a dependence upon the di-jet relative momentum $\PT$, which plays the role of a hard resolution scale. As we shall see, this scale controls both the transverse and the longitudinal resolutions for the gluon emissions concerned by the CSS evolution. This is related to the fact that the CSS equation which naturally emerges from the CGC calculations of di-jet production is a special (``diagonal'') version of the general CSS equation~\cite{Collins:2011zzd,Boussarie:2023izj}, in which the ultraviolet renormalisation scales  and the rapidity scales are identified with each other and with the hardest transverse momentum in the problem, which is $\PT$~\cite{Caucal:2024bae}. The CSS equation must be solved as a boundary value problem with the boundary condition formulated at a scale $P_\perp$ of order $\KT$. This boundary condition involves both the tree-level DTMD (including its BK evolution when appropriate) and the solution to the DGLAP equation. The above discussion also shows that the various evolutions are connected with each other. %, via the respective boundary/initial conditions.

The tree-level gluon diffractive TMD has been extensively studied in the literature~\cite{Iancu:2021rup,Hatta:2022lzj,Iancu:2022lcw}, including its high-energy (BK) evolution with increasing the rapidity gap $\YP$.  Also, one has presented numerical solutions to the DGLAP equation for the diffractive gluon PDF with first-principles initial conditions determined by saturation~\cite{Hauksson:2024bvv}. But the CSS evolution of the DTMD has never been studied from first principles. It is our main purpose in this paper to fill this gap. In this process, we shall clarify the interplay between the CSS and the DGLAP evolutions, the role played by gluon saturation in the context of these evolutions, and the relation between different representations for the CSS equation (transverse momentum vs. transverse coordinate; see below) --- an issue that has been only rarely addressed in the literature (see e.g.~\cite{Ellis:1997ii}). Our analysis will also shed light on the validity and the limitations of preliminaries studies of the diffractive TMD evolutions, like the resummation of Sudakov logarithms in Ref.~\cite{Shao:2024nor}, which relied on the traditional TMD techniques, as developed for {\it inclusive} processes at moderate values of $x$. As we shall discover, the case of diffraction at small $\xP$ is in fact quite special and in particular is better suited for first-principles calculations in pQCD. This is due to the enhanced role played by gluon saturation in the context of elastic scattering.

% It is the main purpose of this paper to fill this gap. 

As previously emphasised, the CSS evolution is generally connected to the other types of evolution (BK/JIMWLK and DGLAP),  via its boundary condition. To simplify the numerical problem and also to disentangle the effects of the various evolutions, we will ignore the effects of the high-energy evolution in this work. This is legitimate so long as the rapidity gap is not too large, such that $\alpha_s \YP\ll 1$. On the other hand, we cannot ignore the DGLAP evolution: as already explained, this becomes important when $\KT \gg Q_s$ and we would like to include this regime in our analysis (while preserving the condition $\KT\ll \PT$, of course).

We will employ three versions of the CSS equation: the ``$\bt$-\,space'', the ``$\KT$-\,space'', and the ``DLA'' respectively. The ``$\bt$-\,space'' is the version of the CSS equation in the transverse coordinate representation, as obtained after a Fourier transform from $\bK_\perp$ to $\bb_\perp$. This is the most widely used version of this equation in the traditional literature on TMD factorisation at moderate $x$~\cite{Boer:2011fh,Collins:2011zzd,Boussarie:2023izj}. It has the virtue that the CSS equation is local in $\bt$ (while it was non-local in $\bK_\perp$), which allows for simple analytic solutions. Yet, these solutions must be transformed back to $\KT$-\,space, which generally introduces complications: the inverse Fourier transform is sensitive to large values of $\bt$ where the QCD dynamics becomes non-perturbative. In general, this problem is solved by introducing prescriptions (like ``non-perturbative Sudakov factors'') which remove the contributions from very large $\bt$. Here however we shall find that such prescriptions are not necessary for the problem at hand: due to gluon saturation, the integral over $\bb_\perp$ is rapidly convergent --- the Fourier transform of the tree-level gluon DTMD decreases as the large power $1/\bt^8$ for $\bt \gg 1/Q_s$ --- and hence insensitive to non-perturbative phenomena. % Still for the $\bt$-\,space calculation, we will provide an original construction for the boundary condition, which combines saturation physics with the DGLAP evolution.

Albeit only rarely used in the literature (see however~\cite{Shi:2023ejp,vanHameren:2025hyo}), the $\KT$-\,space version of the CSS equation is the one where the evolution involves directly  the physical variable $\KT$. It is moreover the version which most naturally emerges (together with its boundary condition) from the CGC calculations at NLO~\cite{Caucal:2024bae,Caucal:2024vbv,Hauksson:2024bvv,Caucal:2025mth}. For these reasons, we shall start our  presentation in Sect.~\ref{sec:evol} with this equation, cf. Sect.~\ref{sec:CSS}. Then in Sect.~\ref{sec:bspace} we shall explicitly verify that, after a Fourier transform from $\bK_\perp$ to $\bb_\perp$, we indeed recover the expected equation in $\bt$-\,space. Yet, these two versions of the CSS equation are not strictly equivalent with each other, since they  must be solved as boundary value problems and the boundary conditions do not ``commute'' with the Fourier transform\footnote{A boundary condition is a constraint which is local in the relevant variable (either $\KT$, or $\bt$, depending upon the representation), but the Fourier transform is a non-local mapping between the two spaces. Hence, the boundary condition in $\bt$-\,space is not simply the Fourier transform of that in $\KT$-\,space; see also the discussion in Sect.~\ref{sec:bspace}.}. In what follows, we shall use corresponding approximations to separately construct the boundary conditions in the two representations, with results which are physically equivalent  to each other (to the accuracy of interest) and also numerically close, as we will later check in Sect.~\ref{sec:sols}. 

Finally, the DLA version of the CSS equation, to be discussed in Sect.~\ref{sec:dla}, is an approximate version of the equation in $\KT$-\,space which neglects transverse momentum conservation at the gluon splitting vertices: successive emissions are assumed to be strongly ordered in transverse momentum, as in the DGLAP evolution. This is a double-logarithmic approximation (DLA) since it faithfully captures only  radiative corrections enhanced by both longitudinal and transverse logarithms\footnote{As a matter of fact, this is the equation originally derived within the CGC approach in Refs.~\cite{Caucal:2024bae,Caucal:2024vbv,Hauksson:2024bvv}. The more precise  $\KT$-\,space equation with transverse momentum conservation has been only recently presented, in~\cite{Caucal:2025mth}.}.  While less precise than the two other versions of the CSS equation aforementioned, the DLA version has the advantage to allow for exact analytic solutions, that we shall present here for the first time. It moreover ensures a full compatibility between the CSS and the DGLAP evolutions: the DLA evolution for the TMD implies the DGLAP evolution for the PDF provided the latter is constructed as the integral of the TMD over the gluon transverse momentum up to the hard resolution scale~\cite{Caucal:2024bae} (see also  Refs.~\cite{Ebert:2022cku,delRio:2024vvq} for similar results within the TMD approach at moderate $x$).

In Sect.~\ref{sec:sols} we present our numerical solutions for the CSS and DGLAP equations. We consider all the three versions of the CSS equation aforementioned and discover that the respective predictions are remarkably close to each other. This in particular implies that the ambiguity associated with the choice of a representation ($\KT$-\,space versus $\bt$-\,space) is innocuous in practice. The numerical differences between the two respective solutions can be seen as a measure of the uncertainty in our predictions. The main effect of the CSS evolution is to shift the diffractive gluon distribution from the saturation plateau at $\KT\lesssim Q_s$ towards much larger momenta. The typical value of $\KT$ after the evolution is proportional to the hard scale $\PT$, but with a proportionality coefficient with is parametrically small at weak coupling. For instance,  at DLA  one finds $\langle \KT\rangle\simeq (2\alpha_sN_c/\pi)Q$. There is moreover an interesting interplay between the CSS and the DGLAP evolutions: the latter modifies the number of sources for emitting a gluon with a given  $\KT$, thus adding a contribution to the boundary condition for the CSS equation. Yet, it turns out that, for sufficiently large values of $\KT$ and/or $x$, this contribution becomes {\it negative} and so does the CSS solution for the gluon DTMD. This behaviour is clearly unphysical, as it would correspond to a negative cross-section. In our opinion, this problem reflects an intrinsic limitation of the TMD factorisation, which should be applied only for sufficiently small $\KT\ll \PT$ and  not too large $x$. The experience with TMD factorisation at moderate values of $x$ offers a strategy for overcoming this difficulty: when $\KT$ is relatively large (a sizeable fraction of the hard scale $\PT$), the contribution to the cross-section given by TMD factorisation must be supplemented with the so-called ``$Y$-term'', that is, the sum of the would-be suppressed power corrections (the perturbative corrections proportional to powers of $\KT^2/\PT^2$) as computed to fixed order in $\alpha_s$~\cite{Collins:2011zzd,Boussarie:2023izj}. As a first attempt in that sense, in Sect.~\ref{sec:sols} we present a partial resummation of power-suppressed corrections which indeed restores positivity for any $\KT\le \PT$ and for $x\le 0.5$.

%Here we cannot rely on this strategy since the NLO corrections to diffractive (2+1)-\,jet production have not been computed. That said, i

%It would be interesting to check whether this negativity problem gets postponed after including higher order corrections to the collinear evolutions.

\section{TMD factorisation for diffractive di-jets at tree-level}
\label{sec:tree}

%\subsection{Brief review of 2+1 jets with a soft gluon}
%\label{sec:soft_gluon}

In this section we shall briefly review the tree-level calculation of the cross-section for the diffractive production of a pair of back-to-back jets in the high-energy collision between an electron and a heavy nucleus with nuclear mass number $A\gg 1$. Our emphasis will be on the emergence of TMD factorisation from the CGC effective theory.

We work in a frame where the exchanged virtual photon $\gamma^*$ is an ultrarelativistic right mover whose 4-momentum in light-cone coordinates is $q^{\mu} = (q^+, q^- = -Q^2/2q^+,\bm{0}_{\perp})$ with $q^+ \gg Q$. The nucleus is a left mover with 4-momentum $P_N^{\mu} = (0,P_N^-,\bm{0}_{\perp})$ per nucleon, where the nucleon mass has been neglected. When $x_{\scriptscriptstyle \rm Bj}\equiv Q^2/(2q^+ P_N^-) \ll 1$, the coherence time $\tau_{\gamma} \sim 2 q^+/Q^2$ of the virtual photon is much larger than the longitudinal extend of the boosted nucleus and it becomes convenient to use the color dipole picture together with the Color Glass Condensate (CGC) effective theory. One expands the virtual photon wavefunction into its Fock space partonic components, which subsequently scatter eikonally off the nuclear target, i.e.~by keeping their transverse coordinates fixed. Such a scattering is described by a suitable Wilson line involving the color gauge field of the target. In turn, one must average over all possible target configurations using an appropriate weight-function, which at high energies and/or for large nuclei is given by the CGC. This takes into account the possibility that color fields get large, non-linear phenomena take place and gluon occupation numbers saturate. A scale, called the saturation momentum $Q_s$, is dynamically generated and the scattering becomes strong when the projectile partons probe modes of the nucleus which have transverse momenta of the order of, or smaller than, $Q_s$.   

At lowest order in the strong coupling $\alpha_s$, the Fock state component of the $\gamma^*$ wavefunction is an off-shell quark-antiquark ($q\bar{q}$) pair, which is eventually put on-shell after its scattering with the nucleus. Let $\bm{k}_1$ and $\bm{k}_2$ be the final transverse momenta of $q$ and $\bar{q}$ and $z_1 = k_1^+/q^+$ and $z_2=k_2^+/q^+=1-z_1$ the corresponding longitudinal momentum fractions. (Here and from now on, we remove the $\perp$ lower index on 2-dimensional vectors in the transverse space, to simplify notations.)
We shall be interested in the case in which the magnitude of the relative transverse momentum $\bm{P} \equiv z_2 \bm{k}_1 - z_1 \bm{k}_2$ is much larger than the saturation momentum $Q_s$. In coherent diffraction, i.e.~when the nucleus remains intact despite the collision, and if the nucleus is homogeneous, the transverse momentum transfer to the projectile system is of the order of $1/2R_A$, with $R_A$ the nuclear radius, and thus negligible. Thus $\bm{k_1} \simeq - \bm{k}_2 \simeq \bm{P}$, that is, the two forward jets are produced back-to-back in the transverse plane. The cross section for this process decreases rather fast, like $1/P_{\perp}^6$, due to color transparency: the $q\bar{q}$ pair has a small size $r \sim 1/P_{\perp}$, and hence interacts only weakly, approximately via the exchange of two gluons, with the target nucleus.

\begin{figure}
	\begin{center}
	\includegraphics[align=c,width=0.45\textwidth]{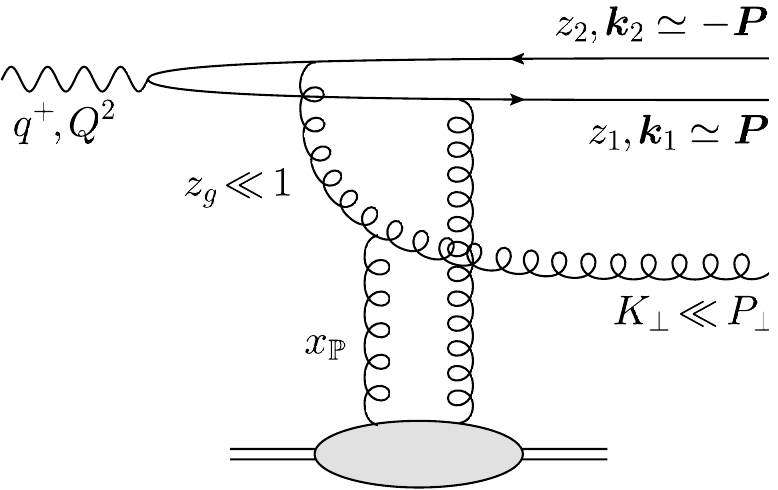}
	\hspace*{0.08\textwidth}
	\includegraphics[align=c,width=0.4\textwidth]{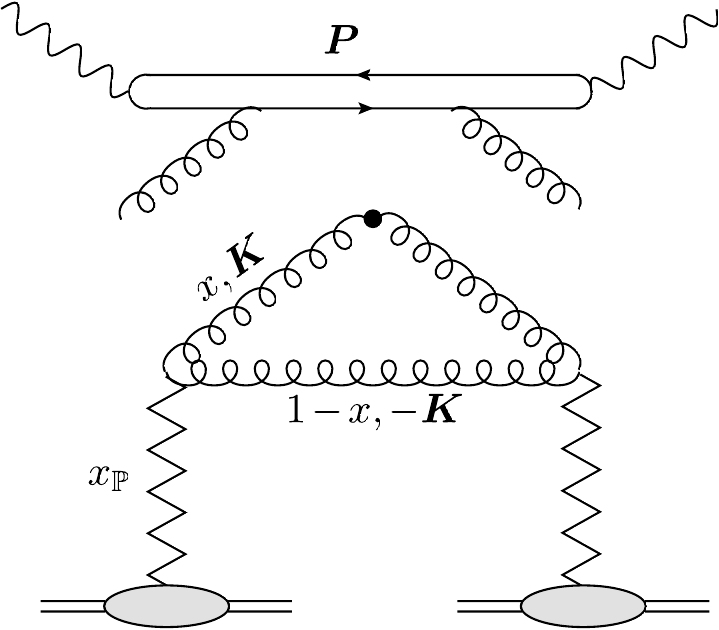}
	\end{center}
	\caption{\small Left panel: Diffractive scattering with a final state composed of a hard $q\bar{q}$ dijet and a semi-hard gluon. The gluon is emitted either from the quark  or from the antiquark and at a large  transverse distance $R\sim 1/K_\perp$  from the small $q\bar{q}$ pair (with transverse size $r\sim 1/P_\perp$). Right panel: the TMD factorisation of the di-jet cross-section. The semi-hard gluon is now radiated by the Pomeron represented by the zig-zag line.}
\label{fig:3jets}
\end{figure}

At next-to-leading order (NLO) in the QCD coupling, the $\gamma^*$ wavefunction includes a $q\bar{q}g$ fluctuation. Despite the fact that the corresponding cross section is suppressed by a factor of $\alpha_s$, a pair of hard $q\bar q$ jets with relative transverse momentum $P_{\perp}\gg Q_s$ can be produced in an efficient way, provided that the gluon carries a semi-hard transverse momentum of the order of $Q_s$. When this happens, the overall transverse size of the $q\bar{q}g$ configuration is quite large, $R\sim 1/Q_s$, and then the scattering is strong. (The left panel in Fig.~\ref{fig:3jets} shows a Feynman amplitude contributing to this 2+1 jet process.) Thus, although one produces a hard dijet in the final state, the suppression due to colour transparency is evaded and one finds that the cross section falls only like $1/P_{\perp}^4$~\cite{Iancu:2021rup}. A pair of hard quark-gluon jets can be similarly produced provided the antiquark is semi-hard~\cite{Hauksson:2024bvv}. For simplicity, here we shall consider only the first scenario: a hard $q\bar q$ di-jet accompanied by a semi-hard gluon. The gluon is not measured in the final state, yet its kinematics can be inferred from that of the hard di-jet, as we shall shortly argue.

Let $\bm{k}_g$ and $z_g\equiv k_g^+/q^+$ denote the gluon transverse momentum and its longitudinal momentum fraction w.r.t. the photon. The dominant contribution to the di-jet cross-section at large $P_\perp$ (the leading power) is obtained when the gluon is  semi-hard, i.e.~$k_{g\perp} \sim Q_s$, and also soft,
 in the sense that $z_g\sim k_{g\perp}^2/Q^2 \ll 1$. This can be easily inferred from a formation time argument: to interact with the nuclear target, the gluon formation time $\tau_g \sim 2 k_g^+/k_{g\perp}^2$ should comparable with the photon coherence time $\tau_{\gamma} \sim 2 q^+/Q^2$, which indeed yields $z_g \sim k_{g\perp}^2/Q^2$. Note that for parametric estimates we shall take $Q$ to be of the order of the relative di-jet momentum $P_{\perp}$ and $z_1\simeq 1-z_2$ not very different from 1/2. In practice though, our results remain valid for much smaller values $Q\ll P_\perp$ (including the photo-production limit $Q^2\to 0$) so long as the condition $P_{\perp}^2\gg Q_s^2$ remains satisfied.
 
 For coherent diffraction, the transverse momentum transfer from the target is negligible, hence the di-jet imbalance $\bm{K} \equiv \bm{k}_1+\bm{k}_2$ is controlled by the recoil of the gluon, $\bm{K} \simeq -\bm{k}_g$, and thus is itself semi-hard: $K_{\perp}  \sim Q_s \ll P_{\perp}$. Accordingly, the two hard jets propagate nearly back to back in the transverse plane. For most purposes, we can still write $\bm{k_1} \simeq - \bm{k}_2 \simeq \bm{P}$, as for the zeroth-order process where the gluon is absent.
 
 As discussed in the Introduction, this coherent process is characterised by a rapidity gap is $Y_{\mathbb P} = \ln(1/x_{\mathbb P})$, with $x_{\mathbb P}$ is the fraction of the (minus) longitudinal momentum $P_N^-$ of a nucleon from the target carried by the Pomeron (the colorless object exchanged between the target and the projectile). Its value is determined by the condition that the three final partons (the produced ``jets'' in this leading-order approximation) be on their mass-shell, that is,
  \begin{align}
 	\label{xpom}
 	x_{\mathbb{P}}=
 	\frac{M_{q\bar{q}g}^2+Q^2}{2q^+ P_N^-}
 	=
 	\frac{1}{2q^+ P_N^-}
 	\left(\frac{k^2_{1\perp}}{z_1} 
 	+\frac{k^2_{2\perp}}{z_2} 
 	+\frac{k^2_{g\perp}}{z_g} +Q^2\right) =
 	\frac{1}{2q^+ P_N^-}
 	\left( 
 	\frac{P_{\perp}^2}{z_1 z_2}+ \frac{k^2_{g\perp}}{z_g} + Q^2
 	\right),
 \end{align}
with  $M_{q\bar{q}g}^2$ the invariant mass squared  of the diffractive system.
Similarly, the ``minus'' longitudinal fraction transmitted to the hard $q\bar{q}$ pair only is
\begin{align}
	\label{xqq}
	 x_{q\bar q} =\frac{1}{2q^+P_N^-}
	 \left( 
 	\frac{P_{\perp}^2}{z_1 z_2}+ Q^2
 	\right).
\end{align}
This is smaller than, but comparable, to $x_{\mathbb P}$, as clear from the fact that $z_g \sim k_{g\perp}^2/Q^2$ and $Q^2\sim P_\perp^2$.

To unveil TMD factorisation for this diffractive process, it is necessary to move from the above picture for the projectile wavefunction to a target picture in which the semi-hard gluon is rather seen as a constituent of the Pomeron. To this end, it is convenient to make a change of variable from $z_g$ to $x$, the fraction of $x_{\mathbb P}$ transferred to the hard pair. This is uniquely determined in terms of the diffractive gap and the hard jets kinematics, more precisely
\begin{align}
	\label{x}
	 x= 
	 \frac{x_{q\bar q}}{x_{\mathbb P}}
	 %\beta\, \frac{x_{q\bar q}}{x_{\rm \scriptscriptstyle Bj}}
	 \quad \Longrightarrow \quad
	  z_g \simeq\frac{x}{1-x}\,\frac{\KT^2}{Q^2+ P_\perp^2/(z_1z_2)}\,,
	 %\,\simeq\,\beta\, \frac{\bar Q^2+P_\perp^2}{\bar Q^2}\,,
\end{align}
where we have also used Eqs.~\eqref{xpom}--\eqref{xqq} together with $\KT\simeq \kg$.
 As previously mentioned, $x_{q\bar q}$ and $\xP$ are comparable with each other for the interesting kinematics, so their ratio $x$ is neither very small, nor very close to one. 
 
 %Notice that, via \eqn{x}, moderate values $x\sim 1/2$ for the ``minus'' fraction w.r.t. the Pomeron correspond indeed to very small values $z_g \sim \KT^2/\PT^2$ for the ``plus'' fraction w.r.t. the photon projectile. This condition $z_g\ll 1$ is essential to be able to transfer a parton from the wavefunction of the projectile to that of the target~\cite{Iancu:2022lcw,Caucal:2024bae}.
 
From the perspective of the target, the process looks quite different: the Pomeron emits a pair of gluons in a colour singlet state and with zero total transverse momentum. The $t$-channel gluon with transverse momentum $\bm{K}$ and longitudinal fraction $x$  w.r.t. the Pomeron is  absorbed by the $q\bar{q}$ pair. The other, $s$-channel, gluon emerges in the final state with transverse momentum $\bm{k}_g = -\bm{K}$ and a longitudinal fraction $1-x$  w.r.t. the Pomeron. 

The clear separation of both transverse and longitudinal scales between the semi-hard gluon and the hard quark-antiquark pair leads to significant simplifications, and notably to the emergence of   TMD factorisation for the cross-section for the diffractive  production of the hard $q\bar q$ di-jet
~\cite{Iancu:2022lcw} (see the right  panel in Fig.~\ref{fig:3jets}):
\begin{align}
	\label{sigma2plus1}
	\frac{\rmd \sigma
	^{\gamma_{T}^* A
	\rightarrow q\bar q g A}}
  	{\rmd z_1
  	\rmd z_2
  	\rmd^{2}\!\bm{P}
  	\rmd^{2}\!\bm{K}
  	\rmd Y_{\mathbb P}} = 
  	{\alpha_{em}\alpha_s}
	\Big(\sum e_{f}^{2}\Big) \,
	\delta_z  
	\left(z_1^{2} + z_2^{2}\right)
	\frac{P_{\perp}^4 + \bar{Q}^4}
	{(P_{\perp}^2 + \bar{Q}^2)^4}\,
  	\mcal{F}^{(0)}_g(x, x_{\mathbb{P}}, K_\perp^2),
 \end{align}
 with $\bar Q^2\equiv z_1 z_2 Q^2$.
In the above we have employed the compact notation $\delta_z \equiv \delta(1-z_1-z_2)$, $e_f$ is the fractional electric charge of the quark with flavor $f$ and the strong coupling $\alpha_s$ should be evaluated at the hard scale $P_{\perp}$.  \eqn{sigma2plus1} refers to a photon with transverse polarisation; in the longitudinal case, one just has to replace  $(z_1^{2} + z_2^{2})(P_{\perp}^4 + \bar{Q}^4)$ $\to$ $8 z_1 z_2 P_{\perp}^2 \bar{Q}^2$ in the hard factor. 

The hard factor in \eqn{sigma2plus1} describes the formation of the hard $q\bar{q}$ pair and its coupling to the $t$-channel gluon emitted by the Pomeron. As anticipated, it exhibits a $1/P_{\perp}^4$ tail when $P_{\perp}^2$ and $\bar{Q}^2$ are of the same order. 
The semi-hard factor $\mcal{F}^{(0)}_g(x, x_{\mathbb{P}}, K_\perp^2)$ is the (tree-level estimate for the) gluon DTMD; that is, it stands for the number of gluons in the Pomeron which carry a transverse momentum $\bm{K}$ and a fraction $x$ of the Pomeron minus longitudinal momentum. It is given by~\cite{Iancu:2021rup,Iancu:2022lcw,Hatta:2022lzj} 
\begin{align}
	\label{Ftree}	
	\mcal{F}^{(0)}_g(x, x_{\mathbb{P}}, K_\perp^2)
	=
	\frac{S_\perp N_g}{4\pi^3}\,  \frac{
	[\mcal{G}_{\mathbb{P}}(x,x_{\mathbb P}, K_{\perp}^2)]^2}{2\pi(1-x)} 
\end{align}
where $S_{\perp}$ is the transverse area of the target assumed to be homogeneous, $N_c$ is the number of quark colors,  $N_g = N_c^2 - 1$ is the number of gluon colors and
\begin{align}
	\label{gp}
	\mcal{G}_{\mathbb{P}}(x,x_{\mathbb P}, K_{\perp}^2)=
	\mcal{M}^2 \int_0^\infty 
	\rmd R\, R\, 
	J_2(K_{\perp} R) 
	K_2(\mcal{M} R) \mcal{T}_g(R,Y_{\mathbb P})
	\quad \mathrm{with} \quad
	\mcal{M}^2 = \frac{x}{1-x}\,K_{\perp}^2.
\end{align}
The quantity $\mcal{M}^2$ is the gluon virtuality in the light-cone wavefunction of the virtual photon. Via the function $K_2(\mcal{M} R) $, it limits the separation $R$ between the gluon and the $q\bar q$ pair. Equivalently, it limits the transverse size of the effective gluon-gluon dipole built with the actual gluon and with the small $q\bar q$ pair ($r\sim 1/\PT$) which looks point-like on the transverse scale that is resolved by the scattering. The limit of a very soft gluon ($x\to 0$), which is quasi-real ($\mcal{M}^2\to 0$), is obtained by replacing $\mcal{M}^2 K_2(\mcal{M} R) \to 2/R^2$.

The whole information about the scattering is encoded in the  amplitude $\mcal{T}_g(R,Y_{\mathbb P})$ for the elastic scattering between the effective gluon-gluon dipole of size $R$ and the nucleus. This amplitude also contains (via its high-energy evolution) the whole dependence upon $x_{\mathbb P}$, hence upon the rapidity gap. For a large nucleus ($A\gg 1$) and a moderately large rapidity gap, such that $(\alpha_s N_c/\pi)\YP\ll 1$, one can neglect the effects of the high-energy evolution and estimate the amplitude $\mcal{T}_g(R)$ using the  McLerran-Venugopalan (MV) model~\cite{McLerran:1993ni,McLerran:1993ka}, which gives
\begin{align}
	\label{tgMV}
	\mcal{T}_g(R) = 
	1- \exp\!
	\left(
	-\frac{Q_A^2 R^2}{4}\ln\frac{4}{R^2 \Lambda^2}
	\right).
\end{align}
This involves two separate scales; the non-perturbative scale $\Lambda^2$, which is of the order of the QCD scale $\Lambda_{\rm QCD}^2$ and the scale $Q_A^2$ which is proportional to the transverse color charge density squared and thus grows like $A^{1/3}$. We often need to replace one of these two scales with the saturation momentum $Q_s^2$, defined through the requirement that $\mcal{T}_g(R=2/Q_s)$ is a number of the order of 1/2. For analytical estimates, it is convenient and customary to choose this number as $1-e^{-1}$, so that
\begin{align}
	\label{QsQA}
	Q_s^2 = Q_A^2 \ln \frac{Q_s^2}{\Lambda^2},
\end{align}
which in turn means that $Q_s^2$ must grow like $A^{1/3} \ln A$. It is also important to notice that the scale $Q_A^2$ corresponding to a $gg$ dipole is larger than the corresponding scale for a $q\bar q$ dipole by a factor $N_c/C_F$ (the ratio of the Casimir factors for the respective colour representations). The  amplitude in \eqn{tgMV} admits the following piecewise approximation 
\begin{align}
	\label{Tpiece}
	\mcal{T}_g(R)
	\simeq 
    \begin{cases}    
    \displaystyle{R^2 Q_s^2}
    & \quad \text{for \,\, $R\!\ll 1/Q_s$},
    \\*[0.1cm]
    \displaystyle{\ 1} &
    \quad \text{for \,\, $R\!\gtrsim 1/Q_s$},
    \end{cases}
\end{align}
where the first line corresponds to a small dipole which undergoes weak scattering (``colour transparency''), while the second one shows that a large dipole suffers strong scattering in the black disk limit.
Using the MV model for the dipole amplitude, one can easily deduce a similar, piecewise, approximation for the tree-level gluon DTMD; one finds
\begin{align}
	\label{Ftreelimits}
	\mcal{F}^{(0)}_g(x, x_{\mathbb{P}}, K_\perp^2)
	\simeq 
	\frac{S_\perp N_g}{4\pi^3}\,
	\frac{1-x}{2\pi}
    \begin{cases}    
    \displaystyle{1}
    &\quad \text{for} \quad
    K_\perp^2\!\ll \tilde Q_{s}^2(x, Y_{\mathbb P}),
    \\*[0.2cm]
    \displaystyle{\frac{\tilde Q_{s}^4(x)}{K_\perp^4}} 
    &\quad \text{for} \quad
    K_\perp^2\!\gg \tilde Q_{s}^2(x),
    \end{cases}
\end{align}
where $\tilde Q_{s}^2(x)= (1-x) \, Q_{s}^2$ is an effective saturation momentum which takes into account the effects of the gluon virtuality (the quantity $\mcal{M}^2$ in \eqn{gp})~\cite{Iancu:2021rup,Iancu:2022lcw}. It is generally smaller than $Q_{s}^2$ because, when $x$ is not too small, the gluon formation time is comparable to, or even larger than, the photon coherence time, so the gluon doesn't have enough time to arrive at a distance $R \sim 1/K_{\perp}$ from the hard pair by the time of scattering: its actual transverse separation is merely $R(x)$ with $R^2(x) \sim (1-x)/\KT^2$.
%This dependence has been studied in detail in Ref.~\cite{Iancu:2022lcw} via analytic arguments together with numerical solutions to the BK equation. Its main effect is the predict a power-like rise with $1/\xP$

% comes only via theIf $x_{\mathbb P}$ takes values in the ballpark of $x_0 \sim 10^{-2}$ one usually relies on the MV model to determine the amplitude, whereas for even smaller $x_{\mathbb P}$ one must evolve $\mcal{T}_g$ in the additional rapidity interval from $x_0$ down to $x_{\mathbb P}$ using the BK/JIMWLK equation (in their collinearly improved form). Whether or not evolution in $x_{\mathbb P}$ is taken into account, we shall refer to Eqs.~\eqref{Ftree} and \eqref{gp} as the tree level gluon DTMD and we denote it by the upper index (0). In particular, it does not depend on the hard scale $P_{\perp}^2 \sim Q^2$. In Sect.~\ref{sec:evol} we shall see how and why such a dependence can arise. 

\begin{figure}
	\begin{center}
		\includegraphics[width=0.45\textwidth]{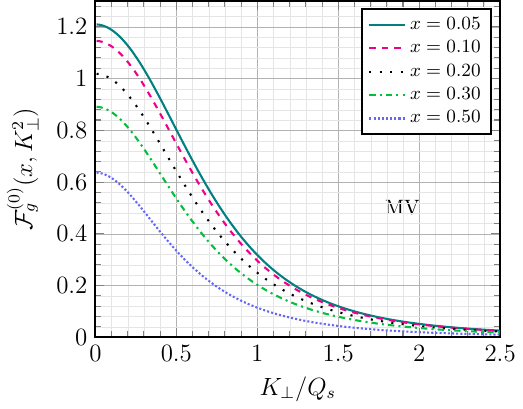}\qquad
		\includegraphics[width=0.45\textwidth]{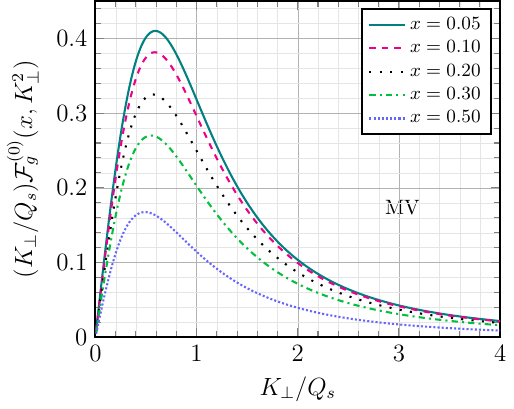}
	\end{center}
	\caption{\small Left: the tree-level gluon DTMD, as computed from Eqs.~\eqref{Ftree}--\eqref{gp} together with the MV model for the dipole amplitude $\mcal{T}_g(R)$ (with gluon saturation scale $Q_s^2=2$~GeV$^2$), is plotted as a function of the dimensionless ratio $\KT/Q_s$ for different values of $x$. (We omit an overall factor ${S_\perp}/{4\pi^3}$, so the plotted function $\mcal{F}^{(0)}_g(x, K_\perp^2)$ is dimensionless.)
	Right: the same as in the left plot, except that the gluon DTMD is now multiplied with the measure factor $\KT$, to better emphasise the tail at large $\KT\gg Q_s$. 
}
\label{fig:tree}
\end{figure}

\begin{figure}
	\begin{center}
		\includegraphics[width=0.45\textwidth]{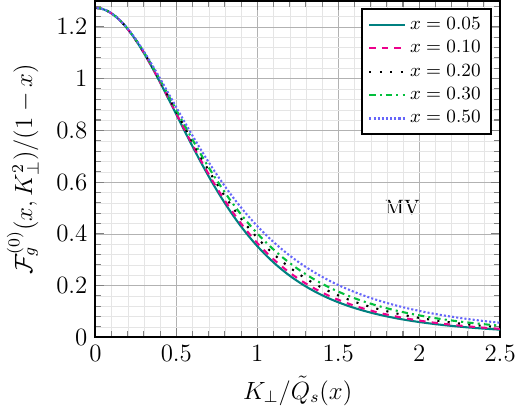}\qquad
		\includegraphics[width=0.45\textwidth]{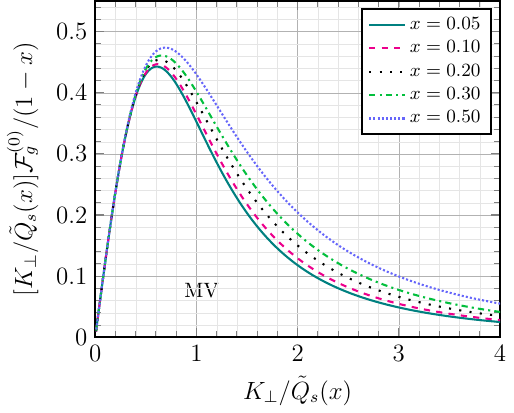}
	\end{center}
	\caption{\small The plots in Fig.~\ref{fig:tree} are re-drawn in such a way to emphasise the (approximate)
	scaling of the   tree-level gluon DTMD as a function of the variables $\KT$ and $x$.}
\label{fig:tree_scaled}
\end{figure}

The qualitative behaviour shown in \eqn{Ftreelimits} can be recognised in the numerical plots based Eqs.~\eqref{Ftree}--\eqref{gp} together with the MV model for the dipole amplitude $\mcal{T}_g(R)$, as shown in Fig.~\ref{fig:tree}.
 Interestingly, the piecewise approximation in \eqn{Ftreelimits} exhibits {\it geometric scaling}: the quantity $\mcal{F}^{(0)}_g(x, x_{\mathbb{P}}, K_\perp^2)/(1-x)$ depends upon the three variables $\KT$, $\xP$ and $x$ only via the scaling variable $\KT/\tilde Q_{s}^2(x, \YP)$. This scaling is only approximately respected by the complete result in  Eqs.~\eqref{Ftree}--\eqref{gp}: as shown by the plots in Fig.~\ref{fig:tree_scaled} for the case of the MV model (no $\xP$-dependence), the scaling is quite good in the saturation region at $\KT\lesssim \tilde Q_{s}(x, \YP)$, but it degrades at larger momenta  $\KT\gg \tilde Q_{s}^2(x, \YP)$.

Integrating over the transverse momentum up to a ``resolution'' scale $Q^2$  yields the tree-level approximation for the gluon diffractive PDF (DPDF):
\begin{align}
	\label{xGP}	
	xG^{(0)}(x, x_{\mathbb{P}}, Q^2)
	\equiv 
	\pi 	
	\int_0^{Q^2}
	\dif K_\perp^2\,
	\mcal{F}^{(0)}_g(x, x_{\mathbb{P}}, K_\perp^2).
\end{align} 
For a hard resolution scale $Q^2\gg  \tilde{Q}_s^2(x,Y_ {\mathbb P})$, the above integral is controlled by 
$K_\perp\lesssim  \tilde{Q}_s$ (so it is only weakly sensitive to its upper limit $Q^2$)
and is roughly of order $ \tilde{Q}_s^2(x,Y_ {\mathbb P})$, as clear from the piecewise estimate \eqref{Ftreelimits}. One can write~\cite{Iancu:2022lcw}
\begin{align}
	\label{xGPhigh}
	xG_{\mathbb{P}}(x, x_{\mathbb{P}}, Q^2\to \infty)
	=
	\frac{S_\perp N_g}{4\pi^3}\,
	\kappa(x)
	(1-x)\, \tilde{Q}_s^2(x,Y_ {\mathbb P}),
%	\quad
%	\mathrm{for}
%	\quad
%	P_{\perp} \gg \tilde{Q}_s(x,Y_ {\mathbb P}), 
\end{align} 
with $\kappa(x)$ a slowly varying function. So, roughly, the gluon DPDF at large $Q^2$ scales like $(1-x)^2$.

The aforementioned qualitative features of the diffractive gluon distributions (the DTMD and the DPDF) are preserved by the high-energy evolution, which in this case refers to the evolution of the dipole scattering amplitude $\mcal{T}_g(R,Y_{\mathbb P})$ with increasing $\YP$. This has been demonstrated via extensive numerical studies~\cite{Iancu:2022lcw,Hauksson:2024bvv} using the collinearly-improved version of the BK equation with running coupling~\cite{Beuf:2014uia,Iancu:2015vea,Ducloue:2019ezk,Boussarie:2025mzh}. The main effects of this evolution are the rise of the saturation momentum $Q_s(\YP)$ and a change in the power-law tail at large $\KT$, which is modified to  
$[\tilde{Q}_s^2(x,\YP)/K_{\perp}^2]^{2\gamma}$ with $0 <1-\gamma< 1/2$ an anomalous dimension ($\gamma \simeq 0.63$ for the leading logarithmic evolution and for asymptotically high $\YP$). 
In what follows, we will ignore the BK evolution and instead concentrate on the CSS evolution with increasing the hard transverse momentum scale, that we shall study here for the first time.

\section{CSS evolution: momentum versus coordinate representations}
\label{sec:evol}

 The leading order result for the gluon DTMD, as shown in Eqs.~\eqref{Ftree} and \eqref{gp}, is independent of the hard scale: the whole dependence of the cross section \eqref{sigma2plus1} upon $P_{\perp}^2 \sim Q^2$ is encoded in the hard factor. Starting with NLO, this hard scale also acts as a resolution scale for the radiative corrections, thus introducing a dependence on $Q^2$ in the gluon TMD,  to be denoted as $\mcal{F}_g(x, x_{\mathbb P},K_{\perp}^2,Q^2)$ from now on. Notice another important point on notations: henceforth we shall often denote the hard scale as $Q$ (rather than $\PT$). Indeed our subsequent study refers to the gluon DTMD {\it per se}, independently of the particular context of di-jet production that  motivates this study. But it is understood that, for diffractive di-jets, the {\it physical} hard scale is the relative momentum $\PT$ of the two jets, while the photon virtuality $Q^2$ can be of the same order as $\PT^2$, or (much)  smaller.
 
 As well understood in the literature (see Refs.~\cite{Mueller:2012uf,Mueller:2013wwa,Altinoluk:2024vgg,Caucal:2024vbv} for various processes at small-$x$), the dominant $Q^2$-dependence introduced by the NLO corrections to a TMD is that associated with the Sudakov (double and single) logarithms and can be resummed to all orders by solving the CSS equation. From the projectile perspective (within the colour dipole picture), the Sudakov logarithms are generated by gluon emissions in the final state, i.e.~after the collision, and at large angles, i.e.~outside the cones defining the two measured jets. In the target picture, where TMD factorisation is manifest, the CSS equation describes the change in the gluon TMD due to soft, unmeasured, gluon emissions in the $s$-channel. Such emissions cannot change the longitudinal momentum of the emitter, hence the CSS equation is local in the longitudinal  fraction $x$. But they affect its distribution in transverse momentum and, in general, the  CSS equation is non-local in $\bK$ (see Sect.~\ref{sec:CSS} below). Nevertheless, a {\it local} equation can be obtained by either making a Fourier transform  to the transverse coordinate representation ($\bK\to \bb$) --- this leads to the usual representation of the CSS equation, to be reviewed in Sect.~\ref{sec:bspace} --- or by staying in momentum space but working in a double logarithmic approximation, which assumes that successive gluon emissions are strongly ordered in transverse momenta, like for the DGLAP evolution --- this approximation will be discussed in Sect.~\ref{sec:dla}.
 
 %We begin in Sect.~\ref{sec:dla} by discussing this simpler scenario, to which we shall refer as double logarithmic approximation (DLA), since the logarithmic accuracy applies to both longitudinal and transverse momenta. This is interesting since, as we shall see, it allows for explicit analytic solutions with a transparent physical interpretation. Moreover it captures all the salient features of the more complete equation to be presented in Sect.~\ref{sec:CSS}, with which is also in very good agreement even at the quantitative level.

\subsection{CSS evolution in transverse momentum space}
\label{sec:CSS}

In this section we present the CSS equation for the gluon TMD in the transverse momentum representation, as emerging from NLO calculations of back-to-back di-jet production in the CGC effective theory. The relevant NLO calculations have been truly performed~\cite{Caucal:2024bae,Caucal:2025mth} for the case of {\it inclusive} di-jet production in $eA$ DIS, yet the ensuing equation is universal and it applies to {\it diffractive} di-jets as well --- that is, it also holds for the diffractive gluon TMD\footnote{For the case of the diffractive {\it quark} TMD, the CSS and DGLAP evolutions have been partially unveiled in Ref.~\cite{Hauksson:2024bvv}, via NLO calculations of the diffractive single inclusive jet production in DIS.}. This universality is well known in the traditional context of TMD factorisation at moderate values of $x$, where the coefficients of the CSS equation are related to the (gluon) cusp and PDF anomalous dimensions~\cite{Boussarie:2023izj}. It is also apparent in the CGC framework, where the CSS evolution is associated with {\it final-state} emissions of relatively soft gluons by the hard di-jets (the quark-antiquark pair). The calculation of these emissions is independent of the details of the scattering between the $q\bar q$ pair and the target. Their effects can be {\it a posteriori} interpreted in the target picture, as the CSS evolution of the gluon TMD --- regardless of the fact that this gluon was exchanged with the target as a whole (the case of  inclusive di-jets), or with the Pomeron (for diffractive di-jets). By the same token, the CSS equation is independent of high-gluon density effects, like gluon saturation. This is quite obvious in the projectile picture, where the gluon is emitted after the collision with the nuclear target, but it may look surprising from the viewpoint of the target picture, where saturation effects play a prominent role for the diffractive di-jet production already at tree-level. Yet, in this particular context, the saturation effects refer to the internal structure of the Pomeron, hence they matter for the high-energy (BK/JIMWLK) evolution with decreasing  $x_{\mathbb P}$,  but not also for the CSS and the DGLAP evolutions to be discussed in what follows. Indeed, the collinear evolutions rather refer to the unique gluon exchanged (in the target picture) between the Pomeron and the $q\bar q$ projectile, so they are naturally {\it linear}. Saturation effects are still important, but they enter these evolutions only via initial or boundary conditions.

To motivate the structure of the CSS equation in transverse momentum space ($\KT$\,-space), let us briefly review the underlying NLO calculation (see~\cite{Caucal:2024bae,Caucal:2025mth} for more details).
Soft gluon emissions in the final state are relatively simple in that they factorise from the rest of the amplitude, hence they quasi-automatically preserve TMD factorisation, with the same hard factor as at tree-level, cf. \eqn{sigma2plus1}. Their effect at NLO is to add a contribution to the gluon diffractive TMD, which reads
 \begin{align}
	\label{DeltaFS}
\Delta \mcal{F}_g^{\rm fin}(x, \xP, \bK, Q^2)=\,&\,\frac{\alpha_sN_c}{\pi}\int\frac{\rmd^2\bell}{\pi\lt^2}\,
\int_{z_m}^{z_M}\frac{\rmd z_\ell}{ z_\ell}
\nonumber\\*[0.1cm]&\,
\left\{\mcal{F}_g^{(0)}\left(x^\prime( z_\ell),  \xP, \bK+\bell\right) - \Theta(Q^2-\lt^2)\mcal{F}_g^{(0)}(x, \xP, \bK)\right\}
 \,. \end{align}
 In this equation $\bell$ and $ z_\ell=\ell^+/q^+$ denote the transverse momentum of the additional gluon and its longitudinal momentum fraction w.r.t. to the photon. (This additional gluon must be distinguished from that appearing at tree-level, that has been transferred to the target and is implicitly included in the leading order DTMD $\mcal{F}_g^{(0)}$.) As before, $ x= {x_{q\bar q}}/{x_{\mathbb P}}$ denotes the fraction of the Pomeron ``minus'' longitudinal momentum taken by the final $q\bar q$ pair. The first term within the accolades represents the contribution of a real gluon emission: the respective final state  involves the quark-antiquark-gluon system\footnote{In this section, we use the symbol $\ell$ (instead of $g$) to denote the gluon associated with NLO corrections to avoid any confusion with the gluon $g$ that was emitted already at tree-level, to construct the 2+1 jet configuration, cf. Sect.~\ref{sec:tree}.}
 $q\bar q \ell$, to which the Pomeron must transfer a transverse momentum $\bK+\bell$ and a ``minus''  longitudinal fraction $x'$, evaluated as
 \begin{align}
	\label{xxi}
	 x'=  \frac{x_{q\bar q\ell }}{x_{\mathbb P}}= x \,\frac{x_{q\bar q\ell }}
	{x_{q\bar q}}\simeq x \left(1+\frac{\lt^2/z_\ell}{Q^2+P_\perp^2/(z_1z_2)}\right)\equiv\frac{x}{\xi}
	\,,
\end{align}
where we have also used the fact that $z_\ell\ll 1$ (see below), hence $z_1+z_2\simeq 1$.  
The second term  within the accolades in \eqn{DeltaFS}, with a negative sign, refers to virtual gluon emissions: the respective final state involves just the $q\bar q$ pair. 

The variable $\xi$ defined by the last equality in \eqn{xxi} has a natural interpretation in the target picture, where the gluon with  transverse momentum  $\bell$ is seen as being produced via a splitting $g\to gg$ of the gluon exchanged in the $t$-channel between the  $q\bar q$ pair and the Pomeron: from that perspective, $\xi=x/x'$ represents the splitting fraction taken by the $t$-channel gluon.

 The lower limit $z_m$ on  the integral over $ z_\ell$  in \eqn{DeltaFS} follows from the obvious kinematical constraint $x' < 1$, that can be equivalently written as $\xi > x$. Specifically, \eqn{xxi} implies
 \begin{align}
	\label{zxi}
z_\ell = \frac{\xi}{1-\xi}\frac{\lt^2}{Q^2+P_\perp^2/(z_1z_2)}\,>\,z_m\equiv 
\frac{x}{1-x}\frac{\lt^2}{Q^2+P_\perp^2/(z_1z_2)}\,\sim\,\frac{\lt^2}{Q^2}\, ,\end{align}
where the last estimate for $z_m$ holds parametrically. (Recall that we work under the assumptions that $P_\perp^2\sim Q^2$ and none of the longitudinal fractions $x$, $z_1$ and $z_2$ is parametrically small.)

The upper limit $z_M$ has a more profound physical origin: it comes from the condition that the gluon ($z_\ell,\,\bell$) be emitted at large angles, outside the jets associated with the measured  $q\bar q$ pair\footnote{This condition of large-angle emissions has been implicitly used in constructing the overall colour factor $N_c$ in \eqn{DeltaFS}: this arises as  $N_c=2C_F+ 1/N_c$, with the 2$C_F$ piece coming from direct emissions by the quark and the antiquark, and the $1/N_c$ from the interference terms.}. Indeed, gluon emissions inside the jets cannot change the structure of the final state, so they do not contribute to the di-jet cross-section. (In the NLO calculation, their effects cancel between real and virtual corrections.) In the small angle approximation, we can evaluate the angles made by the jets and by the gluon w.r.t. the collision axis as $\theta_{1,2}\simeq P_\perp/(z_{1,2}q^+)$ and, respectively, $\theta_\ell\simeq \lt/(z_{\ell}q^+)$. Then the condition $\theta_\ell \gtrsim \theta_{1,2}$ implies, parametrically, $z_{\ell}\lesssim   \lt/P_\perp \sim \lt/Q$, leading to $z_M\simeq\lt/Q\ll 1$. As we shall see, the typical values of $\lt$ contributing to the integral in \eqn{DeltaFS} obey $K_\perp\lesssim \lt \ll Q$. This underscores the importance of the condition $K_\perp^2\ll Q^2$ for the validity of our present approximations. This is moreover consistent with our focus on hard, back-to-back, di-jets.

At this point, it should be clear that the integral over $z_{\ell}$ in \eqn{DeltaFS} has a logarithmic domain at $
 \lt^2/Q^2\ll z_{\ell}\ll  \lt/Q$. Within this domain, one can approximate $x'\simeq x$ (or $\xi\simeq 1$) and  \eqn{DeltaFS} yields the following, leading-logarithmic, contribution (here, in the sense of a longitudinal, or rapidity, logarithm):
 \begin{align}
	\label{DeltaCSS}
\Delta \mcal{F}_g^{\rm CSS}(x, \xP, \bK, Q^2)=\,&\,\frac{\alpha_sN_c}{2\pi}
\int\frac{\rmd^2\bell}{\pi\lt^2}\,\ln\frac{Q^2}{\ell_\perp^2}
%\nonumber\\*[0.1cm]&\,
\left\{\mcal{F}_g^{(0)}\left(x,  \xP, \bK+\bell\right) - \Theta(Q^2-\lt^2)\mcal{F}_g^{(0)}(x, \xP, \bK)\right\}
 \,. \end{align}
 For the physical interpretation of this result, it is useful to observe that the aformentioned logarithmic domain in $z_{\ell}$ corresponds to $1 \gg 1-\xi \gg \xi_0\equiv \lt/Q$ in terms of the $\xi$ variable. Within the target picture, $1-\xi$ is the splitting fraction of the unmeasured gluon emitted in the $s$-channel. Hence, \eqn{DeltaCSS} expresses the effect of a very soft gluon emission in the $s$-channel, which can modify the 
 transverse momentum of the gluon measured in the $t$-channel, but leaves unchanged its  (minus) longitudinal momentum. This is the essence of the CSS evolution,
 
  \eqn{DeltaCSS} can be promoted into an evolution equation by taking a derivative w.r.t. the logarithm $\ln Q^2$. More precisely, the complete version of the CSS equation emerging from the CGC approach to the order of interest reads
\begin{align}
	\label{CSSKT}
	\frac{\del \mcal{F}_g(x, \xP,  \bK, Q^2)}{\del \ln Q^2}
	= &\,\frac{N_c}{2\pi}
	\int\frac{\rmd^2\bell}{\pi\lt^2}\,\alpha_s(\lt^2)
	\Big[\mcal{F}_g(x, \xP, \bK+\bell, Q^2) - 
	\Theta(Q-\lt)\mcal{F}_g(x, \xP, \bK, Q^2)\Big]
	\nonumber\\*[0.2cm]
	& + % \Theta(Q,\mu_0)
	\beta_0\,\frac{\alpha_s(Q^2)N_c}{\pi}\,
	\mcal{F}_g(x, \xP, \bK, Q^2)\,.
\end{align}
As compared to \eqn{DeltaCSS}, this also includes running coupling effects~\cite{Caucal:2024bae}, that are visible at two levels: the scale dependence of the QCD coupling $\alpha_s$, as evaluated at one-loop level,
\begin{align}
	\label{runcoup}
	\alpha_s(Q^2) = \frac{\pi}{\beta_0 N_c \ln (Q^2/\Lambda^2_{\rm QCD})}
	\qquad \textrm{with} \qquad
	\beta_0=\frac{11 N_c - 2 N_f}{12 N_c}
\end{align}
(with $N_f$ the number of active flavours)
 and the additional term explicitly proportional to $\beta_0$, which expresses the anomalous dimension of the gluon PDF.  This additional $\beta_0$--piece is not naturally generated by the NLO calculation
in the dipole picture \cite{Taels:2022tza,Caucal:2023nci,Caucal:2023fsf}, because of the classical approximation used for the scattering~\cite{Xiao:2017yya,Hentschinski:2021lsh}. In the target picture, this piece is well known to arise via one-loop  corrections to the gluon exchange in the $t$-channel  \cite{Ayala:1995hx,Zhou:2018lfq,Mueller:2018llt}. 
 
 %Notice that the scale in the running coupling in the various terms in Eq.~\eqref{CSSKT} is determined by t

%where now $\bK + \bell$ is the momentum of the parent gluon. The latter splits to the ``measured'' gluon with momentum $\bK$ and to the unmeasured one (i.e.~the one in the $s$-channel for the real term or in the loop for the virtual term) with momentum $\bell$. Like with the DLA version, this equation is valid only for $Q^2\gg \KT^2$ and must be solved with a BC at $Q^2= \KT^2$, cf.~\eqn{IC}. \red{[Numerically we solve this CSS equation with ``modified steps'', which is equivalent to multiplying the first term in the above with $\Theta(Q,\mu_0)$, with the latter given in Eq.~\eqref{theta_smooth}].} 

Eq.~\eqref{CSSKT} is well-defined both in the infrared and in the ultraviolet. Indeed, at  low momenta $\lt^2 \ll \KT^2$, real and virtual contributions cancel other since they are approximately weighted by the same TMD $\mcal{F}_g(x,\KT^2,Q^2)$. Besides, for the real emissions with very large momenta $\lt^2 \gg \KT^2$ we have $\mcal{F}_g(x, \bK+\bell, Q^2)\simeq \mcal{F}_g(x, \bell, Q^2)$ and we recall that the gluon DTMD falls like a power of $1/\lt$ for $\lt^2 \gg Q_s^2$. This power is four at tree-level, it reduces to two after including DGLAP and/or CSS evolutions, and it can be further modified when including the BK/JIMWLK evolution. It is clear that the ensuing integral over $\lt^2$ is convergent, independently of the exact aforementioned power fall-off.

One can gain some more insight into the physical consequences of the CSS equation~\eqref{CSSKT}  by dividing the virtual emissions in two regimes, $\lt^2 < \KT^2$ and $\lt^2 > \KT^2$, and then combining the former with the real term; we thus obtain
\begin{align}
	\label{CSSKT2}
	\hspace{-0.5cm}
	\frac{\del \mcal{F}_g(x, \bK, Q^2)}{\del \ln Q^2}=\, &
	\frac{N_c}{2\pi}
	\int\frac{\rmd^2\bell}{\pi\lt^2}\,\alpha_s(\lt^2)
	\Big[\mcal{F}_g(x, \bK+\bell, Q^2) - 
	\Theta(\KT^2-\lt^2)\mcal{F}_g(x, \bK, Q^2)\Big]
	\nonumber\\*[0.2cm]
	&
	-\frac{N_c}{2\pi}
	\int_{\KT^2}^{Q^2}
	\frac{\rmd \lt^2}{\lt^2}\,
	\alpha_s(\lt^2)\,
	\mcal{F}_g(x, \bK, Q^2)
	+%\Theta(Q,\mu_0)
	\beta_0\frac{\alpha_s(Q^2)N_c}{\pi}\mcal{F}_g(x, \bK, Q^2).
\end{align}
The first line is still infrared finite and is reminiscent of BFKL dynamics, something which perhaps should not come as a surprise, since both equations describe an evolution associated with soft gluon emissions and such that the  transverse momentum is exactly conserved at the emission vertex. The first term in the second line, which contains only the virtual emissions in which the momentum $\KT$ of the $t$-channel gluon is larger than the momentum $\lt$ of the $s$-channel one, generates the double Sudakov logarithms in the solution to \eqn{CSSKT2}. The second term, proportional to $\beta_0$ , gives rise to single Sudakov logarithms. These features will become fully explicit in Sect.~\ref{sec:dla}, where we will consider a suitable, double-logarithmic, approximation to the CSS equation, which allows for exact analytic solutions.

 Even though the potential infrared divergencies due to very soft gluon emissions cancel between real and virtual terms, we still face the issue of the Landau pole in Eqs.~\eqref{CSSKT} and \eqref{CSSKT2}. We can regularize such an infrared singularity by freezing the coupling at a scale above $\Lambda_{\rm QCD}^2$. A suitable way to do this is to let $\lt^2 \to \lt^2 + m_0^2$, with $m_0^2 > \Lambda_{\rm QCD}^2$, in the argument of $\alpha_s(\lt^2)$, so that Eqs.~\eqref{CSSKT}, \eqref{CSSKT2} and those to appear in the following sections are well-defined. At a first glance this seems to introduce an additional uncertainty, however our results should not depend on the precise choice of $m_0^2$, so long as $m_0^2 \ll Q_s^2$. Indeed, it is easy to check that gluon emissions with very low transverse momenta $\lt\ll Q_s$ do not contribute to the CSS evolution, Eq.~\eqref{CSSKT}, even when the external momentum $\KT$ is itself very small ($\KT\ll Q_s$); see also the discussion in Sect.~\ref{sec:dla}.

\subsection{The boundary condition for the CSS equation: adding the DGLAP evolution}
\label{sec:BC}

Eq.~\eqref{CSSKT} is a first order differential equation, hence it needs to be supplemented with an additional (initial or boundary) condition  in order for its solution to be uniquely defined. As repeatedly emphasised, this equation describes an evolution with increasing $Q^2$ above $\KT^2$, hence it looks natural to specify the  ``value'' of the solution (actually, its functional form) at $Q^2=\KT^2$. Since moreover $\KT$ is itself a running scale, it should be clear that this procedure amounts to imposing a {\it boundary condition} (BC), rather than an {\it initial} one.  We shall generically write this BC as
\beq\label{BC}
\mcal{F}_g(x, \xP,  \bK, Q^2=\KT^2) = \mcal{F}_0(x, \xP, \bK)\,.\eeq
Then the main question is, what should we use for the boundary function $\mcal{F}_0(x, \xP, \bK)$? To the leading-order accuracy of interest, it may look natural to identify  $\mcal{F}_0$ with the tree-level expression $\mcal{F}_g^{(0)}$ discussed in Sect.~\ref{sec:tree}. Yet, that would be correct only in the absence of other evolutions, which themselves matter at leading order. We already know that for a sufficiently large rapidity gap $\alpha_s\ln(1/\xP)\gtrsim 1$, one needs to take into account the high-energy (BK/JIMWLK) evolution of the Pomeron. The effects of this evolution on the diffractive TMDs have been discussed at length in previous publications~\cite{Iancu:2022lcw,Hauksson:2024bvv} (and briefly reviewed in Sect.~\ref{sec:tree}), and will be ignored in what follows. This is strictly correct so long as $\xP$ is not {\it too} small, say in the ballpark of $10^{-2}$, where one can rely on the MV model to estimate the dipole amplitude in \eqn{gp}. Under this assumption, the gluon DTMD becomes independent of $\xP$. That said, most of the subsequent results hold for generic values of $\xP$, including very small values for which the BK/JIMWLK evolution {\it is} important and the boundary condition $\mcal{F}_0(x, \xP, \bK)$ has a non-trivial dependence upon $\xP$. Indeed, in
the context of the CSS evolution, the variable $\xP$ merely plays the role of an external parameter, that will be often omitted  to simplify writing.

%\footnote{We temporarily restore the diffractive variable $\xP$ in our notations, for more generality: \eqn{F0} also holds at very small $\xP$, where the $\xP$-dependence introduced by the BK/JIMWLK evolution becomes important.} 

Yet, there is another type of quantum evolution which is generally important for the di-jet problem at hand: when the di-jet imbalance $\KT$ is relatively hard, within the range $Q_s \ll \KT\ll Q\sim \PT$, one finds 
NLO corrections enhanced by the transverse logarithm $\ln(\KT^2/Q_s^2)$, which signal the DGLAP evolution of the gluon PDF $xG(x, \KT^2)$. As shown for various examples in Refs.~\cite{Hauksson:2024bvv,Caucal:2024bae,Caucal:2025mth,Caucal:2024vbv}, these corrections can be resummed  to all orders by including the DGLAP evolution in the boundary condition for the CSS equation. 
%A similar strategy is used in the more traditional context  of TMD factorisation at moderate values of $x$, although the respective analysis is generally performed in the transverse {\it coordinate} space~\cite{Boussarie:2023izj} (see also the next section).
Like for the CSS evolution discussed in the previous section, the NLO corrections responsible for the DGLAP evolution are first computed in the {\it projectile} picture, as soft ($z_\ell\ll 1$) gluon emissions by the partons ($q,\bar q$ and $g$) from the photon LCWF, and then reinterpreted in the target picture, as a (hard) splitting of the gluon exchanged in the $t$-channel between the Pomeron and the $q\bar q$ pair. But unlike for the CSS evolution, the derivation of the DGLAP evolution also involves gluon emissions occurring in the {\it initial} state (prior to the collision with the shock wave). Moreover these two evolutions are based on different approximations. To isolate the DGLAP evolution, one needs to consider gluon emissions with relatively large transverse momenta, $\KT\simeq \lt\gg Q_s$, whose recoil controls the di-jet imbalance: $\bK \simeq -\bell$. This kinematics allows one to simplify the transverse momentum structure of the emission vertex (in particular, one can neglect saturation effects), but the ``minus'' longitudinal momentum must be exactly conserved: the emitted gluon is ``soft'' w.r.t. the projectile (as we shall see, it typically has $z_\ell\sim \KT^2/Q^2$), but its ``minus'' splitting fraction $\xi$ can take generic values (recall \eqn{zxi}).

Within the CGC approach, the interplay between the DGLAP and the CSS evolutions has been explicitly demonstrated only for {\it inclusive} di-jets~\cite{Caucal:2024bae}. The corresponding argument for {\it diffractive} (2+1)-\,jets is {\it a priori} more complicated because of the presence of an additional gluon already at tree-level (cf. Sect.~\ref{sec:tree}). This ``tree-level'' gluon not only can act as an additional source for the ``quantum'' gluon, but it also introduces new topologies for the  gluon emissions by the quark, or the anti-quark. A complete proof of the DGLAP evolution in this diffractive context goes beyond our preseent purposes. Yet, we shall present a partial proof, which refers to gluon emissions in the final state alone and reproduces the singular part of the relevant DGLAP splitting function. This is based on the NLO contribution presented in \eqn{DeltaFS}, which also contains a DGLAP piece, as we now explain.

To start with, it is useful to change the longitudinal integration variable, from the projectile-oriented variable $z_\ell$ to the target-oriented variable $\xi$, with the help of \eqn{zxi}. One finds 
\begin{align}
	\label{DeltaFxi}
\Delta \mcal{F}_g^{\rm fin}(x, \bK, Q^2)=\,&\,\frac{\alpha_sN_c}{\pi}\int\frac{\rmd^2\bell}{\pi\lt^2}\,
\int_{x}^{1-\xi_0}\rmd\xi\left(\frac{1}{1-\xi}+\frac{1}{\xi}\right)
\nonumber\\*[0.1cm]&\quad
\left\{\mcal{F}_g^{(0)}\left(\frac{x}{\xi},  \bK+\bell\right) - \Theta(Q^2-\lt^2)\mcal{F}_g^{(0)}(x, \bK)\right\}
 \,, \end{align}
 Here $\xi_0\equiv \lt/Q\ll 1$ and in writing the upper limit as $1-\xi_0$ we neglected terms suppressed by higher powers of $\lt/Q$, which were anyway not under control in  \eqn{DeltaFS}. The rapidity regulator $\xi_0$ is only needed for the term proportional to $1/(1-\xi)$, whose $\xi$-integral can be evaluated as (for the real piece only and with simplified notations)
 \begin{align}\label{plus}
	\int_{x}^{1-\xi_0}\frac{\rmd\xi}{1-\xi}\,\mcal{F}_g^{(0)}\left(\frac{x}{\xi}\right)&\, = \ln\frac{1}{\xi_0}\,\mcal{F}_g^{(0)}(x)+\int_{0}^{1-\xi_0}\frac{\rmd\xi}{1-\xi}\,\left(\Theta(\xi-x)\mcal{F}_g^{(0)}\left(\frac{x}{\xi}\right)
	- \mcal{F}_g^{(0)}(x)\right)
	\nonumber\\*[0.1cm]&\,\simeq\,\ln\frac{1}{\xi_0}\,\mcal{F}_g^{(0)}(x)+\int_{x}^{1}
	\frac{\rmd\xi}{(1-\xi)_+}\,\mcal{F}_g^{(0)}\left(\frac{x}{\xi}\right).
	 \end{align}
	 In writing the second line, we have used the definition of the ``plus'' prescription and neglected power-suppressed corrections, i.e. terms of higher order in $\xi_0$. The term proportional to $\ln(1/\xi_0)$ can be now combined with the respective contribution of the virtual piece to recover the CSS equation \eqref{CSSKT}. After these operations, the remaining part from the original equation \eqref{DeltaFxi}
reads as follows
 \begin{align}
	\label{DeltaFDG}
\Delta \mcal{F}_{_{\rm DGLAP}}^{\rm fin}(x, \bK)=\,&\,\frac{\alpha_sN_c}{\pi}\int\frac{\rmd^2\bell}{\pi\lt^2}\,
\int_{x}^{1}
	\frac{\rmd\xi}{\xi(1-\xi)_+}\,\mcal{F}_g^{(0)}\left(\frac{x}{\xi}, \bK+\bell\right)
	 \,.\end{align}
As suggested by its notation, this particular NLO correction includes a piece of the DGLAP evolution. This can be isolated by considering the regime where $\lt\simeq \KT$ and $k_{g\perp}\equiv |\bK+\bell|\ll \KT$. Clearly, this regime exists only when $\KT$ is much larger than $Q_s$. (Indeed, $k_{g\perp}$ is the momentum transferred by the Pomeron, hence it is typically of order $Q_s$.) In this regime, one can approximate \eqn{DeltaFDG} as
\begin{align}
	\label{DeltaFrecoil}
\Delta \mcal{F}_{_{\rm DGLAP}}^{\rm fin}(x, \bK)&=\,\frac{\alpha_sN_c}{\pi}\,\frac{1}{\KT^2}
%\Theta\left(\frac{1-x}{x}-\xi_0\right)
\int_{\Lambda^2}^{\KT^2} \rmd k_{g\perp}^2
\int_{x}^{1}\frac{\rmd\xi}{\xi(1-\xi)_+}\,
\mcal{F}_g^{(0)}\left(\frac{x}{\xi}, k_{g\perp}\right) 
 \nonumber\\*[0.1cm]
&\,\simeq \frac{\alpha_s}{2\pi^2}\,\frac{1}{\KT^2} \int_{x}^1\rmd\xi\,\frac{2N_c}{\xi(1-\xi)_+}\,\frac{x}{\xi} 
G^{(0)}\left(\frac{x}{\xi}, \KT^2\right),
 \end{align}
where in the second line we identified the (diffractive) gluon PDF according to \eqn{xGP}. As anticipated, the last expression can be recognised as one real step in the DGLAP evolution, but with a simplified version of the splitting function which includes only its singular pieces.
From the experience with the inclusive di-jet cross-section in the back-to-back configuration~\cite{Caucal:2024bae,Caucal:2025mth}, and also with diffractive and inclusive SIDIS~\cite{Hauksson:2024bvv,Caucal:2024vbv},  we expect the full splitting function to be reconstructed after adding the effects of gluon emissions in the initial state (prior to the collision).

The physical interpretation of \eqn{DeltaFrecoil} is transparent in the target picture: a gluon with relatively large transverse momentum $\KT\gg Q_s$, that would have very little probability to be directly produced by the Pomeron, can nevertheless be created via the hard splitting, $g\to gg$, of any of the gluons with much lower transverse momenta $k_{g\perp}\ll \KT$, whose total number is measured by the gluon DPDF $xG^{(0)}\left(x/\xi, \xP, \KT^2\right)$. At the order of interest, the parent gluons are directly produced by the Pomeron, hence they have typical momenta $k_{g\perp}\sim Q_s$ (with a rapidly decaying tail $\propto 1/k_{g\perp}^4$ at larger $k_{g\perp}$). Yet, the daughter gluons produced by the DGLAP splitting have larger transverse momenta  $\KT\gg Q_s$ and follow the standard bremsstrahlung tail $\propto 1/\KT^2$. Hence, although formally a NLO correction, the DGLAP-like contribution in  \eqn{DeltaFrecoil} dominates over the tree-level piece $\mcal{F}_g^{(0)}(x, K_{\perp}^2)$ at sufficiently large $\KT$. This discussion makes clear that the inclusion of the DGLAP evolution is crucial in order to properly describe the gluon distribution at large transverse momenta $\KT\gg Q_s$. As we shall demonstrate in Sect~\ref{sec:dla}, the gluon distribution at large $\KT$ is further modified by the CSS evolution.

The previous discussion shows that the boundary condition (BC) $\mcal{F}_0(x,\xP, \bK)$ for the CSS equation at $Q^2=\KT^2$ must be the sum of the tree-level contribution (expected to dominate at $\KT\lesssim Q_s$) and of the DGLAP-like contribution (responsible for the perturbative tail at high $\KT$):
\begin{align}
	\label{F0}
	\mcal{F}_0(x,  \xP, \KT^2)=
	\mcal{F}_g^{(0)}(x, \xP, K_{\perp}^2) + 
	%\Theta(\KT,\mu_0)\,
	\frac{\alpha_s(K_\perp^2) }{2\pi^2}\,
	\frac{1}{K_\perp^2}
	\int_{x}^{1}\rmd\xi\,{P}^{(+)}_{gg}(\xi)\,
	\frac{x}{\xi}G\left(\frac{x}{\xi}, \xP,  K_\perp^2\right)\,,
\end{align}
with $P_{gg}^{(+)}$ the ``real'' piece of the relevant DGLAP splitting function,  that is 
 \begin{align}
 	\label{Pggreg}
 	P_{gg}^{(+)}
 	\left(\xi\right)=
 	2N_c\left[\frac{\xi}{\big(1-\xi \big)_+}
 	+\frac{1-\xi}{\xi} +\xi(1-\xi)\right].
 \end{align}
The DGLAP piece in \eqn{F0} generalises \eqn{DeltaFrecoil} by including the non-singular pieces of the splitting function and the scale for the running coupling, and by replacing the tree-level gluon DPDF with the all-order solution $ xG\left(x, \xP, \KT^2\right)$ to the DGLAP equation, to be presented shortly.

Before we proceed, let us bring a ``small'' but important modification to \eqn{F0}. The tree-level piece there applies for all values $\KT$ of the transverse momentum, although it rapidly dies away at large values $\KT\gg Q_s$. The DGLAP piece on the other is only correct for large momenta $\KT\gg Q_s$, as quite clear from our derivation. If abusively extrapolated at lower momenta $\KT\lesssim Q_s$, it would predict an explosive growth that would dominate over the saturated contribution. So, we cannot use  \eqn{F0} as it stands when $\KT\lesssim Q_s$. To cope with that, we cut off the DGLAP piece at low momenta, by introducing a ``theta-function'' $\Theta(\KT, \mu_0)$ in front of it.  Here, $\mu_0$ is a transverse scale of order $Q_s$, which marks the onset of the DGLAP evolution, while the ``theta-function'' is truly a function which smoothly interpolate between 0 and 1 at $\KT\ll \mu_0$ and $\KT\gg \mu_0$, respectively, with a rapid variation around $\KT= \mu_0$. Both the precise value of $\mu_0$ and the functional form of $\Theta(\KT, \mu_0)$ are not unique and should be seen as sources of scheme dependence.

%In principle, this expression could be improved by replacing the DGLAP estimate with a more accurate calculation of the NLO contributions (but still to leading power in $1/Q$, to preserve TMD factorisation), that would be valid for arbitrary $\KT$. Such a calculation has recently been reported for the case of {\it inclusive} di-jets~\cite{Caucal:2025mth}, but the result is rather complicated and not easy to use in practice. Moreover, we expect the deviation from the tree-level result at low momenta  $\KT\lesssim Q_s$ to  be a small (pure $\order{\alpha_s}$) effect, unlike the DGLAP contribution which is the dominant contribution at large momenta. In view of that, we shall stick with the relatively simple result \eqref{F0}, but 

Finally, we need to specify the DGLAP equation obeyed by the gluon DPDF $xG(x,\xP, Q^2)$. This is {\it a priori} a standard equation, yet there are a couple of differences which are specific to the small-$x$ problem at hand: \texttt{(i)} The initial condition, as formulated at $Q^2=\mu_0^2$, is presently known from first principles: this is the tree-level DPDF $xG^{(0)}(x, \xP, \mu_0^2)$, possibly supplemented with the BK/JIMWLK evolution for the $\xP$-dependence. \texttt{(ii)} The r.h.s. of the equation also involves a source term, represented by the tree-level gluon TMD $\mcal{F}_g^{(0)}(x, \xP, K_{\perp}^2)$ with $K_{\perp}^2=Q^2$: indeed, this tree-level contribution has unlimited support in $K_{\perp}$, although it  rapidly falls at large momenta $K_{\perp}\gg Q_s$. Hence, for our purposes, we will use the following version of the DGLAP equation,\begin{align}
	\label{DGLAP}
	\frac{\del xG  (x, \xP, Q^2)}{\del \ln Q^2}\,=\,\pi Q^2 
	\mcal{F}_g^{(0)}(x,\xP, Q^2)+
	\Theta(Q,\mu_0)\,
	\frac{\alpha_s(Q^2) }{2\pi}
	\int_{x}^{1}\rmd\xi\,P_{gg}\left(\xi\right)
	\,\frac{x}{\xi}
	G\left(\frac{x}{\xi}, \xP, Q^2\right),
\end{align}
where $P_{gg}(\xi)$ is the complete $g\to gg$ DGLAP splitting function, including the virtual piece proportional to $\beta_0$:
\begin{align}
	\label{Pggful}
	P_{gg}\left(\xi\right)=
	2N_c
	\left[\frac{\xi}{\big(1-\xi \big)_+}
	+\frac{1-\xi}{\xi} +\xi(1-\xi)+\beta_0\delta(1-\xi)
	\right].
\end{align}
In writing \eqn{DGLAP}, we have inserted the (smooth) ``theta-function'' $\Theta(\KT, \mu_0)$ in front of the DGLAP piece in order for the solution to this equation to be a smooth function for all values of $Q^2$, both small and larger than $Q_s^2$. (Without this  ``theta-function'', that is, if one were to solve this equation as a strict initial value problem, the derivative of the solution would be discontinuous at $Q^2=\mu_0^2$.)

\subsection{CSS evolution in the transverse coordinate representation}
\label{sec:bspace}

%\red{To be updated.}

In the previous sections, we focused on the CSS equation in the transverse {\it momentum} representation
($\KT$-\,space), which naturally emerges from the CGC approach. Yet, in the traditional approach to TMD factorisation at moderate $x$, one  rather privileges the version of this equation in the transverse {\it coordinate} representation ($b_{\perp}$-\,space), as obtained via a Fourier transform (FT) from $\bK$ to $\bb$:
\begin{align}
	\label{FTTMD}
	\tF(x, \bb, Q^2)\,
	\equiv\int \frac{\rmd^2\bK}{(2\pi)^2}\,
	\rme^{i\bK\cdot\bb}\, 
	\mcal{F}_g(x, \bK, Q^2)\,.	
\end{align}
The variable $\bb$ is generally referred to as ``the impact parameter'', but should be more properly interpreted as the {\it difference} $\Delta\bb=\bb-\bar{\bb}$ between the impact parameters of the hard di-jet in the direct amplitude and the complex conjugate amplitude, respectively.  In the problem at hand, there is no dependence upon the azimuthal angles made by the vectors $\bK$ and $\bb$, so we shall often ignore the vector orientations and write the respective variable as $\KT^2$ and $\bt^2$, respectively. In the presence of saturation, the gluon DTMD $\mcal{F}_g(x, \KT^2, Q^2)$ approaches a constant value when $\KT\to 0$, hence there is no need to insert an infrared cutoff in the above integration over $\KT$. The FT is roughly controlled by transverse momenta $\KT\lesssim 1/b_{\perp}$. Since in momentum space our approach is valid for $\Lambda^2 \ll \KT^2 \ll Q^2$,  the relevant range in $b_{\perp}$ is  $1/\Lambda^2\gg b_{\perp}^2 \gg 1 /Q^2$.

As we will shortly demonstrate, the FT of Eq.~\eqref{CSSKT} yields the expected version of the CSS equation in $b_{\perp}$-\,space, as found in the literature~\cite{Boussarie:2023izj}. Yet, this does not automatically imply that the respective solutions are equivalent with each other: their equivalence could be broken by the respective boundary conditions\footnote{The equivalence between the two representations would have been easy to enforce  if we had an {\it initial value problem}, i.e.~if we had to solve Eq.~\eqref{CSSKT} knowing the TMD at a fixed (independent of $\KT^2$) scale $Q_0^2$, and similarly in $b_{\perp}$-\,space.}, which are formulated in different spaces and cannot be simply connected by a FT. One of our main tasks in what follows is to construct the boundary condition in $b_{\perp}$-\,space in such a way to preserve the physical equivalence between the two representations to the extent of possible.

Assuming the coupling to be fixed, one can easily take the FT of Eq.~\eqref{CSSKT}. Making the shift $\bK \to \bK - \bell$ in the real term in Eq.~\eqref{CSSKT} and doing the azimuthal integration, the first two terms give
\begin{align}
	\label{FTfirstline}
	& \frac{N_c}{2\pi}\,
	\tF(x, \bt^2, Q^2)
	\int_0^{\infty}
	\frac{\rmd \lt^2}{\lt^2}\,
	\alpha_s
	\Big[J_0(b_{\perp} \lt) -  \Theta(Q^2-\lt^2)\Big]
	= - \frac{N_c}{2\pi}\,
	\tF(x, \bt^2, Q^2)\,
	\alpha_s \ln\frac{Q^2}{\mu_b^2} 
	\nonumber \\*[0.1cm]
	&\qquad
	=- \frac{N_c}{2\pi}\,
	\tF(x, \bt^2, Q^2)\,
	\int_{\mu_b^2}^{Q^2}
	\frac{\rmd \lt^2}{\lt^2}\,
	\alpha_s, 
\end{align}
where we have defined 
\begin{align}
	\label{mub}
	\mu_b^2 \equiv \frac{c_0^2}{b_{\perp}^2}
	\quad \textrm{with} \quad 
	c_0 = 2 e ^{-\gamma_{\rm E}} \simeq 1.123.
\end{align}
We emphasise that this is an exact result. (To demonstrate this,  one can introduce an infrared regulator in the integral over $\lt$, perform the integrals of the two terms separately, and then take the limit of a vanishing regulator.) In the second line, we have rewritten the result as a logarithmic integration, to allow for a running coupling: $\alpha_s \to \alpha_s (\lt^2)$. After also adding the $\beta_0$ term (whose FT is trivially done), we arrive at the $b_{\perp}$-\,space CSS equation  
\begin{align}
	\label{CSSimpact}
	\hspace{-0.1cm}
	\frac{\del{\tF}(x, b_{\perp}^2, Q^2)}{\del\ln Q^2}
	= -\frac{N_c}{\pi}
	\left[\frac{1}{2}
	\int_{\mu_b^2}^{Q^2} \frac{\rmd\lt^2}
	{\ell_\perp^2}\, 
	%\Theta(\lt,\mu_0)
	\alpha_s(\lt^2)
	-
	%\Theta(Q,\mu_0)
	\beta_0
	\alpha_s(Q^2)
	\right]
	\tF(x, \bt^2, Q^2).
\end{align}
When comparing this equation with the original equation \eqref{CSSKT} in $K_{\perp}$-\,space, it looks like the real term has disappeared after going to $b_{\perp}$-\,space. But this is not true: the real term in  $K_{\perp}$-\,space was in fact responsible for the lower limit $\mu_b^2$ on the integral over $\lt^2$ in \eqn{CSSimpact}. The precise value of this lower limit (in particular, the numerical factor $c_0$) is important to ensure that the r.h.s. of Eq.~\eqref{CSSimpact} is correct to {\it single} logarithmic accuracy, like Eq.~\eqref{CSSKT}.

What makes the $b_{\perp}$-\,space version of the CSS equation particularly appealing is the fact that is local in $b_{\perp}$, so it can be easily solved in analytic form. The general solution reads
\begin{align}
	\label{CSSimpactsol}
	{\tF}(x, b_{\perp}^2, Q^2)=
	\tilde{\mcal{F}}_0(x, \mu_b^2)
	\exp \!\left[-\rmS(\mu_b^2,Q^2)\right],
\end{align}
where the ``Sudakov exponent''  $\rmS(\mu_b^2,Q^2)$ encodes the effects of the CSS evolution, 
\begin{align}
	\label{sudakov}
	\rmS(\mu_b^2,Q^2)
	\equiv
	\frac{N_c}{\pi}
	\int_{\mu_b^2}^{Q^2} 
	\frac{\rmd\lt^2}{\lt^2}\,
	%\Theta(\lt,\mu_0)\,
	\alpha_s(\lt^2)
	\left(\frac{1}{2}
	\ln\frac{Q^2}{\lt^2}
	-\beta_0\right),
	%\equiv \ & \rmS_d(\KT^2, Q^2) + \rmS_s(\KT^2, Q^2),
\end{align}
and the (functional) ``prefactor'' $\tilde{\mcal{F}}_0(x, \mu_b^2)$ encodes the
boundary condition (BC) at $Q^2 = \mu_b^2$:
\begin{align}
\label{ICimpact}
	 \tilde{\mcal{F}}_0(x, \mu_b^2)={\tF}(x, b_{\perp}^2, Q^2= \mu_b^2).
\end{align}
%This boundary function $\tilde{\mcal{F}}_0(x, \mu_b^2)$ to be later specified. 
The gluon DTMD is finally obtained as the inverse FT of the $b_{\perp}$-\,space solution:
\begin{align}
	\label{IFTTMD}
	\mcal{F}_g(x, \KT^2, Q^2)
	= \int {\rmd^2\bb}\
	\rme^{-i\bK\cdot\bb}\,\tilde{\mcal{F}}_0(x, \mu_b^2)
	\exp \!\left[-\rmS(\mu_b^2,Q^2)\right].	
\end{align}
A priori, the above integral explores all the values of $b_\perp$, including non-perturbatively large values $b_\perp\sim 1/\Lambda$ where our approach fails to apply. Yet, as we shall shortly argue, the physics of gluon saturation effectively cuts off the integral to values  $b_\perp\lesssim 1/Q_s$ and thus allows one to compute the gluon diffractive TMD from first principles, without the need for non-perturbative regulators.

With respect to \eqn{CSSimpactsol}, the saturation effects are fully encoded in the BC \eqref{ICimpact}, that we have not yet specified. Since the Sudakov exponent vanishes for $Q^2 = \mu_b^2$, it is quite clear that the boundary function $\tilde{\mcal{F}}_0(x, \mu_b^2)$ must encode all the physical information except for the CSS evolution. To construct this function, let us consider first the simpler situation where one ignores the DGLAP evolution. As explained in the previous sections, this is appropriate for relatively low momenta $\KT\lesssim Q_s$, which via the FT implies a similar condition on $\mu_b$. (Indeed, the inverse FT in \eqn{IFTTMD} roughly selects values $b_{\perp}\sim 1/\KT$, meaning $\mu_b\sim \KT$.) Under these assumptions the boundary condition is determined by the tree-level approximation, that is,
\begin{align}
	\label{F0b0}
	\tilde{\mcal{F}}_0(x, \mu_b^2) = \tilde{\mcal{F}}_g^{(0)}(x, \mu_b^2)\equiv 
	\int \frac{\rmd^2\bK}{(2\pi)^2}
	\,\rme^{i\bK\cdot\bb}\, \mcal{F}_g^{(0)}(x, \KT^2)\qquad\mbox{when $\mu_b\lesssim Q_s$}\,.
\end{align}
One may think that the solution thus obtained is equivalent --- in the sense of providing exactly the same result for the gluon DTMD $\mcal{F}_g(x, \bK, Q^2)$ --- to that obtained by solving the CSS equation  in $\KT$-\,space with the tree-level  BC $ \mcal{F}_g^{(0)}(x, \KT^2)$. Yet, this is {\it not} the case, as we now demonstrate. To that aim, it is enough to consider the special value $\mcal{F}_g(x, \bK, Q^2=\KT^2)$ (the would-be boundary value when working  in $\KT$-\,space) as given by the two approaches.  In transverse momentum space, we have, by construction,
\begin{align}
	\label{BCK}
	\mcal{F}_g(x, \KT^2, Q^2=\KT^2)\Big |_{\KT-{\rm space}}= \mcal{F}_g^{(0)}(x, \KT^2)
	= \int {\rmd^2\bb}\
	\rme^{-i\bK\cdot\bb}\,\tilde{\mcal{F}}^{(0)}_g(x, \mu_b^2).
	\end{align}
But when  solving the CSS equation in $b_{\perp}$-\,space (with the BC in \eqn{F0b0}) and then performing the inverse FT, cf.  \eqn{IFTTMD}, one rather finds
\begin{align}
	\label{BCb}
	\mcal{F}_g(x, \KT^2, Q^2=\KT^2)\Big |_{\bt-{\rm space}}
	= \int {\rmd^2\bb}\
	\rme^{-i\bK\cdot\bb}\,\tilde{\mcal{F}}_g^{(0)}(x, \mu_b^2)\,
	\exp \!\left[-\rmS(\mu_b^2,\KT^2)\right],
\end{align}
which is not the same as the r.h.s. of \eqn{BCK}. Hence, even though the evolution equations and respective boundary conditions are related to each other via Fourier transforms, the final predictions are still different for the calculations performed in the two spaces. This seems to imply an ambiguity in our prediction for the solution to the CSS equation, which depends upon the representation. That said, this ambiguity is limited in practice by the quasi-locality of the FT:   the inverse FT  in \eqn{BCb} is controlled by values $b_{\perp}\lesssim 1/\KT$, or $\mu_b\gtrsim \KT$, whereas the Sudakov exponent has support only for $\mu_b < \KT$. Hence, the Sudakov effects are expected to be negligible when evaluating the r.h.s. of \eqn{BCb}, in which case the expressions in Eqs.~\eqref{BCK} and \eqref{BCb} are indeed close to each other. We will later check this expectation by comparing the respective numerical results.

Keeping this in mind, let us explore some physical consequences of the CSS solution in $b_{\perp}$-\,space corresponding to the tree-level BC in \eqn{F0b0}. The fundamental property of this BC for our present purposes is the fact that, at large transverse separations $b_\perp\gg 1/Q_s$, it falls sufficiently fast to ensure the rapid convergence of the inverse FT  in \eqn{BCb}.

As we will explain in more detail  in Appendix~\ref{app:ft_tree}, the large\,-$\bt$ limit and the small-$x$ limit of \eqn{F0b0} do not commute with each other. The most interesting limit for us here is the case where we take $\bt$ to be  large, in the sense that $b_\perp\gg 1/Q_s$, for a fixed value of $x$. In that case, in Appendix~\ref{app:ft_tree} we deduce the following estimate for the asymptotic behaviour of $\tilde{\mcal{F}}_g^{(0)}(x, \bt^2)$ at large $\bt$:
\begin{align}
	\label{f0_large_b}
	\tilde{\mcal{F}}_g^{(0)}(x, \bt^2) \propto \,{x^2(1-x)}\,\frac{\tilde Q_s^2(x)}{[\bt^2 \tilde Q_s^2(x)/8]^4}
	\qquad \mathrm{for} \quad \bt^2 \tilde Q_s^2(x)/8 \gg 1, 
	\end{align}	
with $\tilde Q_s^2(x)=(1-x)Q_s^2$. (For more clarity, we have preferred to write this expression as a function of the coordinate variable $\bt$, instead of the related momentum variable $\mu_b$.) As announced, this shows a rapidly decaying power tail $1/\bt^8$ which ensures rapid convergence. This behaviour is independent upon a specific model for the dipole amplitude\footnote{That said, the proportionality factor not shown in  \eqn{f0_large_b} does actually depend upon the functional form of 
 $\mcal{T}_g(R)$; see  Appendix~\ref{app:ft_tree}.} $\mcal{T}_g(R)$. It is merely a consequence of gluon saturation  (which translates into a dipole amplitude which rapidly approaches the unitarity limit $\mcal{T}_g(R)=1$ when $RQ_s\gg 1$) and of the elastic nature of the diffractive process (which implies that the diffractive TMD must saturate at a constant value when $\KT\ll Q_s$). This can be best appreciated by comparing with the standard (non-diffractive) gluon TMD, and more precisely the Weizs\"{a}cker-Williams gluon TMD, which enters the TMD factorisation of inclusive back-to-back di-jet production in DIS at small $x$~\cite{Dominguez:2010xd,Dominguez:2011wm,Metz:2011wb}. In that case, gluon saturation is important too, but it leads only to a power-like decay $\propto 1/\bt^2$ at large distances. This corresponds to the fact that the WW gluon TMD is still growing, albeit only logarithmically, when decreasing $\KT$ below  $Q_s$ (see Appendix~\ref{app:ft_tree} for details). 

Our estimate \eqref{f0_large_b}  for the asymptotic behaviour vanishes in the limit $x\to 0$. More precisely, our analysis  in Appendix~\ref{app:ft_tree} shows that power-law fall-off shown in \eqn{f0_large_b} applies only so long as $\ln(1/x) \lesssim \bt^2 \tilde{Q}_s^2/16$. In the asymptotic region at $\bt^2 \tilde Q_s^2(x)/8 \gg 1$, this condition is satisfied even for relatively small values $x\ll 1$. If one is interested in even smaller values\footnote{Of course, this case is rather academic, since our approach is valid only so long as $\alpha_s\ln(1/x)\ll 1$.}, one can reverse the above procedure --- that is, start by letting $x\to 0$ in the general results for the tree-level DTMD and study the  large-\,$\bt$ behaviour only after. By doing that, we  find that the FT of the tree-level gluon DTMD vanishes faster than any power of $1/\bt^2$. In particular, when using the MV model for the dipole scattering amplitude, one can analytically show that it falls like a Gaussian (up to power corrections):
\begin{align}
	\label{f0_x0}
	\tilde{\mcal{F}}_g^{(0)}(x=0, \bt^2) \propto \frac{Q_s^2}{[\bt^2 Q_s^2/8]^2}
	%\frac{S_\perp N_g}{4 \pi^3}\frac{64}{\bt^4 Q_s^2}
	\,\exp\left( - \frac{\bt^2 Q_s^2}{8} \right)
	\quad \mathrm{for} \quad \bt^2 Q_s^2/8 \gg 1. 
\end{align}
The physics reason for this change in the asymptotic behaviour is not fully clear to us, but it is likely related to the fact that the gluon virtuality $\mcal{M}^2$ introduced in \eqn{gp} vanishes  when $x\to 0$.

To summarise the above argument, the solutions to the CSS equation in  $b_{\perp}$-\,space and in the physical context at hand (diffraction plus gluon saturation) are under control in the CGC  perturbation theory, and in particular do not require the introduction of a ``non-perturbative Sudakov exponent'', as generally done in the traditional context of TMD factorisation at moderate values of $x$~\cite{Boussarie:2023izj}.

Next, we would like to also include the DGLAP evolution in the BC in \eqn{ICimpact}. This becomes necessary when the transverse momentum of the measured gluon is large enough,  $\KT\gg Q_s$. In $b_{\perp}$-\,space, this means that the scale $ \mu_b$ is much larger than $Q_s$. We will argue that the appropriate BC in this regime reads
\begin{align}
	\label{FOFTsmallb}
	\tilde{\mcal{F}}_0(x, \mu_b^2) \simeq 
	\frac{xG(x,\mu_b^2)}{(2\pi)^2}\qquad\mbox{when  $\mu_b^2\gg Q_s^2$},
\end{align}
where the gluon DPDF  $xG(x,\mu_b^2)$ is computed as the solution to the DGLAP equation \eqref{DGLAP}. To see this, let us evaluate the boundary value ${\tF}(x, b_{\perp}^2, Q^2= \mu_b^2)$ at large $Q^2\gg Q_s^2$ with the help of the 
FT in \eqn{FTTMD}; one can successively write
\begin{align}
	\label{FTBC}
	\tF(x, b_{\perp}^2, Q^2= \mu_b^2) &\,
	=\int \frac{\rmd^2\bK}{(2\pi)^2}\,
	\rme^{i\bK\cdot\bb}\, 
	\mcal{F}_g(x, \KT^2, Q^2= \mu_b^2)\nonumber\\*[0.1cm]
&\,\simeq
	\int \frac{\rmd^2\bK}{(2\pi)^2}\,\Theta(\mu_b^2-\KT^2)
	%\rme^{i\bK\cdot\bb}\, 
	\mcal{F}_g(x, \KT^2, \mu_b^2)\simeq	\frac{xG(x,\mu_b^2)}{(2\pi)^2}\,,
\end{align}
where the approximate equalities hold in the leading logarithmic approximation in the sense of the large transverse logarithm $\ln(\KT^2/Q_s^2)$. Indeed, after including the DGLAP evolution, the gluon TMD $\mcal{F}_g(x, \KT^2, Q^2)$ develops a $1/\KT^2$ tail at large momenta $\KT\gg Q_s$, as manifest in \eqn{DeltaFrecoil}.
Hence to leading logarithmic accuracy the first integral in \eqn{FTBC} is controlled by transverse momenta within the range $Q_s\ll \KT\ll \mu_b=Q$, or $\KT\bt \ll 1$. It then becomes appropriate to neglect the Fourier phase, but instead cut off the integral at $\KT\sim \mu_b$, as we did in the second line. Finally, still to leading log accuracy, one can construct the diffractive gluon PDF from the respective TMD according to
\begin{align}
	\label{PDF}
	xG(x, Q^2) \,\equiv\, 
	\pi\int^{Q^2}_{\Lambda^2} 
	{\rmd \KT^2}\,
	\mcal{F}_g(x, \KT^2, Q^2)
	\,.
\end{align}
This explains our last equality in  \eqn{FTBC} and justifies the estimate \eqref{FOFTsmallb} for the BC.

\eqn{PDF} is the natural generalisation of \eqn{xGP} beyond the tree-level approximation. As we shall demonstrate in Sect.~\ref{sec:dla}, this relation is indeed consistent with the DGLAP and CSS evolutions, provided the latter is evaluated in a ``double logarithmic approximation'' (DLA), in which the successive gluon emissions are strongly ordered in transverse momenta, as for the DGLAP evolution. 

It is furthermore interesting to notice that, in this ``hard'' regime at large $\mu_b\gg Q_s$ the BC in $b_{\perp}$-\,space does not coincide with the FT of the respective BC in  $\KT$-\,space.
%\footnote{This should be contrasted with the situation at $\mu_b\lesssim Q_s$, or large distances $\bt\gtrsim 1/Q_s$, where the BCs in the two spaces are both given by the tree-level approximation and are related via FT,  cf. \eqn{F0b0}.}. 
Indeed, the first integral in \eqn{FTBC} involves the function $\mcal{F}_g(x, \KT^2, Q^2)$ for generic  values $\KT^2$ and $Q^2$ (since we take $Q^2= \mu_b^2$ and $\mu_b^2$ itself is generic), and not just its boundary value $\mcal{F}_0(x, \KT^2)$ at $Q^2=\KT^2$. Moreover, the discussion following \eqn{FTBC}  shows that the integral over $\KT$ is {\it not} controlled by values $\KT\sim \mu_b$ (but rather by $\KT\ll \mu_b$), hence the identification between $\tilde{\mcal{F}}_0(x, \mu_b^2)$ and the FT of $\mcal{F}_0(x, \KT^2)$ does not hold not even in a loose sense. Yet, by itself, this does not imply a genuine mismatch between the respective predictions for the gluon TMD. In Sect.~\ref{sec:dla}, we will demonstrate that, in the  double logarithmic approximation, the CSS solutions obtained by working in the two spaces are in fact equivalent with each other. Beyond DLA, there are small differences related to the non-locality of the FT, as already explained.

\begin{figure}
	\begin{center}
		\includegraphics[width=0.45\textwidth]{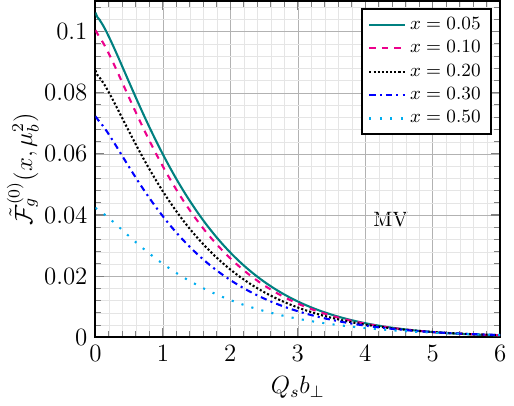}\qquad
		\includegraphics[width=0.45\textwidth]{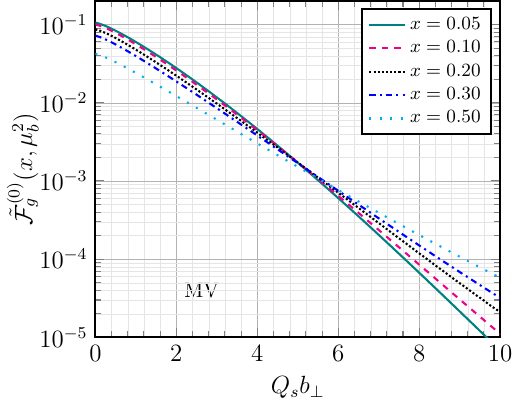}
	\end{center}
	\caption{\small The FT of the tree-level gluon DTMD, as numerically obtained from Eq.~\eqref{Ftree} and after dividing out the dimensionless factor  $Q_s^2 S_\perp/(4\pi^3)$. In the right plot we use a logarithmic scale on the vertical axis. }
\label{fig:F0b}
\end{figure}

\begin{figure}
	\begin{center}
		\includegraphics[width=0.45\textwidth]{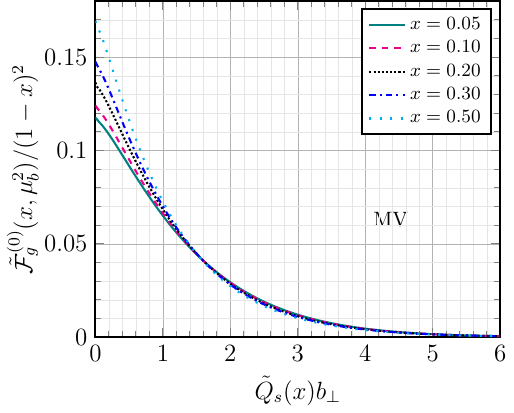}\qquad
		\includegraphics[width=0.45\textwidth]{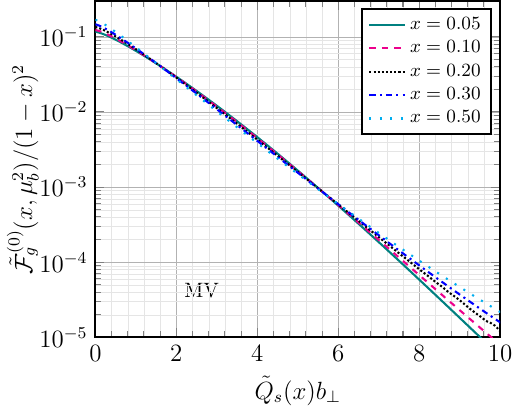}
	\end{center}
	\caption{\small	The same as in Fig.~\ref{fig:F0b}, but after dividing out the dominant $x$-behaviour of  $\tilde{\mcal{F}}_g^{(0)}(x, \mu_b^2)$ as suggested by  Eqs.~\eqref{Ftreesmallb} and \eqref{xGPhigh}, and in terms of the ``scaling'' ($x$-dependent) variable $\tilde Q_s(x)\bt $.}
\label{fig:F0b_scaled}
\end{figure}

Together, Eqs.~\eqref{F0b0} and \eqref{FOFTsmallb} provide a piecewise approximation to the BC in $\bt$-\,space, namely
\begin{align}
	\label{BCbpws}
	\tilde{\mcal{F}}_0(x, \mu_b^2)=\Theta(\mu_0^2- \mu_b^2) \tilde{\mcal{F}}_g^{(0)}(x, \mu_b^2)+
	\Theta(\mu_b^2- \mu_0^2) \frac{xG(x,\mu_b^2)}{(2\pi)^2}\,,\end{align}
with $\mu_0$ a scale of order $Q_s$. To the accuracy of interest, this can be replaced with the following, global, approximation, which is a smooth function:
\begin{align}
	\label{BCbgl}
	\tilde{\mcal{F}}_0(x, \mu_b^2)=\,\frac{xG(x,\mu_b^2)}{xG^{(0)}(x,\mu_b^2)}\,
	 \tilde{\mcal{F}}_g^{(0)}(x, \mu_b^2)\,.
	 \end{align}
Indeed, when $ \mu_b^2\lesssim \mu_0^2$, the  DGLAP evolution is unimportant (recall the ``theta-function'' $\Theta(\KT, \mu_0)$ in \eqn{DGLAP}), hence $xG(x,\mu_b^2)=xG^{(0)}(x,\mu_b^2)$ and \eqn{BCbgl} reduces to the tree-level approximation, as in \eqn{F0b0}. On the other hand, for $ \mu_b^2\gg \mu_0^2$, we can write 
\begin{align}
	\label{Ftreesmallb}
	\tilde{\mcal{F}}_g^{(0)}(x, \mu_b^2) \equiv 
	\int \frac{\rmd^2\bK}{(2\pi)^2}
	\,\rme^{i\bK\cdot\bb}\, \mcal{F}_g^{(0)}(x, \KT^2) \simeq 
	\frac{xG^{(0)}(x,\mu_b^2)}{(2\pi)^2}\qquad\mbox{when  $\mu_b^2> \mu_0^2$},
\end{align}
since the integral is controlled by $\KT\sim \mu_0$ (because of saturation), whereas $\bt \sim 1/\mu_b  \ll 1/\mu_0$. Hence the Fourier phase is negligible and the integral over $\KT$ up to the relatively large value $\bt \gg Q_s$ reconstructs the tree-level DPDF, as in \eqn{xGP}.
We thus recover the expected form \eqref{FOFTsmallb} of the BC at large $\mu_b$ (or small $\bt$).

%\eqn{Ftreesmallb} also predicts the ``geometric scaling'' properties of the boundary function

\eqn{IFTTMD} with the BC in \eqn{BCbgl} and the Sudakov exponent in \eqn{sudakov} express our final result for the gluon TMD computed from CSS solutions in $\bt$-\,space.

We conclude this section with a few plots exhibiting the $\bt$-dependence and the scaling properties of the FT $\tilde{\mcal{F}}_g^{(0)}(x, \mu_b^2)$ of the tree-level gluon TMD. This quantity is {\it a priori} dimensionless, but it includes an overall factor $S_\perp/(4\pi^3)$ which involves the nuclear area $S_\perp$ (recall  Eq.~\eqref{Ftree}).
%a non-perturbative quantity, which is not fixed by our calculation.
In our plots, we prefer to omit that factor, as we did for the respective plots in $\KT$-\,space (cf. Fig.~\ref{fig:tree}), and at the same time divide $\tilde{\mcal{F}}_g^{(0)}$ by a factor $Q_s^2$ --- so that the plotted function is still dimensionless. In the plots, we still use the notation $\tilde{\mcal{F}}_g^{(0)}(x, \mu_b^2)$ for this dimensionless function. Similarly, the transverse separation $\bt$ is measured in units of $1/Q_s$. Our results are shown in Fig.~\ref{fig:F0b}, for various values of $x$ between 0.05 and 0.5, and up to $\bt Q_s=10$.

Also, we shall present plots which illustrate the quality of geometric scaling in terms of the variables $x$ and $\bt$. 
To that aim, it is preferable to measure $\bt$  in units of $1/\tilde Q_s(x)$, with $\tilde Q_s^2(x)=(1-x)Q_s^2$. The expected scaling properties are easily inferred from our previous analysis.
At relatively small  $\bt  \tilde Q_s(x)\lesssim 1$, we can rely on Eqs.~\eqref{Ftreesmallb} and \eqref{xGPhigh}
to conclude that $\tilde{\mcal{F}}_g^{(0)}(x, \mu_b^2)$ scales roughly like $(1-x)^2$. For very large  $\bt  \tilde Q_s(x)\gg 1$, we rather expect $\tilde{\mcal{F}}_g^{(0)}(x, \mu_b^2)\sim x^2(1-x)^2$, cf.  \eqn{f0_large_b}. Yet, that asymptotic scaling is very slowly approached when increasing $\bt\tilde Q_s(x)$, as suggested by our analysis in  Appendix~\ref{app:ft_tree} and confirmed by the numerical results. And indeed, the curves shown in Fig.~\ref{fig:F0b_scaled} demonstrate that the would-be ``small-$\bt$'' scaling, cf. \eqn{Ftreesmallb}, is in fact  well satisfied up to relatively large separations $\bt \tilde Q_s(x)\sim 6$.

\begin{figure}
	\begin{center}
		\includegraphics[width=0.55\textwidth]{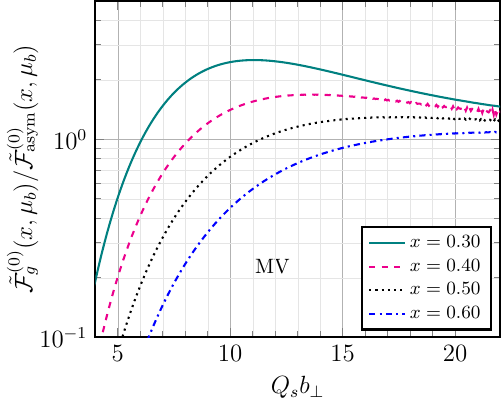}
	\end{center}
	\caption{\small	A numerical test of the asymptotic behaviour predicted by \eqn{f0_large_b} for the Fourier transform  $\tilde{\mcal{F}}_g^{(0)}(x, \mu_b^2)$ of the tree-level gluon TMD at large separations $\bt\gg 1/\tilde Q_s(x)$. We test this by plotting the ratio between the numerical result for  $\tilde{\mcal{F}}_g^{(0)}(x, \mu_b^2)$ (as obtained with the MV model) and the (normalised) asymptotic behaviour analytically constructed   in Appendix~\ref{app:ft_tree} and exhibited in \eqn{a:f0_tilde_x}. (The numerical coefficient $c_3$ is computed according to \eqn{a:cn}, which yields $c_3\simeq 7.58$ for the MV model.)
	}
\label{fig:F0b_tail}
\end{figure}

Finally in Fig.~\ref{fig:F0b_tail} we test asymptotic behaviour at large $\bt$, cf. \eqn{f0_large_b}. The approach towards the power law $1/(\bt \tilde Q_s(x))^8$ is clearly seen in the numerical results, albeit the numerical simulations become extremely tedious for small values of $x$ (which explains why, in this figure, we considered relatively large values $x\ge 0.3$). In fact, even the normalisation of our asymptotic approximation, as obtained  in Appendix~\ref{app:ft_tree}, cf. \eqn{a:f0_tilde_x}, appears to be well under control: the ratio $\tilde{\mcal{F}}_g^{(0)}(x, \mu_b^2)/\tilde{\mcal{F}}_{\rm asym}^{(0)}(x, \mu_b^2)$ is seen to approach unity when increasing $\bt Q_s$, albeit very slowly.

\subsection{CSS evolution in the double logarithmic approximation}
\label{sec:dla}

In this section, we shall present an approximate version of the CSS equation in transverse momentum space, in which successive gluon emissions are treated in a double logarithmic approximation (DLA): not only the gluons emitted in the $s$-channel are assumed to be soft, thus generating the rapidity logarithm $\ln(1/\xi_0)= (1/2)\ln(Q^2/\KT^2)$, but they are also strongly ordered in transverse momenta, as for the DGLAP evolution; this additional condition will allow us to explicitly isolate contributions enhanced by the transverse logarithms $\ln(Q^2/\KT^2)$ and $\ln(\KT^2/Q_s^2)$ in the r.h.s. of the CSS equation.

This additional approximation has two attractive features. First, it allows for an unambiguous matching between the CSS and the DGLAP evolutions; in particular, the BC at $Q^2=\KT^2$, as built with the DGLAP solution, is known with the same accuracy as the subsequent CSS evolution. Second, it allows for explicit analytic solutions which render the physical interpretation more transparent.

The subsequent manipulations  are in fact quite similar to those already developed in Sect.~\ref{sec:BC}, in our partial derivation of the DGLAP equation. The only difference is that these  manipulations will now be applied to the CSS equation \eqref{CSSKT}. Taking a fixed coupling for simplicity,  the integral over $\lt$ in the r.h.s. of \eqn{CSSKT} features two logarithmic regimes:

\texttt{(i)} When $\KT^2 \ll \lt^2\ll Q^2$, the $\lt$-integral multiplying the virtual piece clearly develops a logarithm  $\ln(Q^2/\KT^2)$, but the real term is suppressed since  $\mcal{F}_g(\bK+\bell) 
\simeq \mcal{F}_g(\bell)$ falls  away at least as $1/\lt^2$ when $\lt \gg \KT\gtrsim Q_s$. In the above, we implicitly assumed that $\KT$ is larger than $Q_s$. When this is not the case, that is, when $\KT\ll Q_s$, then one must separately consider the contributions from $\KT \ll \lt <  Q_s$ and those obeying $\lt \gg Q_s$. For the former, we can write $\mcal{F}_g(\bell)\simeq \mcal{F}_g(\bK)$, since both $\lt$ and $\KT$ are below $Q_s$ and the gluon DTMD is roughly flat at saturation. Hence, the real and virtual terms mutually cancel in this range. For the latter, i.e. for $\lt \gg Q_s$, the real term is again power-suppressed and we are left with the virtual contribution alone.

To summarise, this first logarithmic domain is truly restricted to  $K_0^2 \ll \lt^2\ll Q^2$, with  $K_0\equiv \max[\KT,\mu_0]$, and it yields a purely virtual contribution, multiplied by the following integral:
\begin{align}
	\label{VLog}
	\int_{K_0^2}^{Q^2}\frac{\rmd \lt^2}{\lt^2} \alpha_s(\lt^2) = \int_{\KT^2}^{Q^2}\frac{\rmd \lt^2}{\lt^2} 
	\,\Theta(\lt, \mu_0) \alpha_s(\lt^2)
\end{align}
where we restored the running coupling, we have replaced the fixed scale $Q_s$ by the flexible scale $\mu_0$ that can be varied around $Q_s$, and we have inserted the  ``theta-function'' $\Theta(\lt, \mu_0)$ inside the second integral, in order to replace $K_0\to \KT$ in its lower limit. The meaning of this ``theta-function'' is the same as previously discussed after \eqn{F0} (recall also the DGLAP equation \eqref{DGLAP}).

\texttt{(ii)} The other logarithmic domain corresponds to $Q_s\ll k_{g\perp}\equiv |\bK+\bell|\ll \KT$ and  refers to the real term alone. In  the projectile picture, this corresponds to the situation where the imbalance is controlled by the recoil of the emitted gluon and is relatively hard: $\KT\simeq \lt\gg Q_s$.  In the target picture, the $t$-channel gluon $\bk_{g\perp}$ taken from the Pomeron (with $k_{g\perp}\sim Q_s$) undergoes a hard splitting into a pair of gluons, $\bK$  and $\bell =\bk_{g\perp}-\bK\simeq -\bK$, which are nearly back-to-back in the transverse plane. This is the same transverse kinematics as for a DGLAP splitting, but the  $t$-channel gluon with transverse momentum $\bK$ is assumed to have a splitting fraction
$\xi\simeq 1$; so, this splitting truly contributes to the CSS evolution. Via manipulations similar to those leading to \eqn{DeltaFrecoil}, one finds that the real term in \eqn{CSSKT} simplifies to
\begin{align}\label{RLog}
	\Theta(\KT, \mu_0)\,\frac{\alpha_s(\KT^2)N_c}{2\pi}\,\frac{1}{\KT^2}
%\Theta\left(\frac{1-x}{x}-\xi_0\right)
\int_{\Lambda^2}^{\KT^2} \rmd k_{g\perp}^2 \mcal{F}_g\left(x, k_{g\perp}^2, Q^2\right).
\end{align}
We have inserted the ``theta-function'' $\Theta(\KT, \mu_0)$ to emphasise that this contribution only exists for $\KT > Q_s$.
The logarithmic enhancement is generated by the interval $Q_s^2\ll k_{g\perp}^2  \ll \KT^2$, where the gluon TMD (assumed to also include the effects of the DGLAP evolution)  develops a power-tail $1/\KT^2$. 
%Notice also that, unlike what we did in  \eqn{DeltaFrecoil}, here we do not identify 

After putting together Eqs.~\eqref{VLog} and \eqref{RLog}, and also adding the virtual piece proportional to $\beta_0$  in \eqn{CSSKT}, we find the following, simplified, version of the CSS equation, valid at DLA~\cite{Caucal:2024bae}:
\begin{align}
	\label{CSS}
	\hspace{-0cm}
	\frac{\del \mcal{F}_g(x, K_\perp^2,Q^2)}{\del \ln Q^2} =\, &
	\, \frac{N_c}{2\pi}
	\bigg[\Theta(\KT,\mu_0)\,\frac{\alpha_s(K_\perp^2) }{K_\perp^2}
	\int^{K_\perp^2}_{\Lambda^2}\rmd \ell_\perp^2\,
	\mcal{F}_g(x,\ell_{\perp}^2, Q^2)
	\nonumber\\*[0.2cm]
	-\int_{K_\perp^2}^{Q^2}
	\frac{\rmd \ell_\perp^2}{\ell_\perp^2}\,&
	\Theta(\ell_{\perp},\mu_0)\,
	\alpha_s(\ell_\perp^2)\mcal{F}_g(x, K_\perp^2, Q^2)\bigg]+
	\Theta(Q,\mu_0)\,
\beta_0\frac{\alpha_s(Q^2)N_c}{\pi}\mcal{F}_g(x, K_\perp^2, Q^2).
	\end{align}
This equation has to be solved with a boundary condition	of the type \eqref{BC} and we anticipate that the boundary function $\mcal{F}_0(x, \xP, \bK)$ takes the form shown in \eqn{F0}, that here will be re-derived from a slightly different perspective. Notice that, although the BC is formulated at $Q^2=\KT^2$, the CSS evolution truly starts at the scale $Q^2=K_0^2\equiv \max[\KT^2,\mu_0^2]$. When $\KT< \mu_0$, the gluon TMD is fixed to the boundary condition so long as $Q^2$ takes intermediate values
 $\KT^2 < Q^2 < \mu_0^2$ and it starts evolving only for larger values $Q^2 > \mu_0^2$.

As anticipated at the beginning of this section, \eqn{CSS} can be solved in analytic form. To that aim, it is convenient to introduce two ``auxiliary'' functions defined as 
\begin{align}
	\label{PDFaux}
	xG(x, Q^2) \,\equiv\, 
	\pi\int^{Q^2}_{\Lambda^2} 
	{\rmd \ell_{\perp}^2}\,
	\mcal{F}_g(x, \ell_{\perp}^2, Q^2)
	\,,\qquad
	xG(x, \KT^2, Q^2) \,\equiv\, 
	\pi \int^{\KT^2}_{\Lambda^2} 
	{\rmd \ell_{\perp}^2}\,
	\mcal{F}_g(x, \ell_{\perp}^2, Q^2)\,.
\end{align}
The first quantity was already defined in \eqn{PDF} and plays the role of the gluon DTMD in the approximations of interest, as we shall shortly argue. The second one can be characterised as  a  ``generalized PDF'',  from which one can deduce both the PDF (as a boundary value) and the TMD (as a derivative). Indeed, we readily see that
 \begin{align}
 	\label{PDFaux2}
	xG(x, Q^2) 
	=xG(x, \KT^2=Q^2, Q^2) 
	\,,\qquad
	\mcal{F}_g(x, K_{\perp}^2, Q^2) 
	= \frac{1}{\pi}\, 
	\frac{\del xG(x,\KT^2, Q^2)}{\del\KT^2}\,.  
\end{align}
It is then straightforward to show that the ``generalized PDF'' obeys an equation similar to Eq.~\eqref{CSS} in which the real term is absent. Indeed, by first renaming  $\KT^2\to q_{\perp}^{2}$ in Eq.~\eqref{CSS} and then integrating over $q_{\perp}^{2}$ from $\Lambda^2$ to $\KT^2$, we obtain
\begin{align}
	\label{CSSxG}
	\hspace{-0.4cm}
	\frac{\del xG(x, K_\perp^2, Q^2)}{\del \ln Q^2}
	=- \frac{N_c}{\pi}
	\left[\frac{1}{2}
	\int_{K_\perp^2}^{Q^2}\frac{\rmd \ell_\perp^2}
	{\ell_\perp^2}\,
	\Theta(\ell_{\perp},\mu_0)
	\alpha_s(\ell_\perp^2)-
	\Theta(Q,\mu_0)
	\beta_0 \alpha_s(Q^2)
	\right]
	xG(x, K_\perp^2, Q^2).
	\end{align}
The solution to the above equation that satisfies the BC in Eq.~\eqref{PDFaux2} reads
\begin{align}
	\label{genPDF}
	xG(x, \KT^2, Q^2)=xG(x, \KT^2)
	\exp\!\left[-\rmS(\KT^2,Q^2)\right],
  \end{align}
with the following expression for the  ``Sudakov exponent'':
\begin{align}
	\label{sudakovK}
	\rmS(\KT^2,Q^2)
	=
	\frac{N_c}{\pi}
	\int_{K_{\perp}^2}^{Q^2} 
	\frac{\rmd\lt^2}{\lt^2}\,
	\Theta(\lt,\mu_0)\,
	\alpha_s(\lt^2)
	\left(\frac{1}{2}
	\ln\frac{Q^2}{\lt^2}
	-\beta_0\right).
	%\equiv \ & \rmS_d(\KT^2, Q^2) + \rmS_s(\KT^2, Q^2),
\end{align}
At this stage, it is easy to deduce the gluon DTMD according to the second relation in \eqn{PDFaux2}:
\begin{align}\label{CSSsol1}
\mcal{F}_g(x, K_\perp^2, Q^2)=\frac{1}{\pi}\frac{\del }{\del K_\perp^2}\left\{
xG(x, \KT^2)\,\exp \left[-\frac{N_c}{\pi}\int_{K_{\perp}^2}^{Q^2} \frac{\rmd\lt^2}{\lt^2}\,
\Theta(\lt,\mu_0)\, \alpha_s(\lt^2)\left(\frac{1}{2}\ln\frac{Q^2}{\lt^2}-\beta_0\right)
 \right]\right\}.
\end{align}

Before we proceed with the study of this result, let us notice that the above expressions are quite similar to those encountered in the discussion of the transverse coordinate representation in Sect.~\ref{sec:bspace}. One can observe 
for instance the similarities between the respective versions of the CSS equation, cf. Eqs.~\eqref{CSSimpact} and \eqref{CSSxG}, and also between the respective Sudakov factors, cf.  Eqs.~\eqref{sudakov} and \eqref{sudakovK}. In spite of such formal resemblances, the two formulations are not fully equivalent with each other --- the $\bt$-\,space formulation of the CSS equation is more precise than the DLA formulation discussed here ---, yet they can be related to each other to double logarithmic accuracy. To see that, let us evaluate the FT in \eqn{FTTMD} using the DLA result \eqref{CSSsol1} for the gluon DTMD:
\begin{align}
	\label{FTDLA}
	\tF(x, \bb, Q^2)\,
	&\, =\int \frac{\rmd^2\bK}{(2\pi)^2}\,
	\rme^{i\bK\cdot\bb}\, \frac{1}{\pi}\frac{\del }{\del K_\perp^2}\left\{
xG(x, \KT^2)\,\exp\!\left[-\rmS(\KT^2,Q^2)\right]\right\}
\nonumber\\*[0.2cm]
&\,\simeq \int^{\mu_b^2}_{\Lambda^2}  \frac{\rmd\KT^2}{(2\pi)^2}\,
\frac{\del }{\del K_\perp^2}\left\{
xG(x, \KT^2)\,\exp\!\left[-\rmS(\KT^2,Q^2)\right]\right\}
\nonumber\\*[0.2cm]
&\,\simeq \frac{xG(x,\mu_b^2)}{(2\pi)^2}\,\exp\!\left[-\rmS(\mu_b^2,Q^2)\right],
\end{align}
where the successive approximations are valid in the DLA sense and for $\mu_b^2\gg Q_s^2$. Modulo these approximations, we have indeed recovered the respective limit of the CSS solution in $\bt$-\,space, cf. Eqs.~\eqref{CSSimpactsol} and \eqref{FOFTsmallb}.

Returning to \eqn{CSSsol1}, one can evaluate the  external derivative to deduce the following expression for the gluon DTMD to DLA:
\begin{align}
  \label{CSSsolTMD}
  \mcal{F}_g(x, K_\perp^2, Q^2)=\, &
  \left[\frac{1}{\pi}
  \frac{\del\,xG(x, \KT^2) }{\del K_\perp^2}
  +\Theta(\KT,\mu_0)\,
  \frac{\alpha_s(\KT^2)N_c}{\pi^2}
  \left(\frac{1}{2}\ln\frac{Q^2}{\KT^2}-\beta_0\right) 
  \frac{xG(x, \KT^2)}{\KT^2}\right]
  \nn*[0.1cm]
  &\times \exp \!\left[-\rmS(\KT^2,Q^2)\right],
\end{align}
where the second term in the square bracket emerged from (minus) the derivative of the Sudakov exponent $\rmS(\KT^2,Q^2)$  w.r.t.~$\KT^2$.
Although exact, this solution may still look formal, since it involves the auxiliary function  $xG(x, \KT^2) $, itself defined as an integral over $\mcal{F}_g(x, \ell_{\perp}^2, Q^2)$, cf.~\eqn{PDFaux}. So, at a first sight, \eqn{CSSsolTMD} looks more like an integral version of equation \eqref{CSS}, rather than its solution. The crucial observation at this point is that the function $xG(x, \KT^2) $  is an ``integral of motion'' for the CSS dynamics --- in the absence of the $\beta_0$ piece it would be conserved by the CSS evolution (see below) --- and hence can be directly computed from the BC in \eqn{BC}. To see this we first take the derivative of $xG(x, Q^2) $ w.r.t.~$Q^2$ in its defining equation \eqref{PDFaux} to get
\begin{align}
	\label{derG}
	\frac{\del \,xG  (x, Q^2)}{\del \ln Q^2}&
	=\pi Q^2 \mcal{F}_g(x, Q^2, Q^2)+
	\pi \int^{Q^2}_{\Lambda^2} 
	{\rmd \ell_\perp^2}\, \frac{\del \mcal{F}_g(x, \ell_{\perp}^2, Q^2)}
	{\del \ln Q^2}.
 \end{align}
 The first term in the r.h.s.~is fixed by the BC in Eq.~\eqref{BC}. The second term gives
\begin{align} 
	\hspace{-0.1cm}
	\pi \int^{Q^2}_{\Lambda^2} {\rmd \ell_\perp^2}\, 
	\frac{\del \mcal{F}_g(x, \ell_{\perp}^2, Q^2)}{\del \ln Q^2}=
	\frac{\del xG(x, K_\perp^2, Q^2)}{\del \ln Q^2}\bigg |_{\KT^2=Q^2}
	=\Theta(Q,\mu_0)
	\beta_0\,\frac{\alpha_s(Q^2)N_c}{\pi}\,xG(x,Q^2)\,, 
\end{align}
where we have used the definition  \eqref{PDFaux} for the generalised PDF together 
with \eqn{CSSxG}. Thus we arrive at the following, {\it closed}, equation for the gluon PDF
\begin{align}
	\label{DLAxG}
	\frac{\del \,xG  (x, Q^2)}{\del \ln Q^2}\,=\,
	\pi Q^2 \mcal{F}_0(x, Q^2)+
	\Theta(Q,\mu_0)
	\beta_0\,\frac{\alpha_s(Q^2)N_c}{\pi}\,xG(x,Q^2)\,,
\end{align}
 in which the boundary function $\mcal{F}_0(x, \KT^2)$ plays the role of a source term. The above is readily solved to give 
 \begin{align}
	\label{PDFDLA}
	xG(x, Q^2)&\, =\pi  \int^{Q^2}_{\Lambda^2} {\rmd \ell_\perp^2}\,
	\exp 
	\!\left[ \frac{\beta_0 N_c}{\pi}\int^{Q^2}_{\lt^2}
	\frac{\rmd\mu^2}{\mu^2}\,
	\Theta(\mu,\mu_0)
	\alpha_s(\mu^2)\right]\mcal{F}_0(x, \lt^2)\,.
\end{align}
It is interesting to notice that in the limit $\beta_0\to 0$, the gluon PDF  is not  modified by the CSS dynamics ---  it is equal to the integral of BC  $\mcal{F}_0(x, \lt^2)$. This shows that, for $\beta_0=0$, the ``gain'' (real) and the ``loss'' (virtual) terms in \eqn{CSS} merely redistribute the gluons in transverse momentum space, without modifying their total number, i.e.~the PDF. The latter can only be changed by the virtual term proportional to $\beta_0$, which is the DGLAP anomalous dimension for gluons.

Using Eq.~\eqref{DLAxG} to replace the derivative of $xG$ in \eqn{CSSsolTMD}, one sees that the terms in the prefactor which are proportional to $\beta_0$ cancel each other and we finally arrive at the following expression for the gluon TMD
\begin{align}
	\hspace{-0.6cm}
	\label{DLAsol}
	\mcal{F}_g(x, K_\perp^2, Q^2)= \left[\mcal{F}_0(x,  \KT^2)
	+ \Theta(\KT,\mu_0)\,
	\frac{\alpha_s(\KT^2)N_c}{2\pi^2}\,\frac{xG  (x,\KT^2)}{\KT^2}\,
	\ln\frac{Q^2}{\KT^2}\right]
	\exp \!\left[-\rmS(\KT^2,Q^2)\right].
\end{align}
Together, Eqs.~\eqref{DLAsol} and \eqref{PDFDLA} provide an analytic solution to the DLA version of the CSS equation for a  generic  boundary function $\mcal{F}_0(x,  \KT^2)$. In principle, we would like this function to encode the information about the tree-level gluon DTMD and its DGLAP evolution up to $\KT^2$. This is quite easy to achieve in practice: by comparing  \eqn{DLAxG}  obeyed by the PDF with the DGLAP equation \eqref{DGLAP}, one immediately sees that these two equations become identical if the boundary function $\mcal{F}_0(x,  \KT^2)$ is chosen according to  \eqn{F0}. With this choice, the CSS equation \eqref{CSS} for the gluon TMD and the DGLAP equation \eqref{DGLAP} for the gluon PDF are fully consistent with each other provided the PDF is constructed from the TMD as shown in \eqn{PDFaux}.

\eqn{DLAsol} makes clear that the CSS evolution brings a contribution to the $1/\KT^2$ tail at large transverse momenta, on top of the respective contribution of the DGLAP evolution, as included in the boundary function $\mcal{F}_0(x,  \KT^2)$, cf.  \eqn{F0}. As is should be clear from the discussion in Sect.~\ref{sec:BC}, the NLO corrections contributing to the DGLAP and the CSS evolutions are in fact generated on the same footing~\cite{Caucal:2024bae}, but they are separated from each other for the purposes of the respective resummations.

%Let us finally present an equivalent version of the CSS solution \eqref{DLAsol} that is obtained after inserting \eqn{F0} for  $\mcal{F}_0(x,  \KT^2)$:

\section{Numerical solutions to the CSS and DGLAP equations}
\label{sec:sols}

In this section, we shall present (mostly) numerical solutions to the CSS equation for the gluon DTMD, by using all the three versions of this equation introduced in the previous section: $\KT$-\,space, $\bt$-\,space and DLA. For more clarity, we shall first consider boundary conditions given by the tree-level approximation (as evaluated with the MV model for the gluon dipole amplitude) and then include the effects of the DGLAP evolution, as obtained from numerical solutions to \eqn{DGLAP}.

\subsection{CSS solutions with tree-level boundary conditions}
\label{sec:BC0}

In this section, we present solutions to the three versions of the CSS equation with boundary conditions which are fixed by the tree-level approximation (and hence ignore the DGLAP evolution). As discussed at several places, this approximation is justified so long as we are interested in transverse momenta which are not much larger than the saturation scale,  such that $\alpha_s\ln(\KT^2/Q_s^2)$, for which one can neglect the DGLAP evolution. Here however we shall use this simplified set-up for pedagogical purposes, in order to be better disentangle the effects of the DGLAP evolution that will be added later.
In practice, we will effectively let ${P}^{(+)}_{gg}(\xi)\to 0$ in \eqn{F0}, meaning that the boundary condition in $\KT$-\,space reduces to its tree-level version, $\mcal{F}_0(x,  \KT^2)=	\mcal{F}_g^{(0)}(x, K_{\perp}^2)$, but we will still keep a running coupling and (for consistency) a non-zero value for $\beta_0$.

We start with DLA, where the CSS solution is known in analytic form, cf. \eqn{DLAsol}. When specialised to the tree-level BC, this reduces to %($\mcal{F}_0(x,  \KT^2)=	\mcal{F}_g^{(0)}(x, K_{\perp}^2)$) 
\begin{align}
	\hspace{-0.6cm}
	\label{DLAsol_tree}
	\mcal{F}_g(x, K_\perp^2, Q^2)= \left[\mcal{F}_g^{(0)}(x,  \KT^2)
	+ \frac{\KT^2}{\KT^2 + \mu_0^2}\,
	\frac{\alpha_s(\KT^2+m_0^2)N_c}{2\pi^2}\,\frac{xG_{\beta}^{(0)}(x,\KT^2)}{\KT^2}\,
	\ln\frac{Q^2}{\KT^2}\right]
	\exp \!\left[-\rmS(\KT^2,Q^2)\right].
\end{align}
The gluon DPDF $xG_{\beta}^{(0)}(x,\KT^2)$ is computed according to \eqn{PDFDLA} in which we replace the boundary function $\mcal{F}_0$ by its tree-level expression $\mcal{F}_g^{(0)}$.
 As compared to \eqn{DLAsol}, we have now introduced a specific regularisation of the  $\Theta$-function, that is,
\begin{align}
	\label{theta_smooth}
	\Theta(\KT,\mu_0) = \frac{\KT^2}{\KT^2 + \mu_0^2}.
\end{align} 
We recall that $\mu_0$ is a scale of order $Q_s$ whose variations around $Q_s$ should be seen as a source of scheme dependence. For the purposes of this section, this scheme dependence is less important, hence we shall identify $\mu_0^2$ with $\tilde{Q}_s^2(x)$~\footnote{Incidentally, \eqn{theta_smooth} with $\mu_0^2=\tilde{Q}_s^2(x)$ obeys geometric scaling, i.e. it is a function of the ratio $\KT^2/\tilde{Q}_s^2(x)$.}. Furthermore, we have shifted the argument of the running coupling by $m_0^2 = 0.36$~GeV$^2$ to avoid the Landau pole. Of course, the same prescriptions should be used within the Sudakov exponent \eqref{sudakovK}, which becomes
\begin{align}
	\label{sudKTheta}
	\rmS(\KT^2,Q^2)
	=
	\frac{N_c}{\pi}
	\int_{K_{\perp}^2}^{Q^2} 
	\frac{\rmd\lt^2}{\lt^2+ \mu_0^2}\,
	%\Theta(\lt,\mu_0)\,
	\alpha_s(\lt^2+m_0^2)
	\left(\frac{1}{2}
	\ln\frac{Q^2}{\lt^2}
	-\beta_0\right).
	%\equiv \ & \rmS_d(\KT^2, Q^2) + \rmS_s(\KT^2, Q^2),
\end{align}

\begin{figure}
	\begin{center}
		\includegraphics[width=0.45\textwidth]{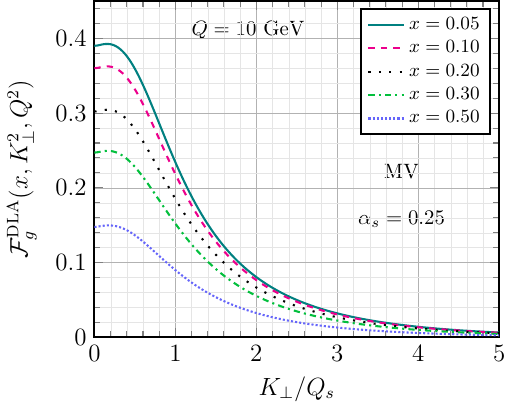}\qquad
		\includegraphics[width=0.45\textwidth]{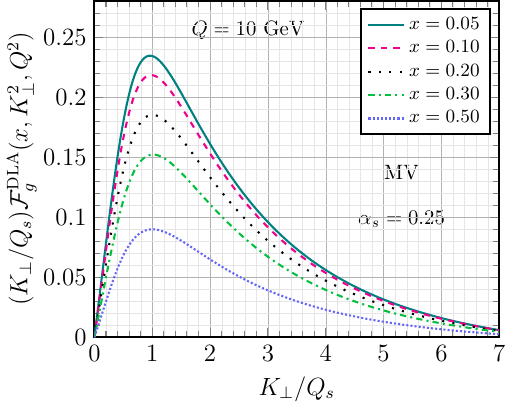}
		\includegraphics[width=0.45\textwidth]{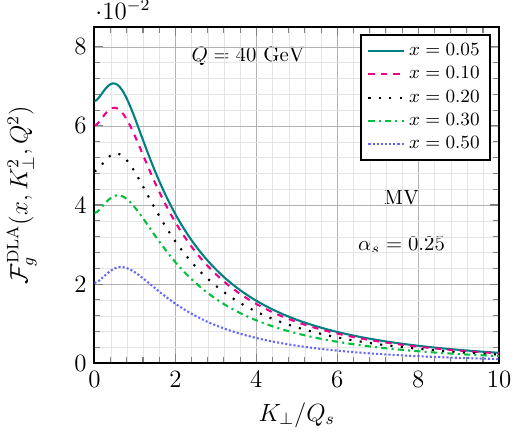}\qquad
		\includegraphics[width=0.45\textwidth]{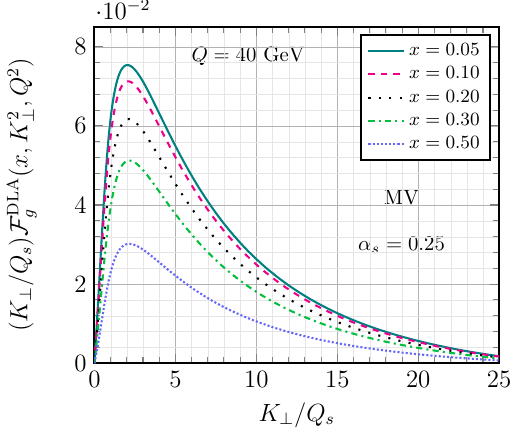}
	\end{center}
	\caption{\small The solution to the DLA version of the CSS equation, for a fixed coupling $\alpha_s=0.25$ and $\beta_0=0$, is plotted as a function of $\KT/Q_s$ for $\KT < Q$, for several values of $x$, and for two values of the hard scale: $Q=10$~GeV (upper row) and $Q=40$~GeV (lower row). We omit the overall factor  $S_\perp/(4\pi^3)$. }
\label{fig:DLA}
\end{figure}

For a qualitative study of the DLA solution, we will first use an even simpler approximation in which the coupling is fixed (we chose $\alpha_s=0.25$) and, for consistency, we also put $\beta_0=0$. Then the  gluon DPDF reduces to its tree-level approximation $xG^{(0)}(x,\KT^2)$, cf. Eq.~\eqref{xGP}. Using this approximation, together with the MV model~\footnote{For details on our numerical implementation of the MV model, we refer to Sect.~6.1 in Ref.~\cite{Hauksson:2024bvv}; see notably Eqs.~(6.6)--(6.7) there.}
with adjoint saturation scale $Q_s^2=2$~GeV$^2$, we have obtained the results shown in Fig.~\ref{fig:DLA}. Specifically, we have plotted  the quantities  $\mcal{F}_g(x, K_\perp^2, Q^2)$ and $(\KT/Q_s)\mcal{F}_g(x, K_\perp^2, Q^2)$ as functions of $\KT$ for several values of $x$ ($x=0.05, 0.1, 0.3, 0.5$) and for  two values ($Q=10$~GeV and $Q=40$~GeV) of the hard resolution scale. As usual, we omit the overall factor $S_\perp/(4\pi^3)$, so the plotted functions are dimensionless. 
We exhibit the results for all the values $\KT$ up to $Q$, but in practice they can be trusted only for sufficiently low values $\KT/Q\ll 1$, for which our approximations are justified.
%\footnote{By construction (due to our formulation of the boundary condition), the effects of the CSS evolution vanish when $\KT\to Q$, but of course such large values of $\KT$ are not properly described by our approximations.}.  
Indeed, both the validity of the TMD factorisation for the di-jet process at hand and the derivation of the CSS equation from the NLO corrections have heavily relied on the assumption that $\KT^2\ll Q^2$ (cf.  Sect.~\ref{sec:CSS}).
%,  we have preserved the NLO corrections enhanced by (single or double) powers of the logarithm $\ln(Q^2/\KT^2)$, while neglecting the other, pure--$\order{\alpha_s}$, corrections. 
That said, all the salient features visible in  Fig.~\ref{fig:DLA} lie at sufficiently low values of $\KT/Q$ to be trustable, as we shall shortly argue.

%
%\begin{figure}
%	\begin{center}
%		\includegraphics[width=0.45\textwidth]{Sec_4_DLA_fc_MV_Q_10_scaled.pdf}\qquad
%		\includegraphics[width=0.45\textwidth]{Sec_4_DLA_fc_MV_Q_10_scaled_tail.pdf}
%	\end{center}
%	\caption{\small	The same as in Fig.~\ref{fig:F0b}, but after dividing out to dominant $x$-behaviour of  $\tilde{\mcal{F}}_g^{(0)}(x, \mu_b^2)$ as suggested by  Eqs.~\eqref{Ftreesmallb} and \eqref{xGPhigh}, and in terms of the ``scaling'' ($x$-dependent) variable $\tilde Q_s(x)\bt $.}
%\label{fig:DLA10_scaled}
%\end{figure}
%
To better appreciate the results in Figs.~\ref{fig:DLA}, it is useful to compare them to the corresponding ones at tree-level, as shown in  Fig.~\ref{fig:tree}. This comparison reveals that the main effect of the CSS evolution is to push the gluons from the saturation plateau at $\KT\lesssim \tilde Q_s(x)$ towards higher momenta  $\KT\gg \tilde Q_s(x)$, where they tend to generate a power tail $1/\KT^2$. 
 %(as described by the ``real'' term within the square brackets in \eqn{DLAsol_tree}). 
 This power tail is however strongly distorted by the virtual terms encoded in the Sudakov factor, hence it cannot be easily recognised in the numerics. With increasing $\KT$ towards $Q$, the effects of the evolution become less and less important (they exactly cancel, by construction, when $\KT=Q$), meaning that the gluons must accumulate at intermediate momenta $ \tilde Q_s(x)\ll \KT\ll Q$. This is indeed visible in the plots for  $(\KT/Q_s)\mcal{F}_g(x, K_\perp^2, Q^2)$ which show a maximum in this intermediate range. One can estimate the location of this maximum by using the following approximation to \eqn{DLAsol_tree}, valid when $\KT\gg  \tilde Q_s(x)$ (with a fixed coupling and  $\beta_0=0$):
\begin{align}
	%\hspace{-0.6cm}
	\label{DLAsol_largeK}
	\mcal{F}_g(x, K_\perp^2, Q^2)= 	\frac{\alpha_sN_c}{2\pi^2}\,
	\ln\frac{Q^2}{\KT^2}\,\frac{xG^{(0)}(x,\infty)}{\KT^2}\,
	\exp \!\left(-\frac{\alpha_sN_c}{4\pi}\ln^2\frac{Q^2}{\KT^2}\right),
\end{align}
with $xG^{(0)}(x,\infty)$ given by Eq.~\eqref{xGPhigh}.
For small $\abar\equiv {\alpha_sN_c}/{\pi}$, one easily finds that the function $(\KT/Q_s)\mcal{F}_g(x, K_\perp^2, Q^2)$ has a maximum at~\footnote{\eqn{DLAsol_largeK} also predicts a maximum for the function $\mcal{F}_g(x, K_\perp^2, Q^2)$, which however lies at the much lower value $\KT^{\rm max}\simeq  Q\,\rme^{-\frac{1}{\abar}}$,  outside the validity range of the  large-$\KT$ approximation leading to \eqn{DLAsol_largeK}. That said, such a maximum is indeed visible, at least marginally, in the plots corresponding to $Q=40$~GeV.}
\beq\label{Kmax}
\KT^{\rm max}\simeq Q\,\rme^{-\frac{1}{2\abar}}\,.\eeq
Clearly, this value $\KT^{\rm max}$ is parametrically smaller than $Q$, hence it lies within the validity limits of our approximations. For $\alpha_s=0.25$ and $N_c=3$, one finds $\KT^{\rm max}\sim 0.1 Q$, in qualitative agreement with the numerical results  in Fig.~\ref{fig:DLA}.

%
%\begin{figure}
%	\begin{center}
%		\includegraphics[width=0.45\textwidth]{Sec_4_DLA_fc_MV_Q_40.pdf}\qquad
%		\includegraphics[width=0.45\textwidth]{Sec_4_DLA_fc_MV_Q_40_tail.pdf}
%	\end{center}
%	\caption{\small The FT of the tree-level gluon DTMD, as numerically obtained from Eq.~\eqref{Ftree} and after dividing out the dimensionless factor  $Q_s^2 S_\perp/(4\pi^3)$. In the right plot we use a logarithmic scale on the vertical axis. }
%\label{fig:DLA40}
%\end{figure}
%

\begin{figure}
	\begin{center}
	 \includegraphics[width=0.45\textwidth]{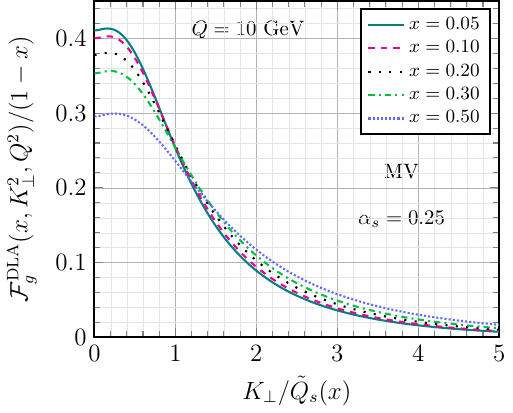}\qquad
		\includegraphics[width=0.45\textwidth]{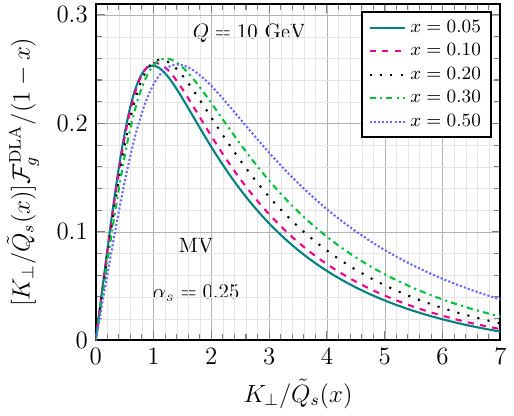}
		\includegraphics[width=0.45\textwidth]{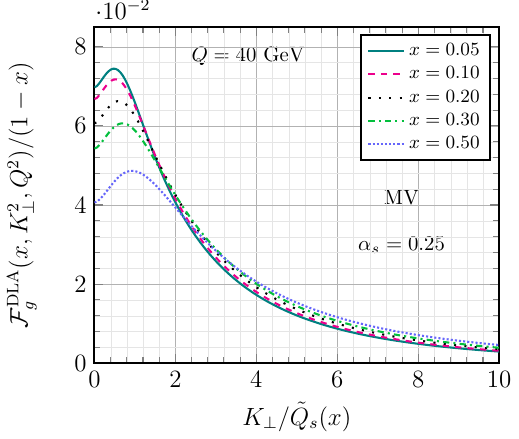}\qquad
		\includegraphics[width=0.45\textwidth]{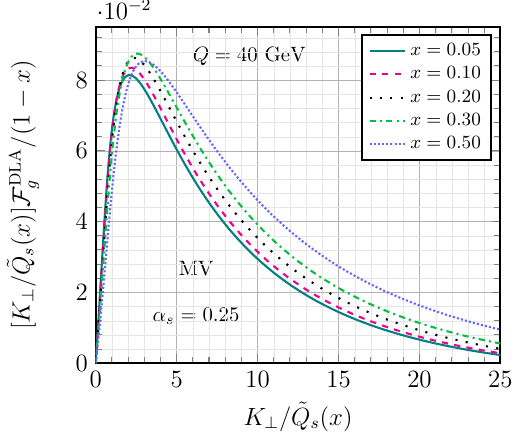}
	\end{center}
	\caption{\small A test of the quality of geometric scaling for the solution to the CSS equation in the
	double-logarithmic approximation (with fixed coupling and $\beta_0=0$). We use the same plotting conventions as in the corresponding test at tree-level, cf. Fig.~\ref{fig:tree_scaled}. By comparing these figures, one can notice that the quality of geometric scaling gets degraded by the CSS evolution.}
\label{fig:DLA_scaled}
\end{figure}

It is useful to study the $\KT$-distribution produced by the CSS evolution in more detail. To that aim, let us introduce  the average transverse momentum, defined as
\begin{align}
	\label{ktave}
	\langle \KT \rangle 
	\equiv 
	\frac{\int \dif^2 \bK\, \KT \mcal{F}_g(x, K_\perp^2, Q^2)}
	{\int \dif^2 \bK\, \mcal{F}_g(x, K_\perp^2, Q^2)}
\end{align}
One finds
\begin{align}
	\label{ktavetoy}
	\langle \KT \rangle_{\rm tree} \sim \tilde{Q}_s(x)
	\qquad \textrm{and} \qquad
	\langle \KT \rangle_{\rm CSS} \simeq 2 \abar Q, 
\end{align}
where the first estimate immediately follows from the piecewise approximation~\eqref{Ftreelimits}, while the second one has been obtained by using the large-$\KT$ expression \eqref{DLAsol_largeK} in the limit $\abar \ll 1$. 
%The integrand in the numerator in \eqn{ktave} involves the function $\KT^2 \mcal{F}_g(x,\KT^2,Q^2)$ which develops a maximum at $\KT=\KT^*$ with
%\begin{align}	\label{ktmax} K^*_{\perp \rm tree} \sim \tilde{Q}_s(x) 
	%\qquad \textrm{and} \qquad
	% K^*_{\perp \rm CSS} \simeq Q %\mkern1mu 
	%\, e^{\textstyle-\frac{1}{\sqrt{2\abar}}}.\end{align}

The above results illustrate the shift in the $\KT$-\,distribution due to the CSS evolution,  from values of order $\tilde{Q}_s(x) $ at tree-level to values proportional to the hard scale $Q$, but with proportionality factors which are parametrically small at weak coupling. Despite the smallness of $\alpha_s$, the typical transverse momenta are much larger after the CSS evolution, notably due to the creation of the tail in $1/\KT^2$. More estimates of this type, including the expectation values of higher powers of  $\KT$ and of $\ln(Q^2/\KT^2)$,  are presented in App.~\ref{app:mm}.

Another effect of this change in the $\KT$-\,distribution is the deterioration of the quality of geometric scaling. Clearly, this scaling is violated by the higher order corrections proportional to powers of $\ln(Q^2/\KT^2)$, as resummed by the CSS evolution. This degradation is indeed visible in Fig.~\ref{fig:DLA_scaled}, that should be compared to the respective results at tree-level, cf. Fig.~\ref{fig:tree_scaled}.
%\red{\it (We should plot exactly the same functions as in  Fig.~\ref{fig:tree_scaled}, in terms of the scaling variable $\bt\tilde Q_s(x)$.)}

\begin{figure}
	\begin{center}
		\includegraphics[width=0.45\textwidth]{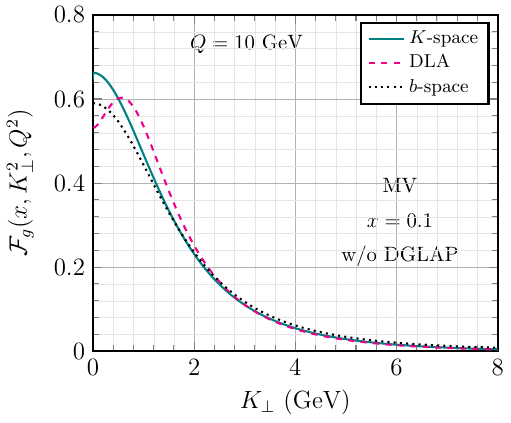}\qquad
		\includegraphics[width=0.45\textwidth]{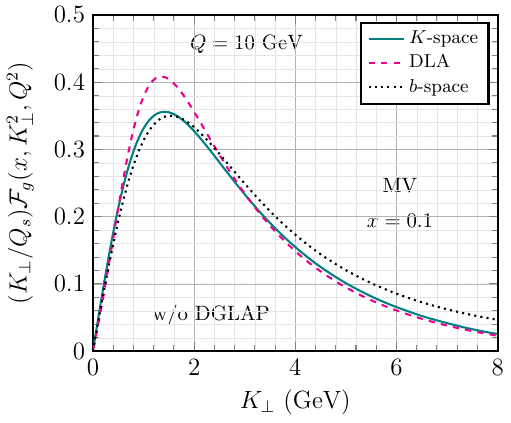}
		\includegraphics[width=0.45\textwidth]{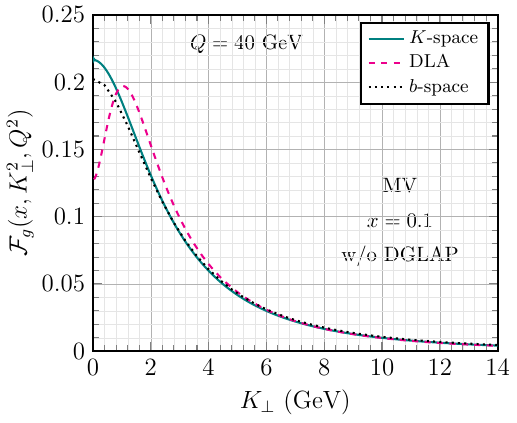}\qquad
		\includegraphics[width=0.45\textwidth]{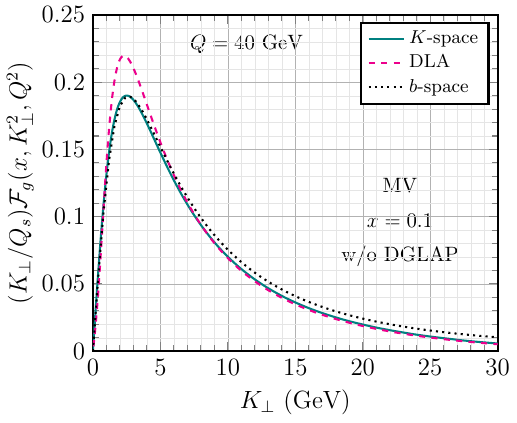}
	\end{center}
	\caption{\small Numerical solutions to the three versions of the CSS equation ($\KT$-\,space, $\bt$-\,space, and DLA) in the absence of the DGLAP evolution (${P}^{(+)}_{gg}(\xi)\to 0$). We use the one-loop running coupling and the physical value for  $\beta_0$ corresponding to $N_f=4$. The results are plotted as a function of $\KT$ for $x=0.1$ and for two values of the hard scale: $Q=10$~GeV (upper row) and $Q=40$~GeV (lower row). We omit the overall factor  $S_\perp/(4\pi^3)$. }
\label{fig:DLAfull}
\end{figure}

After this detailed discussion of the DLA version of the CSS equation, let us now present numerical results for all the three versions of this equation introduced in Sect.~\ref{sec:evol}. To that aim, we restore the running coupling and the physical, non-zero, value for $\beta_0$.
 In order to facilitate the comparison between the respective predictions, we show in Fig.~\ref{fig:DLAfull} the three sets of results in the same plots, as functions of $\KT$ for fixed values of $x$ and $Q$. We consider the same two values for the hard scale as above, i.e. $Q=10$~GeV and $Q=40$~GeV, but we limit ourselves to a single value for $x$, namely $x=0.1$, to avoid a proliferation of plots. We use the one-loop running coupling and the physical value for $\beta_0$ (cf. \eqn{runcoup} with $N_f=4$ and $\Lambda_{\rm QCD} = 0.1535$~GeV), and we choose  $\mu_0^2=Q_s^2=2$~GeV$^2$ for the separation scale which marks the onset of the CSS evolution. As announced, we solve the respective versions of the CSS equation with tree-level boundary conditions, that is,
\begin{align}
&\,\mcal{F}_0(x,  \KT^2) =
	\mcal{F}_g^{(0)}(x, K_{\perp}^2)\quad \mbox{in $\KT$-\,space,}
	\nn*[0.2cm]
&\,\tilde{\mcal{F}}_0(x, \mu_b^2)=\,\frac{xG_{\beta}^{(0)}(x,\mu_b^2)}{xG^{(0)}(x,\mu_b^2)}\,
	 \tilde{\mcal{F}}_g^{(0)}(x, \mu_b^2)\quad \mbox{in $\bt$-\,space}.
\end{align}
Here $xG_{\beta}^{(0)}(x,\mu_b^2)$ is the solution to \eqn{DLAxG} (with $\mcal{F}_0\to \mcal{F}_g^{(0)}$)
evaluated at $Q^2=\mu_b^2$.  It is important to observe that the BC in $\bt$-\,space does not reduce to its tree-level approximation in \eqref{F0b0} despite the fact that we ignore the DGLAP evolution (we set ${P}^{(+)}_{gg}(\xi)\to 0$). This is due to the fact that we use the physical (non-zero) value for the gluon anomalous dimension $\beta_0$, which amplifies the gluon PDF as compared to its naive tree-level estimate for $\beta_0=0$, as manifest in \eqn{PDFDLA}. Notice also that when solving the CSS equation in $\KT$-\,space, \eqn{CSSKT}, we assume that the evolution starts at the largest among the scales $\KT^2$ and $\mu_0^2$; that is, the BC condition is truly formulated at $Q^2=\max[\KT^2,\mu_0^2]$.

When inspecting the various curves in Fig.~\ref{fig:DLAfull}, it is most interesting to compare the results corresponding to the two ``accurate'' versions of the CSS equation --- in $\KT$-\,space and in $\bt$-\,space, respectively --- which enforce transverse momentum conservation. As explained in Sect.~\ref{sec:bspace}, these two equations are exactly mapped onto each each other via the Fourier transform \eqref{FTTMD}, yet their solutions are not guaranteed to be identical, or even close, to each other, because of the respective boundary conditions (which do not ``commute'' with the FT). Yet, the respective numerical results in Fig.~\ref{fig:DLAfull} turn out to be remarkably close to each other. We will return to a discussion of this point in Sect.~\ref{sec:full}, after also including the effects of the DGLAP evolution in the boundary condition.

\subsection{DGLAP evolution and the associated negativity problem}
\label{sec:DGLAP}

In this section, we shall first present numerical solutions to the DGLAP equation \eqref{DGLAP} and then discuss their implications for the boundary condition \eqref{F0} for the CSS evolution. As in the previous section, we use the mass parameter $m_0^2 = 0.36$~GeV$^2$ to screen the Landau pole in the running coupling and we employ the smooth regularisation shown in  \eqn{theta_smooth} for the ``$\Theta$-function'' $\Theta(Q,\mu_0)$ appearing in the r.h.s. of \eqn{DGLAP}. The precise value of the separation scale $\mu_0$ is perhaps less important in the case of the DGLAP evolution, which proceeds via ``hard'' splittings (the transverse momenta corresponding to successive splittings are strongly ordered). In what follows we shall present results obtained with three different values for $\mu_0$, namely $\mu_0^2= (1/2, 1, 2)Q_s^2$, with $Q_s^2=2$~GeV$^2$, in order to verify the associated scheme dependence.

\begin{figure}
	\begin{center}
		\includegraphics[width=0.45\textwidth]{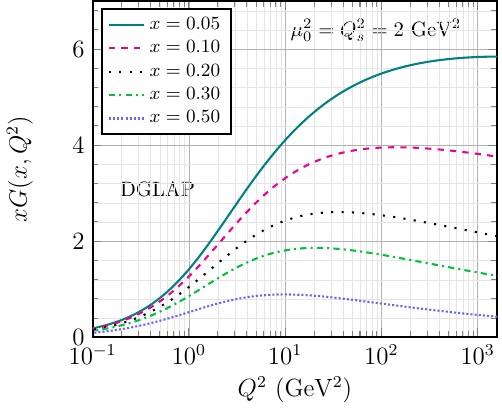}\qquad
		\includegraphics[width=0.46\textwidth]{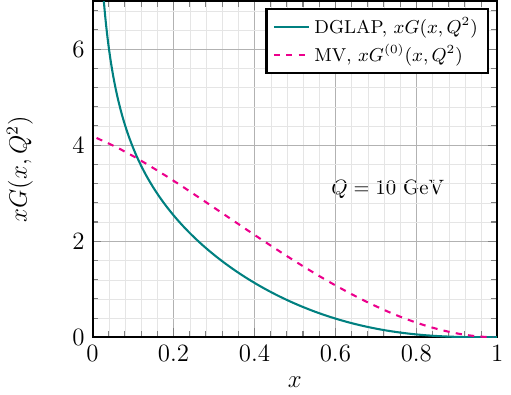}
			\end{center}
	\caption{\small  Left: the solution to the DGLAP equation \eqref{DGLAP} for the gluon DPDF with the source term (the tree-level DTMD)  computed from the MV model with $Q_s^2=2$~GeV$^2$. The separation scale $\mu_0$ is fixed to the value of $Q_s$. Right: a comparison between the DGLAP solution and the tree-level gluon DTMD for a hard scale $Q=10$~GeV. The PDF has been normalised by dividing out the dimensionless factor $Q_s^2S_\perp/(4\pi^3)$. }
\label{fig:PDF}
\end{figure}

\begin{figure}
	\begin{center}
		\includegraphics[width=0.45\textwidth]{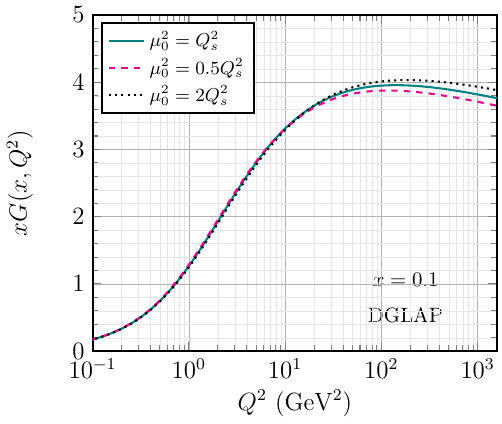}\qquad
		\includegraphics[width=0.45\textwidth]{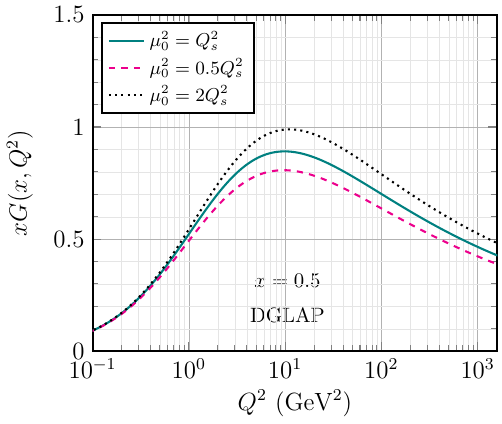}
	\end{center}
	\caption{\small  A study of the scheme dependence of our DGLAP solutions as coming from varying the separation scale $\mu_0^2$ by a factor of 2 around an average value $\mu_0^2=Q_s^2$.  }
\label{fig:PDF_scheme}
\end{figure}

Specifically, in the left plot in  Fig.~\ref{fig:PDF} we fix $\mu_0^2=Q_s^2$ and we present our numerical results for $xG(x, Q^2)$ as a function of $Q^2$  and for five  values  of $x$. In the right plot, we also fix $Q=10$~GeV and compare the  DGLAP solution $xG(x, Q^2)$ to the respective tree-level prediction $xG^{(0)}(x, Q^2)$, Eq.~\eqref{xGP}, plotted as functions of $x$. In Fig.~\ref{fig:PDF_scheme} we compare the  DGLAP solutions obtained for the three values of $\mu_0^2$ aforementioned and for two values of $x$:  $x=0.1$ in the left plot and $x=0.5$ in the right plot.

\begin{figure}
	\begin{center}
		\includegraphics[width=0.46\textwidth]{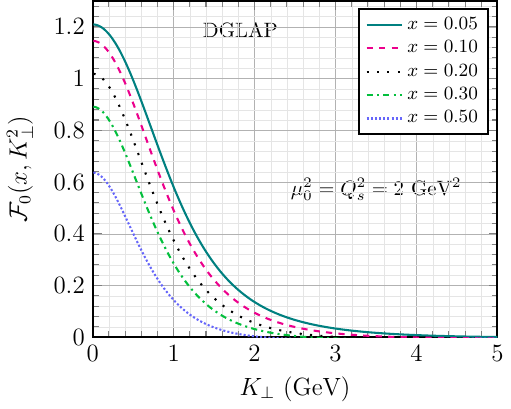}\qquad
		\includegraphics[width=0.45\textwidth]{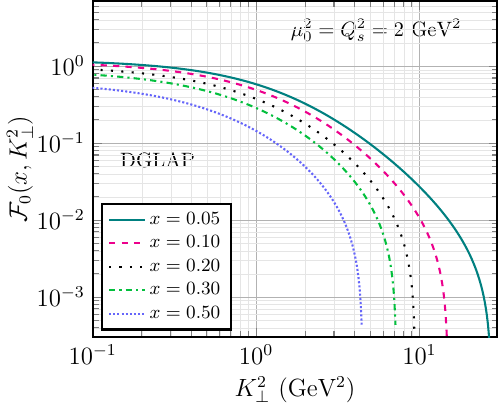}
			\end{center}
	\caption{\small The boundary function $\mcal{F}_0(x, K_\perp^2)$ evaluated from DGLAP solutions according to \eqn{F0}. To better illustrate the negativity problem, we exhibit the same curves twice, first in regular units (left plot) and then in logarithmic units (right plot). As usual, we omit an overall factor $S_\perp/(4\pi^3)$. }
\label{fig:F0}
\end{figure}

Next we use the DGLAP solutions to evaluate the boundary function $\mcal{F}_0(x,  \KT^2)$ according to \eqn{F0}, in which we also insert the ``$\Theta$-function'' $\Theta(\KT,\mu_0)=\KT^2/(\KT^2+\mu_0^2)$ in front of the DGLAP piece. We plot our results in Fig.~\ref{fig:F0} as a function of $\KT$ for $\mu_0=Q_s$ and for the same values of $x$ as before. These plots exhibit a perhaps intriguing trend: due to the DGLAP evolution, the function $\mcal{F}_0(x,  \KT^2)$ becomes negative for sufficiently large $\KT$  (at a given $x$) and for large enough $x$ (at fixed $\KT$). For more clarity we also show in the right plot the same curves in logarithmic units. Before we speculate on the origin of this negativity in our calculations, let us explain that it could have been anticipated by inspection of  \eqn{F0}. Indeed, the r.h.s. there can be equivalently rewritten as (recall Eqs.~\eqref{DGLAP} and \eqref{DLAxG})
\begin{align}
  \label{F02}
  \mcal{F}_0(x, K_\perp^2)=\, 
  \frac{1}{\pi}
  \frac{\del\,xG(x, \KT^2) }{\del K_\perp^2}
  -\beta_0
  \frac{\alpha_s(\KT^2)N_c}{\pi^2}
   \frac{xG(x, \KT^2)}{\KT^2+\mu_0^2}\,,
 \end{align}
where the first term in the r.h.s. --- the derivative of the gluon DPDF --- is bound to become negative with increasing $\KT^2$ and/or $x$, because of the dynamics underlying the DGLAP evolution: via successive (hard) splittings, gluons at large $x$ lose longitudinal momentum and thus moves towards bins at lower values of $x$. 
The second term in \eqn{F02}, which is manifestly negative, amplifies this trend towards negativity. Taken at face value, this behaviour looks deeply unphysical: it suggests that the gluon DTMD, hence the di-jet cross-section, become negative when $\KT^2$ approaches $Q^2$. At this point, one should recall that our approximations are only valid at $\KT^2\ll Q^2$ and hence cannot properly describe this approach. Clearly this issue sheds doubts on our formulation of the CSS evolution, as a boundary value problem with the BC specified at the point $Q^2=\KT^2$ which is not under control.

To gain further insight with this issue, it is useful to recall that the DGLAP and the CSS evolutions where generated {\it together} by the NLO corrections, and that the original NLO contribution encoding both of them is {\it manifestly positive}. The negativity issue is ``artificially'' generated by separating the two contributions and neglecting corrections suppressed by powers of $\KT/Q$. Specifically, the relevant NLO correction (as computed in Ref.~\cite{Caucal:2024bae} in the context of inclusive back-to-back di-jets) reads
\begin{align}\label{CSSreal}
&\ \Delta \mcal{F}_g(x, K_{\perp}^2, Q^2)\,=\, 
 \frac{\alpha_s(K_\perp^2) }{2\pi^2}\frac{1}{K_\perp^2}
\int_{x}^{1/(1+\xi_0)}\rmd\xi\,{P}_{gg}(\xi)\,\frac{x}{\xi}G^{(0)}\left(\frac{x}{\xi}, K_\perp^2\right)
\,  \nn*[0.2cm]
&\quad =\,\frac{\alpha_s(\KT^2)N_c}{2\pi^2}
\,\ln\frac{Q^2}{\KT^2}\,\frac{xG^{(0)}  (x,\KT^2)}{\KT^2}+
 \frac{\alpha_s(K_\perp^2) }{2\pi^2}\frac{1}{K_\perp^2}
\int_{x}^{1}\rmd\xi\,{P}^{(+)}_{gg}(\xi)\,\frac{x}{\xi}G^{(0)}\left(\frac{x}{\xi}, K_\perp^2\right)+\order{\frac{\KT}{Q}}.
\end{align}
Here, $\xi_0=\KT/Q\ll 1$ and ${P}_{gg}(\xi)$ is the unregularised $g\to gg$ splitting function, i.e. \eqn{Pggreg} {\it without} the plus prescription. To obtain the final result we used the following identity
\begin{align}
\int_{x}^{1-\xi_0}\rmd\xi\,\frac{f(\xi)}{1-\xi}&\,=\int_{x}^{1}\rmd\xi\,\frac{f(\xi)}{\big(1-\xi \big)_+}
+f(1)\ln\frac{1}{\xi_0} - \int_{1-\xi_0}^1\rmd\xi\,\frac{f(\xi)-f(1)}{1-\xi}\nn*[0.2cm]
&\,= \int_{x}^{1}\rmd\xi\,\frac{f(\xi)}{\big(1-\xi \big)_+}
+f(1)\ln\frac{1}{\xi_0} + \frac{\rmd f}{\rmd \xi}\bigg |_{\xi=1}\,\xi_0+ \order{\xi_0^2}\,,
\end{align}
%\order{\xi_0}\,,\eeq
and we ignored corrections of $\order{\xi_0}$.  The two lines in
\eqn{CSSreal} can be recognised as a generalisation of our previous manipulations in  Eqs.~\eqref{DeltaFxi}--\eqref{DeltaFrecoil}. As compared to the latter, \eqn{CSSreal} also includes the effects of  real gluon emissions with relatively large transverse momenta $\lt\simeq \KT\gg Q_s$ which occur in the {\it initial} state. The full DGLAP splitting function ${P}_{gg}(\xi)$  has been reconstructed after adding both initial-state and final-state emissions, together  with the interference terms~\cite{Caucal:2024bae}.
 
 The NLO correction in the first line of  \eqn{CSSreal} is positive definite, as anticipated. So long as $\xi_0\ll 1$ (i.e. $\KT\ll Q$), its limited expansion shown in the second line makes full sense. But there is no guarantee that this limited expansion remains positive for larger values of $\KT/Q$. We have already seen that the second term in this expansion, which involves ${P}^{(+)}_{gg}(\xi)$, eventually becomes negative when increasing $x$ and/or $\KT$. As for the first term, proportional to $\ln(Q^2/\KT^2)$, this is clearly positive, but its importance is reduced when increasing $\KT$ towards $Q$. Hence, the final expression in \eqn{CSSreal} will eventually become negative as well. For $\KT\gg Q_s$, this expression is the same as the prefactor within the square brackets in the DLA solution \eqref{DLAsol} to the CSS equation. Hence, the DLA solution is bound to become negative for sufficiently large $x$ and/or $\KT/Q$. This will be confirmed by the numerical results in the next section, which will exhibit a similar negativity issue for the other versions of the CSS equation.

At a first sight, one may think that a possible strategy to avoid the negativity problem consists in replacing the expansion in the second line of  \eqn{CSSreal} by the original expression in the first line: if one does that within the DLA solution \eqref{DLAsol}, then one indeed finds a positive result for any $\xi_0< 1$. Yet, this ``strategy'' would be artificial and inappropriate for several reasons. First, it would be inconsistent with the DGLAP and CSS evolutions, which differently treat the two terms in the expansion  in the second line of  \eqn{CSSreal}. Second, this  ``strategy'' could not be applied to the more accurate versions of the CSS equation, as described in Sects.~\ref{sec:CSS} and \ref{sec:bspace}, which must be numerically solved with the BC \eqref{F0}. Last but not least, the difference between the expressions in the two lines of  \eqn{CSSreal} is not truly under control within the present approximations. Indeed, already the first line there has been obtained as a leading-order contribution in an expansion in powers of $\KT/Q$; hence our whole approach is valid only in the regime where $\KT/Q$ is small enough for the  expressions in the two lines of  \eqn{CSSreal} to be physically equivalent (and numerically close) to each other.

In our opinion, this negativity issue underscores the limitation of the whole TMD approach to problems with widely separated transverse momentum scales, $\KT\ll Q$. We have not been able to identify a criterion for the upper limit on $\KT/Q$ up to which the formalism can be trusted. That said, most of the interesting properties exhibited by the CSS solutions lie at values of $\KT/Q$ which are parametrically small, where our results are expected to be reliable.

\subsection{The interplay between the CSS and the DGLAP evolutions}
\label{sec:full}

%\red{\bf (to be developed)}

In this section, we present our main physical results in this paper, namely the numerical results for the gluon diffractive TMD as obtained from combined solutions to the DGLAP and CSS equations. As before, we shall present numerical solutions to all the three versions of the CSS equation, but the main focus should be on the two more accurate versions of this equation --- one in $\KT$-\,space, cf. \eqn{CSSKT}, the other one in  $\bt$-\,space, cf. \eqn{CSSimpact} --- which obey transverse momentum conservation at the gluon splitting vertex. Within the present approximations, these two versions cannot be physically distinguished from each other, hence it will be important to observe the numerical difference between their solutions. The DLA solution on the other hand involves additional approximations, so it may considerably differ from the other two.

\begin{figure}
	\begin{center}
		\includegraphics[width=0.45\textwidth]{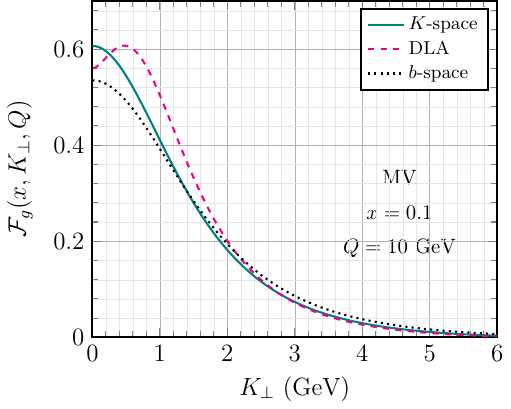}\qquad
		\includegraphics[width=0.45\textwidth]{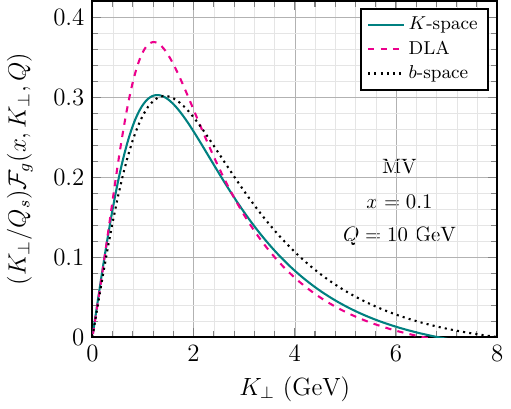}
		\includegraphics[width=0.45\textwidth]{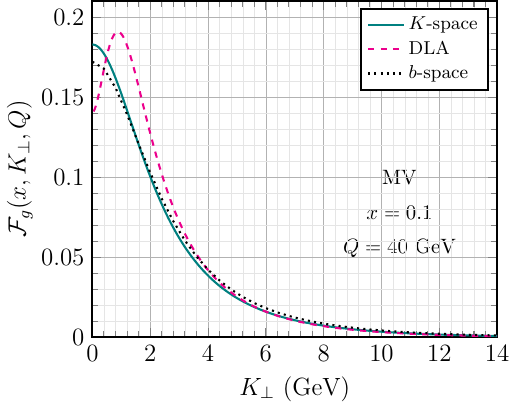}\qquad
		\includegraphics[width=0.45\textwidth]{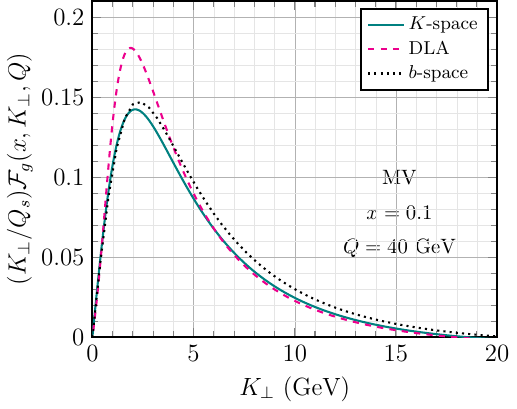}
	\end{center}
	\caption{\small Numerical solutions to the three versions of the CSS equation --- $\KT$-\,space, $\bt$-\,space, and DLA --- including the DGLAP evolution in the boundary condition and with the one-loop running coupling. The results are plotted as a function of $\KT$ for $x=0.1$ and for two values of the hard scale: $Q=10$~GeV (upper row) and $Q=40$~GeV (lower row). We omit the overall factor  $S_\perp/(4\pi^3)$. }
\label{fig:full}
\end{figure}

\begin{figure}
	\begin{center}
		\includegraphics[width=0.45\textwidth]{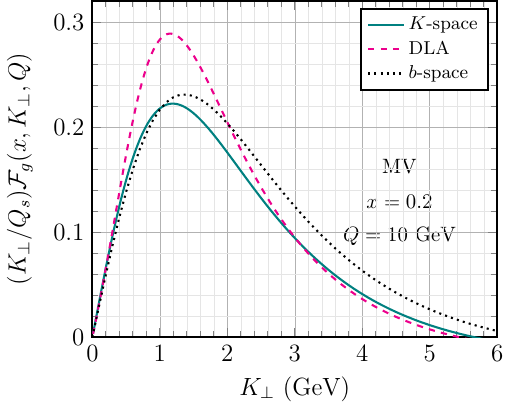}\qquad
		\includegraphics[width=0.44\textwidth]{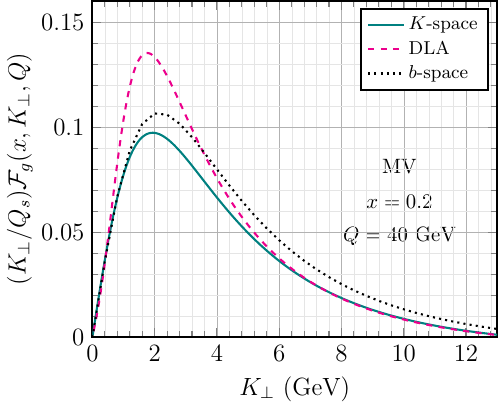}
		\includegraphics[width=0.45\textwidth]{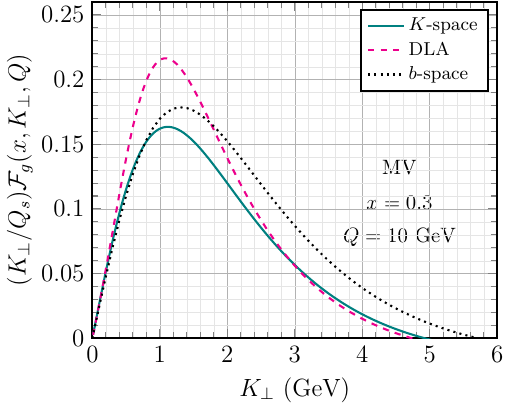}\qquad
		\includegraphics[width=0.44\textwidth]{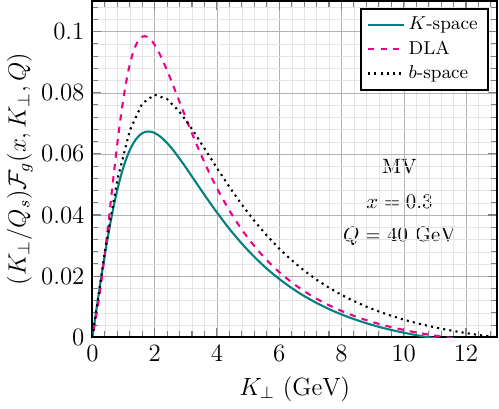}
	\end{center}
	\caption{\small The radial distribution   $(\KT/Q_s)\mcal{F}_g(x, K_\perp^2, Q^2)$ plotted as a function of $\KT$ for $x=0.2$ (upper row) and $x=0.3$ (lower row) and for the two values $Q=10$~GeV and 
	$Q=40$~GeV.
	 }
\label{fig:full2}
\end{figure}

Our results for $x=0.1$ are presented in  Fig.~\ref{fig:full}, which should be compared to the previous results without DGLAP evolution, as shown Fig.~\ref{fig:DLAfull}. Once again, one can observe a shift in the $\KT$-distribution towards momenta larger than the saturation scale and the emergence of a pronounced maximum in the radial distribution   $(\KT/Q_s)\mcal{F}_g(x, K_\perp^2, Q^2)$, with the peak position $\KT^{\rm max}$ increasing with $Q$. Moreover, the distributions eventually become negative when increasing $\KT$, as an effect of the DGLAP evolution --- in agreement with the discussion in Sect.~\ref{sec:DGLAP}. Importantly, this is also the case for the solutions obtained in the transverse coordinate representation, for which the argument  in Sect.~\ref{sec:DGLAP} does not apply as it stands: the boundary condition in $\bt$-\,space, as given by \eqn{BCbgl}, is manifestly positive. In that case, the negativity is introduced by the inverse Fourier transform. This can be heuristically understood via an argument similar to \eqn{FTDLA}: when $\KT\gg Q_s$, the inverse FT essentially acts as a derivative $\del/\del\KT^2$.

\begin{figure}
	\begin{center}
		\includegraphics[width=0.45\textwidth]{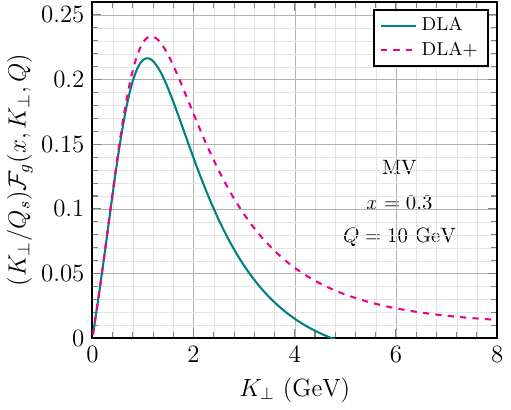}\qquad
		\includegraphics[width=0.45\textwidth]{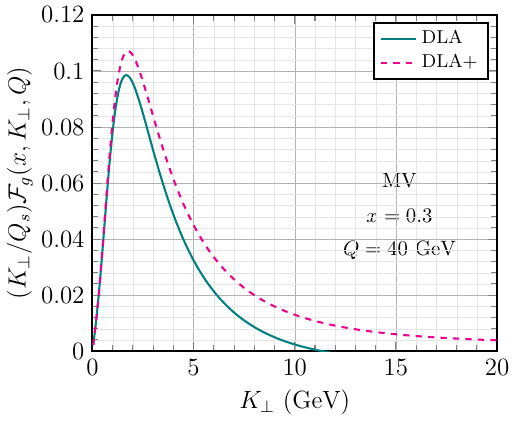}
		\includegraphics[width=0.45\textwidth]{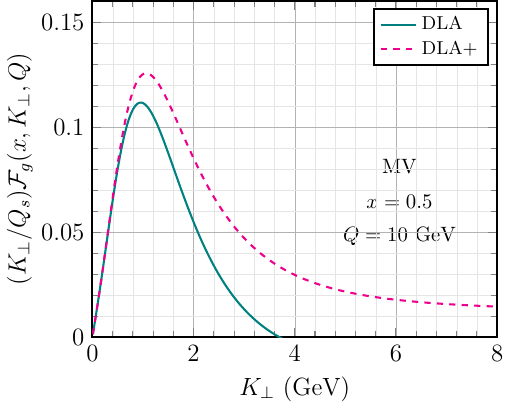}\qquad
		\includegraphics[width=0.45\textwidth]{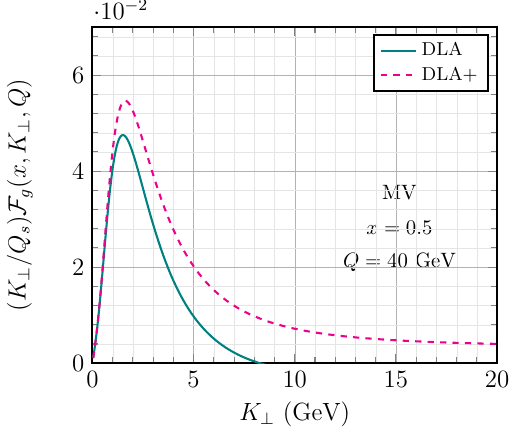}
	\end{center}
	\caption{\small The radial distribution   $(\KT/Q_s)\mcal{F}_g(x, K_\perp^2, Q^2)$ plotted as a function of $\KT$ for $x=0.2$ (upper row) and $x=0.3$ (lower row) and for the two values $Q=10$~GeV and 
	$Q=40$~GeV.
	 }
\label{fig:DLA+}
\end{figure}

For $x=0.1$, the numerical results turn negative at relatively large values $\KT \gtrsim 0.3Q$, where our approximations (more generally, the TMD factorisation) are not justified anymore. With increasing $x$, the problem moves towards lower values of $\KT$, as illustrated in Fig.~\ref{fig:full2} for  $x=0.2$ and $x=0.3$. In fact, we expect these results to become unreliable already before they turn negative. We have no sharp criterion in that sense (this would require an estimate of the power-suppressed corrections that were here ignored), but we can safely assume that our results are physically meaningful so long as $\KT$ remains parametrically smaller than $Q$. This is the case for characteristic values, like the position $\KT^{\rm max}$ of the peak, or the average momentum $\langle \KT \rangle $, cf. Eqs~\eqref{ktave}--\eqref{ktavetoy}.

One important feature of the results shown in  Fig.~\ref{fig:full} is the good agreement between the two ``accurate'' versions of the CSS solutions, as obtained in different representations --- $\KT$-\,space and  $\bt$-\,space, respectively. This is remarkable given the different formulations we used for the respective boundary conditions. In $\KT$-\,space, the boundary function $\mcal{F}_0(x, K_\perp^2)$ in \eqn{F0} is essentially one step in the DGLAP evolution with the $\beta_0$-\,piece subtracted (recall also \eqn{F02}). In $\bt$-\,space, on the other hand, the boundary function $\tilde{\mcal{F}}_0(x, \mu_b^2)$ in \eqn{BCbgl} is proportional to the solution $xG(x, \mu_b^2)$ of the DGLAP equation, including the $\beta_0$-\,piece in the splitting function. This difference has important consequences e.g. for the negativity problem: whereas the $\KT$-\,space boundary function $\mcal{F}_0(x, K_\perp^2)$ becomes negative with increasing $\KT$ and/or $x$, cf. Fig.~\ref{fig:F0}, the respective function $\tilde{\mcal{F}}_0(x, \mu_b^2)$ in $\bt$-\,space is positive definite and the negativity issue only appears after performing the inverse FT to $\KT$-\,space. There are also important differences in the implementation of the saturation effects. In  $\bt$-\,space, they are encoded in the boundary function $\tilde{\mcal{F}}_0(x, \mu_b^2)$, cf. \eqn{BCbgl}, and ensure the rapid decay of the CSS solution at large $\bt\gg 1/\tilde Q_s(x)$. In $\KT$-\,space, they enter too via the boundary condition, but they act in a more subtle way, via cancellations between real and virtual effects at $\KT\lesssim \tilde Q_s(x)$. In view of such differences, the good agreement that we observe between the  CSS solutions obtained in the two spaces is truly remarkable and also reassuring: it demonstrates that the inherent ambiguity associated with the choice of a representation is rather small (presumably even smaller than the neglected higher-order  corrections) and does not dramatically affect the physical content of our predictions.

We shall conclude this section with a little calculation which although incomplete (in the sense of not being systematic)  illustrates the way how the aforementioned negativity problem might be solved in practice by the addition of the NLO corrections suppressed by powers of $\KT/\PT$ (the ``$Y$-term''). Strictly speaking, we do not know the  whole set of such corrections, but only a part of it --- namely that part that is resumed in the expression in the first line of \eqn{CSSreal}. As already mentioned, the first two terms in the expansion of that expression in powers of $\KT/Q$, as shown in the second  line of \eqn{CSSreal}, have been used to infer the emergence of the CSS evolution in this particular context. So, their effects are already resummed in the structure of our CSS solutions. This is particularly clear at the level of the DLA solution \eqref{DLAsol}: the first term in the second line of \eqn{CSSreal} is manifest in \eqn{DLAsol}, while the second term involving ${P}^{(+)}_{gg}(\xi)$ is included via  the boundary function $\mcal{F}_0(x,  \KT^2)$, cf. \eqn{F0}. Our simple but pedagogical suggestion for  ``improving'' our previous CSS solutions consists in adding to them the difference between the first and the second lines in \eqn{CSSreal}. For simplicity, we shall perform this exercice by using the DLA approximation --- this is enough for illustration purposes. Our respective results are shown in Fig.~\ref{fig:DLA+}, where the ``improved''  DLA solution, denoted as ``DLA+'', is compared to the original respective solution for two values of $Q$ (the same as in the previous plots) and for $x=0.3$ and $x=0.5$. Remarkably, the ``improved''  solutions remain positive for all values of $\KT\le Q$ and for both  values of $x$ --- including the relatively large $x=0.5$.

\section{Conclusions and perspectives}
\label{sec:conc}

In this paper we have for the first time studied the CSS evolution of the gluon diffractive TMD as emerging from the CGC effective theory in the context of back-to-back di-jet production. Our study has involved both analytic and numerical aspects. 

On the analytic side, we have clarified the relation between three different versions of the CSS equation that exist in the literature and we employ in our analysis. Two of these versions, the ``$\KT$-\,space'' and the ``DLA'',  are formulated in the transverse momentum representation and naturally emerge from next-to-leading order calculations in the CGC. The third, ``$\bt$-\,space'', version is written in the transverse coordinate representation and is generally used in the more traditional context of TMD factorisation at moderate $x$. None of these versions has been previously used for the evolution of the diffractive (gluon or quark) TMDs. While the mathematical correspondences between these three equations are quite clear --- the $\KT$-\,space and the $\bt$-\,space equations are connected via a Fourier transform, and the DLA equation emerges via controlled approximations from the general equation in $\KT$-\,space ---, the discussion of the respective boundary conditions turns out to be more subtle: the boundary conditions are differently formulated in the two representations ($\KT$-\,space and $\bt$-\,space, respectively), so they are not exactly equivalent with each other. This problem is further complicated by the matching between the  CSS and the DGLAP evolutions, as performed at the level of the boundary conditions: this matching relies on piecewise approximations that are differently implemented in the two representations. That said, we managed to construct smooth approximations which exhibit the correct piecewise behaviour and which agree with each other (to parametric accuracy) between the two representations.

Still on the analytic side, we have constructed an exact solution to the DLA version of the CSS equation, which is particularly illuminating for the physical discussion of our results. All the results and conclusions above mentioned are valid beyond the particular context of diffraction --- they apply to the CSS evolution {\it in general}. But we also have new results which are specific to the gluon diffractive TMD. The most important among them refers to the CSS evolution in $\bt$-\,space and, more precisely, to the fact that saturation effects ensure the rapid convergence of the inverse Fourier transform (from $\bt$-\,space to $\KT$-\,space) without the need for non-perturbative regulators. This is indeed specific to diffraction, since it is only in this case that the Fourier transform of the tree-level gluon DTMD decays very fast, as $1/\bt^8$, when $Q_s \bt\gg 1$. (For the Weizs\"{a}cker-Williams gluon TMD, as occurring in the cross-section for inclusive d-jets,  this decrease is much slower, like $1/\bt^2$.)

On the numerical side, we have presented solutions for all the three versions of the CSS equation aforementioned. To disentangle the effects of the various evolutions, we have considered boundary conditions with and without the DGLAP contribution. In both cases, the main effect of the CSS evolution was to transport the gluons from the saturation plateau at $\KT\lesssim \tilde Q_s(x)$ (where they mostly lie at tree-level) to much larger transverse momenta, with characteristic values proportional to, but parametrically smaller than,  $Q$. The solutions to the $\KT$-\,space and the $\bt$-\,space equations are numerically close to each other and their difference may be seen as a measure of the theoretical uncertainty of our calculation. Of course, it might well be that the next-to-leading order corrections that we have here neglected are even larger and dominate the theoretical uncertainty in practice. 

After adding the DGLAP contribution to the boundary condition, all our numerical solutions turn negative for sufficiently large values of  $\KT$, which become smaller when increasing $x$. This is an unphysical behaviour that  in our opinion reflects an intrinsic limitation of TMD factorisation (which can only be applied for sufficiently small values of $\KT$ and/or $x$). Based on a simple, numerical, example, we have argued that this problem might be cured by the resummation of the power-suppressed corrections (the so-called ``$Y$-terms), as computed in fixed order perturbation theory.

We are confident that this work will open the way to further studies of diffractive jet production in photon-hadron interactions at high energies. On the theory side, it would be interesting to perform a similar study for the quark diffractive TMD, which enters the CGC calculations of diffractive SIDIS and of di-jet production via the quark-gluon channel~\cite{Hatta:2022lzj,Hauksson:2024bvv}. It would be also important to compute the NLO corrections to diffractive (2+1)-jet production within the CGC --- notably for extracting the contributions suppressed by powers of $\KT^2/\PT^2$, which become important when the separation between the ``hard'' scale $\PT$ and the ``semi-hard'' one $\KT$ is not so large anymore. Furthermore, one should consider the effects of the NLO corrections to the collinear evolution equations (DGLAP and CSS). These corrections are well known in the context of collinear factorisation and we expect them to be universal --- hence, to also apply for diffractive phenomena at small $x$. 

On the experimental side, it would be interesting to explore the consequences of our results for the phenomenology of diffractive d-jet production in photon-mediated processes, like electron-nucleus ($eA$) DIS at the future EIC~\cite{Accardi:2012qut,Aschenauer:2017jsk,AbdulKhalek:2022hcn} and ultra-peripheral nucleus-nucleus collisions ($AA$ UPCs) at the LHC~\cite{ATLAS:2022cbd,CMS:2022lbi,ATLAS:2024mvt}. The collinear resummations are clearly essential for the UPC experiments at CERN, where the measured jets have large transverse momenta $P_\perp\ge 20$~GeV and the di-jet imbalance was found to be large as well, $\KT\ge 6$~GeV~\cite{CMS:2022lbi}. In this kinematics, both the CSS and the DGLAP evolutions become important and their combined resummation goes beyond the scope of previous  studies~\cite{Iancu:2023lel,Shao:2024nor}. At this level, it is perhaps interesting to notice that our results for the $\KT$--distribution, which show a pronounced peak in the radial distribution   $(\KT/Q_s)\mcal{F}_g(x, K_\perp^2, Q^2)$, cf.  Figs.~\ref{fig:full},~\ref{fig:full2} and~\ref{fig:DLA+}, bear some qualitative resemblance with the experimentally found distributions for the di-jet imbalance (see Fig.~1 in Ref.~~\cite{CMS:2022lbi}). Yet, the position of the peak in our results appears to be at significantly lower values of $\KT$ than in the data (1 to 2 GeV as compared to, roughly, 6~GeV in the experiments). It would be important to understand and resolve this discrepancy.

Last but not least, our methods in this paper can be extended to {\it inclusive} processes involving hard particle production at small $x$ (in both electron-nucleus and proton-nucleus collisions), thus extending previous studies in the literature~\cite{Zheng:2014vka,Albacete:2018ruq,Stasto:2018rci,vanHameren:2016ftb,vanHameren:2023oiq,Cassar:2025vdp} --- notably by the inclusion of the DGLAP evolution. An interesting open question refers to the interplay between gluon saturation and non-perturbative phenomena in the $\bt$-\,space solutions to the CSS equation for inclusive TMDs.

\paragraph{Acknowledgements.}  We would like to thank Paul Caucal for discussions and comments on the manuscript. E.I.~and F.Y.~acknowledge financial support from the France-Berkeley-Fund from University of California at Berkeley. F.Y.~is grateful to Institut de Physique Th\'{e}orique de Saclay for hospitality and support. E.I.~is grateful for the support of the Saturated Glue (SURGE) Topical Theory Collaboration, funded by the U.S.~Department of Energy, Office of Science, Office of Nuclear Physics. The work of F.Y.~is supported in part by the U.S.~Department of Energy, Office of Science, Office of Nuclear Physics, under contract number DE-AC02-05CH11231, under the umbrella of the Quark-Gluon Tomography (QGT) Topical Collaboration with Award DE-SC0023646. S.Y.~Wei is grateful for the support from the European Center for Theoretical Studies in Nuclear Physics and Related Areas. The work of S.Y.~Wei is also supported by the National Science Foundations of China under Grant No.~12405156, the Shandong Province Natural Science Foundation under grant No.~2023HWYQ-011 and No.~ZFJH202303, and the Taishan fellowship of Shandong Province for junior scientists. Ce travail a b\'{e}n\'{e}fici\'{e} d'une aide de l'\'{E}tat au titre de France 2030 (P2I\,-\,Graduate School Physique) portant la r\'{e}f\'{e}rence ANR-11-IDEX-0003. Figure 1 was created with JaxoDraw~\cite{Binosi:2003yf}.

\appendix

\section{The tree-level diffractive TMD in transverse coordinate space}
\label{app:ft_tree}

%It must have been clear in the main text that the FT of the tree level TMD is an important quantity, since together with the corresponding PDF they determine the boundary condition in transverse coordinate space. Here we shall do a comprehensive analytical study of the gluon and quark TMDs, for both the diffractive and inclusive cases, focusing mostly on the behavior at large\,-$\mkern1mu\bt$.

In this appendix we will present a comprehensive analytical study of the Fourier transform  of the tree level gluon diffractive TMD, as computed within the MV model. We will also briefly discuss the corresponding {\it inclusive} quantity (the WW gluon TMD), for the purposes of comparing with the diffractive case.

\paragraph{The gluon DTMD:} For the diffractive DTMDs we shall need to distinguish two different scenarios depending on whether the value of $x$ is small or moderate. We shall be a bit more precise about this condition later on.

\subparagraph{Small-$x$:} When $x \to 0$, \eqn{gp} simplifies to
\begin{align}
	\label{a:gp0}
	\mcal{G}_{\mathbb{P}}(0,x_{\mathbb P}, K_{\perp}^2)=
	2 \int_0^{\infty} \frac{\rmd R}{R}\,
	J_2(K_{\perp} R)\,
	\mcal{T}_g(R,Y_{\mathbb P}).
\end{align}
Let us start by assuming that the tree level gluon DTMD is determined by the MV model, cf. \eqn{tgMV}. If one is interested in momenta $\KT^2 \lesssim Q_s^2$ (notice that the transition regime around saturation is included), then to the order of accuracy it is equivalent to consider the GBW model for which the dipole amplitude is simply a Gaussian, with the scale set by the saturation momentum in \eqn{QsQA}:
\begin{align}
	\label{a:t_gbw}
	\mcal{T}_g(R) = 1- \exp\bigg(\!\!-\frac{R^2 Q_s^2}{4}\bigg).
\end{align}
For this amplitude, the integral in Eq.~\eqref{a:gp0} can be exactly computed and then Eq.~\eqref{Ftree} leads to \cite{Iancu:2022lcw}
\begin{align}
	\label{a:f0_gbw}
	\mcal{F}_g^{(0)}(0,x_{\mathbb P},\KT^2) =
	\frac{S_\perp N_g}{8 \pi^4} \frac{Q_s^4}{\KT^4}
	\left[1 - \exp{\left(-\frac{\KT^2}{Q_s^2}\right)} \right]^2,
\end{align}  
which is consistent with the piecewise expression in Eq.~\eqref{Ftreelimits}. It is possible to perform the FT of the above exactly. Changing to a dimensionless integration variable according to $\KT=\kappa \mkern1mu Q_s$ we have
\begin{align}
\hspace*{-0.2cm}
	\label{a:f0tilde}
	\tilde{\mcal{F}}_g^{(0)}(0,x_{\mathbb P},\bt^2)
	=\,& \frac{S_\perp N_g \mkern1mu Q_s^2}{16 \pi^5}
	\int_0^\infty \frac{\dif \kappa}{\kappa^3}\, J_0(\kappa \mkern1mu \bt \mkern-1mu Q_s) \Big(1 - e^{-\kappa^2}\Big)^2
	\nn*[0.1cm]
	=\, &\frac{S_\perp N_g \mkern1mu Q_s^2}{16 \pi^5}\!
	\left[(1+B)E_1(B)- e^{-B} -(1+2B)E_1(2B) + e^{-2 B}  \right],
	 \quad B\equiv \frac{\bt^2 Q_s^2}{8},
\end{align}
where $E_1(z) = \int_z^{\infty} \dif t\, e^{-t}/t$ is the exponential integral function. In the transition region $\bt^2 Q_s^2/8 \sim 1$ the square bracket in the above is of order one  and we can convince ourselves that momenta such that $\KT^2 \lesssim Q_s^2$  dominate  the integration in the FT\footnote{In general, the FT of a Gaussian of the type considered here is sensitive to momenta up to $K_{\max} = \max(\bt Q_s^2, Q_s)$.}.

At large $\bt\gg 1/Q_s$, $\tilde{\mcal{F}}_g^{(0)}(0,x_{\mathbb P},\bt^2)$ falls exponentially, more precisely
\begin{align}
	\label{a:f0_tilde_large_b}
	\tilde{\mcal{F}}_g^{(0)}(0,x_{\mathbb P},\bt^2)
	\simeq
	\frac{S_\perp N_g \mkern1mu Q_s^2}{16 \pi^5}\bigg(\frac{8}{\bt^2 Q_s^2}\bigg)^2\!
	\exp\left( - \frac{\bt^2 Q_s^2}{8} \right)
	\quad \mathrm{for} \quad \bt^2 Q_s^2/8 \gg 1.
\end{align}
Perhaps one would have expected the low momenta $\KT^2 \lesssim 1/\bt^2$ to determine the FT for large $\bt^2$, due to the oscillations of the Bessel function. However, since the gluon DTMD saturates to a {\it constant} value when $\KT^2$ becomes very small, cf.~Eq.~\eqref{a:f0_gbw}, we readily find that the corresponding result should be proportional to $\int \dif^2 \bK\, e^{i \bK \cdot \bb}= (2\pi)^2 \delta(\bb)$ and thus vanishes. Moreover, it is important to realize that Eq.~\eqref{a:f0_gbw} implies more than that: the gluon DTMD can be written as a power series in $\KT^2$ in which all the coefficients are constants and since $ \KT^2 e^{i \bK \cdot \bb} = -\nabla_{\bb}^2\, e^{i \bK \cdot \bb}$, the FT of any $\KT^{2n}$ term is just proportional to $\big(\nabla_{\bb}^2\big)^n\delta(\bb)$ and therefore vanishes too. Then, straightforward dimensional analysis implies that $\tilde{\mcal{F}}_g^{(0)}(0,x_{\mathbb P},\bt^2)$ vanishes faster than any inverse power of $\bt^2$. This argument is not only consistent with the particular expression in Eq.~\eqref{a:f0_tilde_large_b}, which is specific to the GBW/MV model, but is also generally valid in the small-$x$ limit as we now explain. First we write $\mcal{T}_g = 1- \mcal{S}_g$ and then Eq.~\eqref{a:gp0} gives
\begin{align}
	\label{a:gp0_series}
	\mcal{G}_{\mathbb{P}}(0,x_{\mathbb P}, K_{\perp}^2)
	& = \,1 -
	2 \int_0^{\infty} \frac{\rmd R}{R}\,
	J_2(K_{\perp} R)\,
	\mcal{S}_g(R,Y_{\mathbb P})
	\nn*[0.1cm]
	& =\,1 - 2
	\sum_{n=0}^{\infty} \frac{(-1)^n \KT^{2n+2}}{n! (n+2)!}
	\int_0^{\infty}
	\!\dif R \, R^{2n+1} \mcal{S}_g(R) \equiv 1 - \sum_{n=0}^{\infty} a_n \KT^{2n+2}.
\end{align}
In writing the last equality we have made the assumption that $\mcal{S}_g(R)$ falls faster than any inverse power of $R$, like in the MV model or in the solution to the BK equation, so that the integration is convergent. Then, $\mcal{G}_{\mathbb{P}}(0,x_{\mathbb P}, K_{\perp}^2)$ and therefore $\mcal{F}_g^{(0)}(0,x_{\mathbb{P}},\KT^2)$, can be written as power series in $\KT^2$ with constant (i.e.~$\KT^2$-independent) coefficients and the conditions for our argument are met.

\subparagraph{Moderate-$x$:} Consider first intermediate values for $\bt$, namely such that  $\bt \tilde{Q}_s \sim 1$. In this regime, it is not possible to make significant simplifications to Eq.~\eqref{gp}, nor analytically  perform the subsequent FT. Still, we can get some insight by deriving a scaling relation. Let us recall that the tree level gluon DTMD in momentum space obeys
\begin{align}
	\label{a:fg0_scaling}
	\mcal{F}_g^{(0)}(x,x_{\mathbb{P}},\KT^2)
	\approx (1-x) f_g(\KT/\tilde{Q}_s)
	\quad \mathrm{for} \quad \KT \lesssim \tilde{Q}_s.  
\end{align}
Also notice that these are the momenta which dominate the FT for $\bt \tilde{Q}_s \sim 1$, thus we have (after a change of the integration variable to $\kappa = \KT/\tilde{Q}_s$)
\begin{align}
	\label{a:fg_tilde_scaling}
	\tilde{\mcal{F}}_g^{(0)}(x,x_{\mathbb{P}},
	\bt^2)
	&\approx (1-x) \tilde{Q}_s^2
	\int_0^1 \dif \kappa \, \kappa\,
	J_0(\kappa \mkern1mu \bt \tilde{Q}_s) f(\kappa)
	\nn*[0.1cm]
	&\approx (1-x)^2
	\tilde{f}_g(\bt \tilde{Q}_s)
	\quad \mathrm{for} \quad \bt \tilde{Q}_s \sim 1.
\end{align}
In writing the second equality we used $\tilde{Q}_s^2 =(1-x) Q_s^2$, which gave rise to the additional factor $1-x$. The numerical results in Fig.~\ref{fig:F0b_scaled} confirm  that this scaling relation is indeed satisfied in the transition region around $1/ \tilde{Q}_s$.

Now let us move on to the large $\bt^2$ regime. Similarly to Eq.~\eqref{a:gp0}, we write Eq.~\eqref{gp} is terms of $\mcal{S}_g$, that is
\begin{align}
	\label{a:gp_s}
	\mcal{G}_{\mathbb{P}}(x,x_{\mathbb P}, K_{\perp}^2)
	= \,1 -x -
	\mcal{M}^2\int_0^{\infty}
	\rmd R\,R\,
	J_2(K_{\perp} R)
	K_2(\mcal{M}R)
	\mcal{S}_g(R,Y_{\mathbb P})
\end{align}
and we also recall the expansion of the modified Bessel function
\begin{align}
	\label{a:k2_series}
	\mcal{M}^2 K_2(\mcal{M}R)
	=
	\frac{2}{R^2}
	-\frac{x \KT^2 R^2}{2(1-x)}
	+\frac{x^2\KT^4 R^2}{32 (1-x)^2}
	\left[
	2 \ln \frac{4 x (1-x)}{\KT^2 R^2} + 3 - 4 \gamma_{\rm \scriptscriptstyle E}
	\right]
	+ \mcal{O}\big(\KT^6\big).
\end{align}
The crucial observation is that the second derivative of the above function w.r.t.~$\KT^2$ is not analytic at the origin due to the presence of the logarithm. In turn, as we shall shortly see, this will lead to a non-zero contribution to the FT, dominated by very low momenta such that $\KT^2 \lesssim 1/\bt^2$ (which, by the way, justifies a posteriori our small $\KT^2$ expansion). On the other hand (and following our discussion below Eq.~\eqref{a:f0_tilde_large_b}), we can simply neglect all other terms in Eq.~\eqref{a:k2_series} which are powers of $\KT^2$ with constant coefficients --- they yield zero contribution to the FT. By further expanding the Bessel function according to $J_2(\KT R) \simeq \KT^2 R^2/8$, we find that the term of interest in Eq.~\eqref{a:gp_s} becomes
\begin{align}
	\label{a:gp_k6}
	\mcal{G}_{\mathbb{P}}(x,x_{\mathbb P}, K_{\perp}^2)
	\to \frac{c_3\mkern1mu x^2}{2(1-x)^2}\,
	\frac{\KT^6}{Q_s^6}\,\ln \KT^2,
\end{align}
where we have defined
\begin{align}
\label{a:cn}
	c_n \equiv \frac{Q_s^{2n}}{\Gamma(n)2^{2n-1}} \int_0^\infty
	\!\dif R\,R^{2n-1}\mkern1mu \mcal{S}_g(R,Y_{\mathbb P}).
\end{align}
Notice that the general coefficient $c_n$ is dimensionless and its normalisation is such that $c_n=1$ for the GBW model in Eq.~\eqref{a:t_gbw}. Clearly, when taking the square of $\mcal{G}_{\mathbb{P}}$ to construct the DTMD, the dominant relevant contribution comes from the product of the r.h.s.~in Eq.~\eqref{a:gp_k6} with the leading $1-x$ term in Eq.~\eqref{a:gp_s}, so that Eq.~\eqref{Ftree} gives
\begin{align}
	\label{a:fg_k6}
	\mcal{F}_g^{(0)}(x,x_{\mathbb{P}},\KT^2) \to
	\frac{S_\perp N_g}{4 \pi^3}\,
	\frac{c_3 \mkern1mu x^2}{2\pi (1-x)^2}\,
	\frac{\KT^6}{Q_s^6}\,
	\ln \KT^2.
\end{align}
Let us first give the two-dimensional FT of the logarithm, which can be conveniently evaluated as
\begin{align}
\label{a:ft_log}
	\int
	\frac{\dif^2 \bK}{(2\pi)^2}\,
	e^{i \bK \cdot \bb}\,
	\ln \KT^2 & =
	- \frac{1}{\bt^2}
	\int \frac{\dif^2 \bK}{(2\pi)^2}\,
	\big(\nabla^2_{\bK}
	e^{i \bK \cdot \bb}\big) \ln \KT^2
	\nn
	 &= - \frac{1}{\bt^2}
	\int \frac{\dif^2 \bK}{(2\pi)^2}\,
	e^{i \bK \cdot \bb}
	\underbrace{\big(\nabla^2_{\bK}
	\ln \KT^2\big)}_{\displaystyle 4 \pi \delta(\bK)} =
	- \frac{1}{\pi \bt^2}.
\end{align}
Then the FT of $\mcal{F}_g^{(0)}(x,x_{\mathbb{P}},\KT^2)$ appearing in Eq.~\eqref{a:fg_k6} can be calculated by applying to the above equation three times the Laplacian w.r.t.~$\bb$ and leads to an asymptotic power-law behaviour\footnote{Such a FT can be numerically confirmed by introducing a damping factor for $\KT^2 \gtrsim Q_0^2 \gg 1/\bt^2$ and showing that the result converges fast to the analytical expression as we take the limit $Q_0^2 \to \infty$. This also demonstrates that only rather low momenta up to $\sim 1/\bt$ give a significant contribution to the FT and, in turn, it justifies why the upper limit in Eq.~\eqref{a:cn} can be set equal to infinity: with increasing $R$ the fall-off of $\mcal{S}_g(R,Y_{\mathbb{P}})$ starts at $R \sim 1/Q_s$, a value at which the arguments of the Bessel functions are still very small, since $\KT R \lesssim 1/(\bt Q_s) \ll 1$.}:
\begin{align}
	\label{a:f0_tilde_x}
	 \tilde{\mcal{F}}_{\rm asym}^{(0)}(x, x_{\mathbb{P}},\mu_b^2)
		=
	\frac{S_\perp N_g \mkern1mu Q_s^2}{16 \pi^5}\,
	\frac{9 c_3x^2}{8(1-x)^2}
	\left(\frac{8}{\bt^2 Q_s^2} \right)^4
	\quad \mathrm{for} \quad \bt^2 Q_s^2 \gg 8.
\end{align}
It is trivial to observe that the above scales according to
\begin{align}
	\label{a:fg_tilde_scaling_low}
	\tilde{\mcal{F}}_g^{(0)}(x,x_{\mathbb{P}},
	\bt^2) \propto
	\frac{x^2 (1-x)^2}{\big(\bt^2 \tilde{Q}_s^2/8\big)^4}
	\quad \mathrm{for} \quad \bt \tilde{Q}_s \gg 1.
\end{align}
The explicit dependence on $x$ is not exactly the same as the one found earlier in the transition region in Eq.~\eqref{a:fg_tilde_scaling}, due to the presence of the additional factor $x^2$.

Comparing Eqs.~\eqref{a:f0_tilde_large_b} and \eqref{a:f0_tilde_x} we find that the power-law fall-off is in effect so long as $\ln(1/x) \lesssim \bt^2 \tilde{Q}_s^2/16$. This condition is typically satisfied, since we are interested in values of $x$ which are moderate or small, but not too small, more precisely when $\abar \ln (1/x) \ll 1$.

\paragraph{The gluon inclusive TMD:}

It is instructive to also explore the $\bt$-\,dependence of the Weizs\"{a}cker-Williams (WW) gluon TMD, which enters the TMD factorisation of inclusive back-to-back di-jet production in DIS at small $x$~\cite{Dominguez:2010xd,Dominguez:2011wm,Metz:2011wb}.  For simplicity, we shall rely on the MV model, where the relation between the WW gluon TMD and the dipole scattering amplitude has been explicitly computed and reads~\cite{Iancu:2002xk,Iancu:2003xm}
\begin{align}
	\label{a:fg_inc}
	\mcal{F}_{g,{\rm WW}}^{(0)}(\KT^2) =
	\frac{S_\perp N_g}{4 \pi^4}\,
	\frac{1}{\alpha_s N_c}
	\int \dif^2 \bb\, e^{-i \bK \cdot \bb}\, \frac{\mcal{T}_g(\bt)}{r\bt^2}\,.
\end{align}
From the above we can immediately extract the FT of this gluon TMD for arbitrary $\bt^2$, namely
\begin{align}
	\label{a:fg_inc_ft}
	\tilde{\mcal{F}}_{g,{\rm WW}}^{(0)}(\bt^2)
	=
	\frac{S_\perp N_g}{4 \pi^4}\,
	\frac{1}{\alpha_s N_c}
	\frac{\mcal{T}_g(\bt)}{\bt^2}\,
	\xrightarrow[\bt Q_s \gg 1]{}\,
	\frac{S_\perp N_g}{4 \pi^4}\,
	\frac{1}{\alpha_s N_c}\,
	\frac{1}{\bt^2}\,.
\end{align}
The second, approximate, expression holds at large $\bt^2$ and it exhibits a much slower suppression than the respective diffractive TMD, cf.~Eqs.~\eqref{a:f0_tilde_large_b} and \eqref{a:f0_tilde_x}. It is further useful to understand which momentum modes contribute to such an asymptotic behavior. First, by letting $\mcal{T}_g(\bt)\to 1$ in Eq.~\eqref{a:fg_inc} the integration becomes logarithmic and we arrive at the low momentum behaviour
\begin{align}
	\label{a:fg_inc_low_kt}
	\mcal{F}_{g,{\rm WW}}^{(0)}(\KT^2) \simeq
	\frac{S_\perp N_g}{4 \pi^4}\,
	\frac{1}{\alpha_s N_c}\,
	\ln \frac{Q_{s}^2}{\KT^2}
	\quad \mathrm{for} \quad \KT^2 \ll Q_{s}^2.
\end{align}
Then, our earlier discussion following Eq.~\eqref{a:k2_series} makes it clear that the {\it logarithmic} saturation of the inclusive TMD allows low $\KT^2$ modes to contribute to the FT at large $\bt^2$ and lead to the rather mild suppression in Eq.~\eqref{a:fg_inc_ft}.

\section{Moments and maxima of the CSS distribution at DLA}
\label{app:mm}

Here we shall study some of the characteristic momenta of the large-$\KT$ distribution given by the DLA solution to the CSS equation in Eq.~\eqref{DLAsol_largeK}. First let us consider the maximum of $(\KT^2/Q^2)^n \mcal{F}_g(x,\KT^2,Q^2)$. It is determined by the quadratic equation
\begin{align}
	\label{a:max_n}
	\frac{\abar}{2} \ln^2\frac{Q^2}{\KT^2} 
	+ (n-1)\ln\frac{Q^2}{\KT^2} 
	-1=0,
\end{align}
in which the first term emerges from differentiating the virtual Sudakov factor. The solution reads
\begin{align}
	\label{a:max_n_sol}
	\ln\frac{Q^2}{\KT^2} =
	\frac{1-n + \sqrt{(1-n)^2 + 2 \abar}}{\abar} 
\end{align}
Notice that the above can be written in the piecewise form
\begin{align}
	\label{a:max_n_piece}
	\ln\frac{Q^2}{\KT^2}
	\simeq
    \begin{cases} 
    \displaystyle{\frac{2}{\abar}}
    & \quad \text{for\,\, $n=0$},
    \\*[0.5cm]   
    \displaystyle{\sqrt{\frac{2}{\abar}}}
    & \quad \text{for\,\, $n=1$},
    \\*[0.5cm]
    \displaystyle{\frac{1}{n-1}} &
    \quad \text{for\,\, $n \geq 2$},
    \end{cases}
\end{align}
where the $n \neq 1$ expressions are valid approximately when $\abar \ll 1$.

Next, let us calculate the average value of $(\KT^2/Q^2)^n$, i.e.~the $n$-moment of the distribution in $\KT^2$. Making a change of the integration variable according to $\rho = \ln (Q^2/\KT^2)$, we readily find that
\begin{align}
	\label{a:KTn_ave}
	\left\langle \frac{\KT^{2n}}{Q^{2n}} \right\rangle
	= 
	\frac{\displaystyle \int_0^{\infty} 
	\dif \rho\,\rho\, e^{-n\rho} 
	e^{-\abar\rho^2/4} }{\displaystyle \int_0^{\infty} \dif \rho\,\rho\, 
	e^{-\abar \rho^2/4}}
	\simeq \frac{\abar}{2 n^2}.	
\end{align}
The factor $\abar/2$ is just the normalization factor arising from the denominator. In the numerator the integration is effectively restricted in the regime $\rho \lesssim 1/n$ (i.e. by transverse momenta $\KT$ within the range $Q^2 e^{-1/n} \lesssim \KT^2 \lesssim Q^2$), which allowed us to completely the virtual Sudakov factor to the accuracy of interest. This means that such moments do not probe the resummation of the double logarithms. In particular for $n=1/2$ we recover the result in \eqn{ktavetoy}.
% (like in the calculation of the $R_{pA}$ ratio). 

Instead, let us consider the moments of the distribution in the logarithm of the transverse momentum. They are given by
\begin{align}
	\label{a:lognKT_ave}
	\left\langle \ln^n\frac{Q^2}{\KT^2} \right\rangle
	= 
	\frac{\displaystyle \int_0^{\infty} 
	\dif \rho\,\rho^{n+1} 
	e^{-\abar\rho^2/4} }{\displaystyle \int_0^{\infty} \dif \rho\,\rho\, 
	e^{-\abar \rho^2/4}}
	= \frac{2^n \Gamma\big(\frac{n}{2}+1\big)}{\abar^{n/2}}	
\end{align}
and obviously they are sensitive to the Sudakov resummation.

It is also interesting to notice that for $n=1$ the above becomes
\begin{align}
	\label{a:logKT_ave}
	\left\langle \ln\frac{Q^2}{\KT^2} \right\rangle
	= \sqrt{\frac{\pi}{\abar}},	
\end{align}
which, up to a factor of order one, coincides with the location of the maximum 
of $(\KT^2/Q^2) \mcal{F}_g(x,\KT^2,Q^2)$, cf.~the middle line in Eq.~\eqref{a:max_n_piece}.

\providecommand{\href}[2]{#2}\begingroup\raggedright\endgroup


\begin{thebibliography}{10}

\bibitem{Boer:2011fh}
D.~Boer et~al., \emph{{Gluons and the quark sea at high energies:
  Distributions, polarization, tomography}},
  \href{https://arxiv.org/abs/1108.1713}{{\ttfamily 1108.1713}}.

\bibitem{Accardi:2012qut}
A.~Accardi et~al., \emph{{Electron Ion Collider: The Next QCD Frontier}:
  {Understanding the glue that binds us all}},
  \href{https://doi.org/10.1140/epja/i2016-16268-9}{\emph{Eur. Phys. J. A}
  {\bfseries 52} (2016) 268} [\href{https://arxiv.org/abs/1212.1701}{{\ttfamily
  1212.1701}}].

\bibitem{Aschenauer:2017jsk}
E.~Aschenauer, S.~Fazio, J.~Lee, H.~Mantysaari, B.~Page, B.~Schenke et~al.,
  \emph{{The electron\textendash{}ion collider: assessing the energy dependence
  of key measurements}},
  \href{https://doi.org/10.1088/1361-6633/aaf216}{\emph{Rept. Prog. Phys.}
  {\bfseries 82} (2019) 024301}
  [\href{https://arxiv.org/abs/1708.01527}{{\ttfamily 1708.01527}}].

\bibitem{AbdulKhalek:2022hcn}
R.~Abdul~Khalek et~al., \emph{{Snowmass 2021 White Paper: Electron Ion Collider
  for High Energy Physics}},  in \emph{{2022 Snowmass Summer Study}}, 3, 2022
  [\href{https://arxiv.org/abs/2203.13199}{{\ttfamily 2203.13199}}].

\bibitem{Iancu:2002xk}
E.~Iancu, A.~Leonidov and L.~McLerran, \emph{{The Color glass condensate: An
  Introduction}},  in \emph{{Cargese Summer School on QCD Perspectives on Hot
  and Dense Matter}}, pp.~73--145, 2, 2002
  [\href{https://arxiv.org/abs/hep-ph/0202270}{{\ttfamily hep-ph/0202270}}].

\bibitem{Iancu:2003xm}
E.~Iancu and R.~Venugopalan, \emph{{The Color glass condensate and high-energy
  scattering in QCD}},  in \emph{{Quark-gluon plasma 4}}, R.C.~Hwa and
  X.-N.~Wang, eds., pp.~249--3363 (2003),
  \href{https://doi.org/10.1142/9789812795533_0005}{DOI}
  [\href{https://arxiv.org/abs/hep-ph/0303204}{{\ttfamily hep-ph/0303204}}].

\bibitem{Gelis:2010nm}
F.~Gelis, E.~Iancu, J.~Jalilian-Marian and R.~Venugopalan, \emph{{The Color
  Glass Condensate}},
  \href{https://doi.org/10.1146/annurev.nucl.010909.083629}{\emph{Ann. Rev.
  Nucl. Part. Sci.} {\bfseries 60} (2010) 463}
  [\href{https://arxiv.org/abs/1002.0333}{{\ttfamily 1002.0333}}].

\bibitem{Kovchegov:2012mbw}
Y.V.~Kovchegov and E.~Levin, \emph{{Quantum Chromodynamics at High Energy}},
  vol.~33, Oxford University Press (2013),
  \href{https://doi.org/10.1017/9781009291446}{10.1017/9781009291446}.

\bibitem{Wusthoff:1997fz}
M.~Wusthoff, \emph{{Large rapidity gap events in deep inelastic scattering}},
  \href{https://doi.org/10.1103/PhysRevD.56.4311}{\emph{Phys. Rev. D}
  {\bfseries 56} (1997) 4311}
  [\href{https://arxiv.org/abs/hep-ph/9702201}{{\ttfamily hep-ph/9702201}}].

\bibitem{GolecBiernat:1999qd}
K.J.~Golec-Biernat and M.~Wusthoff, \emph{{Saturation in diffractive deep
  inelastic scattering}},
  \href{https://doi.org/10.1103/PhysRevD.60.114023}{\emph{Phys. Rev.}
  {\bfseries D60} (1999) 114023}
  [\href{https://arxiv.org/abs/hep-ph/9903358}{{\ttfamily hep-ph/9903358}}].

\bibitem{Hebecker:1997gp}
A.~Hebecker, \emph{{Diffractive parton distributions in the semiclassical
  approach}}, \href{https://doi.org/10.1016/S0550-3213(97)00512-9}{\emph{Nucl.
  Phys. B} {\bfseries 505} (1997) 349}
  [\href{https://arxiv.org/abs/hep-ph/9702373}{{\ttfamily hep-ph/9702373}}].

\bibitem{Buchmuller:1998jv}
W.~Buchmuller, T.~Gehrmann and A.~Hebecker, \emph{{Inclusive and diffractive
  structure functions at small x}},
  \href{https://doi.org/10.1016/S0550-3213(98)00682-8}{\emph{Nucl. Phys. B}
  {\bfseries 537} (1999) 477}
  [\href{https://arxiv.org/abs/hep-ph/9808454}{{\ttfamily hep-ph/9808454}}].

\bibitem{Hautmann:1998xn}
F.~Hautmann, Z.~Kunszt and D.E.~Soper, \emph{{Diffractive deeply inelastic
  scattering of hadronic states with small transverse size}},
  \href{https://doi.org/10.1103/PhysRevLett.81.3333}{\emph{Phys. Rev. Lett.}
  {\bfseries 81} (1998) 3333}
  [\href{https://arxiv.org/abs/hep-ph/9806298}{{\ttfamily hep-ph/9806298}}].

\bibitem{Hautmann:1999ui}
F.~Hautmann, Z.~Kunszt and D.E.~Soper, \emph{{Hard scattering factorization and
  light cone Hamiltonian approach to diffractive processes}},
  \href{https://doi.org/10.1016/S0550-3213(99)00568-4}{\emph{Nucl. Phys. B}
  {\bfseries 563} (1999) 153}
  [\href{https://arxiv.org/abs/hep-ph/9906284}{{\ttfamily hep-ph/9906284}}].

\bibitem{Hautmann:2000pw}
F.~Hautmann and D.E.~Soper, \emph{{Color transparency in deeply inelastic
  diffraction}}, \href{https://doi.org/10.1103/PhysRevD.63.011501}{\emph{Phys.
  Rev. D} {\bfseries 63} (2001) 011501}
  [\href{https://arxiv.org/abs/hep-ph/0008224}{{\ttfamily hep-ph/0008224}}].

\bibitem{Golec-Biernat:2001gyl}
K.J.~Golec-Biernat and M.~Wusthoff, \emph{{Diffractive parton distributions
  from the saturation model}},
  \href{https://doi.org/10.1007/s100520100661}{\emph{Eur. Phys. J. C}
  {\bfseries 20} (2001) 313}
  [\href{https://arxiv.org/abs/hep-ph/0102093}{{\ttfamily hep-ph/0102093}}].

\bibitem{Iancu:2021rup}
E.~Iancu, A.H.~Mueller and D.N.~Triantafyllopoulos, \emph{{Probing Parton
  Saturation and the Gluon Dipole via Diffractive Jet Production at the
  Electron-Ion Collider}},
  \href{https://doi.org/10.1103/PhysRevLett.128.202001}{\emph{Phys. Rev. Lett.}
  {\bfseries 128} (2022) 202001}
  [\href{https://arxiv.org/abs/2112.06353}{{\ttfamily 2112.06353}}].

\bibitem{Collins:2011zzd}
J.~Collins, \emph{{Foundations of Perturbative QCD}}, vol.~32 of
  \emph{Cambridge Monographs on Particle Physics, Nuclear Physics and
  Cosmology}, Cambridge University Press (7, 2023),
  \href{https://doi.org/10.1017/9781009401845}{10.1017/9781009401845}.

\bibitem{Boussarie:2023izj}
R.~Boussarie et~al., \emph{{TMD Handbook}},
  \href{https://arxiv.org/abs/2304.03302}{{\ttfamily 2304.03302}}.

\bibitem{Marquet:2009ca}
C.~Marquet, B.-W.~Xiao and F.~Yuan, \emph{{Semi-inclusive Deep Inelastic
  Scattering at small x}},
  \href{https://doi.org/10.1016/j.physletb.2009.10.099}{\emph{Phys. Lett. B}
  {\bfseries 682} (2009) 207}
  [\href{https://arxiv.org/abs/0906.1454}{{\ttfamily 0906.1454}}].

\bibitem{Dominguez:2010xd}
F.~Dominguez, B.-W.~Xiao and F.~Yuan, \emph{{$k_t$-factorization for Hard
  Processes in Nuclei}},
  \href{https://doi.org/10.1103/PhysRevLett.106.022301}{\emph{Phys. Rev. Lett.}
  {\bfseries 106} (2011) 022301}
  [\href{https://arxiv.org/abs/1009.2141}{{\ttfamily 1009.2141}}].

\bibitem{Dominguez:2011wm}
F.~Dominguez, C.~Marquet, B.-W.~Xiao and F.~Yuan, \emph{{Universality of
  Unintegrated Gluon Distributions at small x}},
  \href{https://doi.org/10.1103/PhysRevD.83.105005}{\emph{Phys. Rev.}
  {\bfseries D83} (2011) 105005}
  [\href{https://arxiv.org/abs/1101.0715}{{\ttfamily 1101.0715}}].

\bibitem{Metz:2011wb}
A.~Metz and J.~Zhou, \emph{{Distribution of linearly polarized gluons inside a
  large nucleus}},
  \href{https://doi.org/10.1103/PhysRevD.84.051503}{\emph{Phys. Rev. D}
  {\bfseries 84} (2011) 051503}
  [\href{https://arxiv.org/abs/1105.1991}{{\ttfamily 1105.1991}}].

\bibitem{Dominguez:2011br}
F.~Dominguez, J.-W.~Qiu, B.-W.~Xiao and F.~Yuan, \emph{{On the linearly
  polarized gluon distributions in the color dipole model}},
  \href{https://doi.org/10.1103/PhysRevD.85.045003}{\emph{Phys. Rev. D}
  {\bfseries 85} (2012) 045003}
  [\href{https://arxiv.org/abs/1109.6293}{{\ttfamily 1109.6293}}].

\bibitem{Xiao:2017yya}
B.-W.~Xiao, F.~Yuan and J.~Zhou, \emph{{Transverse Momentum Dependent Parton
  Distributions at Small-x}},
  \href{https://doi.org/10.1016/j.nuclphysb.2017.05.012}{\emph{Nucl. Phys. B}
  {\bfseries 921} (2017) 104}
  [\href{https://arxiv.org/abs/1703.06163}{{\ttfamily 1703.06163}}].

\bibitem{Marquet:2017xwy}
C.~Marquet, C.~Roiesnel and P.~Taels, \emph{{Linearly polarized small-$x$
  gluons in forward heavy-quark pair production}},
  \href{https://doi.org/10.1103/PhysRevD.97.014004}{\emph{Phys. Rev. D}
  {\bfseries 97} (2018) 014004}
  [\href{https://arxiv.org/abs/1710.05698}{{\ttfamily 1710.05698}}].

\bibitem{Altinoluk:2024tyx}
T.~Altinoluk, G.~Beuf, E.~Blanco and S.~Mulani, \emph{{Quark TMDs from
  back-to-back dijet production at forward rapidities in pA collisions beyond
  eikonal accuracy in the CGC}},
  \href{https://arxiv.org/abs/2412.08485}{{\ttfamily 2412.08485}}.

\bibitem{Caucal:2025xxh}
P.~Caucal, M.G.~Morales, E.~Iancu, F.~Salazar and F.~Yuan, \emph{{Unveiling the
  sea: universality of the transverse momentum dependent quark distributions at
  small $x$}},  \href{https://arxiv.org/abs/2503.16162}{{\ttfamily
  2503.16162}}.

\bibitem{Hatta:2022lzj}
Y.~Hatta, B.-W.~Xiao and F.~Yuan, \emph{{Semi-inclusive diffractive deep
  inelastic scattering at small x}},
  \href{https://doi.org/10.1103/PhysRevD.106.094015}{\emph{Phys. Rev. D}
  {\bfseries 106} (2022) 094015}
  [\href{https://arxiv.org/abs/2205.08060}{{\ttfamily 2205.08060}}].

\bibitem{Iancu:2022lcw}
E.~Iancu, A.H.~Mueller, D.N.~Triantafyllopoulos and S.Y.~Wei, \emph{{Gluon
  dipole factorisation for diffractive dijets}},
  \href{https://doi.org/10.1007/JHEP10(2022)103}{\emph{JHEP} {\bfseries 10}
  (2022) 103} [\href{https://arxiv.org/abs/2207.06268}{{\ttfamily
  2207.06268}}].

\bibitem{Hauksson:2024bvv}
S.~Hauksson, E.~Iancu, A.H.~Mueller, D.N.~Triantafyllopoulos and S.Y.~Wei,
  \emph{{TMD factorisation for diffractive jets in photon-nucleus
  interactions}}, \href{https://doi.org/10.1007/JHEP06(2024)180}{\emph{JHEP}
  {\bfseries 06} (2024) 180}
  [\href{https://arxiv.org/abs/2402.14748}{{\ttfamily 2402.14748}}].

\bibitem{Caucal:2024bae}
P.~Caucal and E.~Iancu, \emph{{Evolution of the transverse-momentum dependent
  gluon distribution at small x}},
  \href{https://doi.org/10.1103/PhysRevD.111.074008}{\emph{Phys. Rev. D}
  {\bfseries 111} (2025) 074008}
  [\href{https://arxiv.org/abs/2406.04238}{{\ttfamily 2406.04238}}].

\bibitem{Taels:2022tza}
P.~Taels, T.~Altinoluk, G.~Beuf and C.~Marquet, \emph{{Dijet photoproduction at
  low x at next-to-leading order and its back-to-back limit}},
  \href{https://doi.org/10.1007/JHEP10(2022)184}{\emph{JHEP} {\bfseries 10}
  (2022) 184} [\href{https://arxiv.org/abs/2204.11650}{{\ttfamily
  2204.11650}}].

\bibitem{Caucal:2022ulg}
P.~Caucal, F.~Salazar, B.~Schenke and R.~Venugopalan, \emph{{Back-to-back
  inclusive dijets in DIS at small x: Sudakov suppression and gluon saturation
  at NLO}}, \href{https://doi.org/10.1007/JHEP11(2022)169}{\emph{JHEP}
  {\bfseries 11} (2022) 169}
  [\href{https://arxiv.org/abs/2208.13872}{{\ttfamily 2208.13872}}].

\bibitem{Caucal:2024vbv}
P.~Caucal, E.~Iancu, A.H.~Mueller and F.~Yuan, \emph{{Jet definition and TMD
  factorisation in SIDIS}},  \href{https://arxiv.org/abs/2408.03129}{{\ttfamily
  2408.03129}}.

\bibitem{Balitsky:1995ub}
I.~Balitsky, \emph{{Operator expansion for high-energy scattering}},
  \href{https://doi.org/10.1016/0550-3213(95)00638-9}{\emph{Nucl. Phys.}
  {\bfseries B463} (1996) 99}
  [\href{https://arxiv.org/abs/hep-ph/9509348}{{\ttfamily hep-ph/9509348}}].

\bibitem{Kovchegov:1999yj}
Y.V.~Kovchegov, \emph{{Small-x F2 structure function of a nucleus including
  multiple pomeron exchanges}},
  \href{https://doi.org/10.1103/PhysRevD.60.034008}{\emph{Phys. Rev.}
  {\bfseries D60} (1999) 034008}
  [\href{https://arxiv.org/abs/hep-ph/9901281}{{\ttfamily hep-ph/9901281}}].

\bibitem{JalilianMarian:1997jx}
J.~Jalilian-Marian, A.~Kovner, A.~Leonidov and H.~Weigert, \emph{{The BFKL
  equation from the Wilson renormalization group}},
  \href{https://doi.org/10.1016/S0550-3213(97)00440-9}{\emph{Nucl. Phys.}
  {\bfseries B504} (1997) 415}
  [\href{https://arxiv.org/abs/hep-ph/9701284}{{\ttfamily hep-ph/9701284}}].

\bibitem{JalilianMarian:1997gr}
J.~Jalilian-Marian, A.~Kovner, A.~Leonidov and H.~Weigert, \emph{{The Wilson
  renormalization group for low x physics: Towards the high density regime}},
  \href{https://doi.org/10.1103/PhysRevD.59.014014}{\emph{Phys.Rev.} {\bfseries
  D59} (1998) 014014} [\href{https://arxiv.org/abs/hep-ph/9706377}{{\ttfamily
  hep-ph/9706377}}].

\bibitem{Kovner:2000pt}
A.~Kovner, J.G.~Milhano and H.~Weigert, \emph{{Relating different approaches to
  nonlinear QCD evolution at finite gluon density}},
  \href{https://doi.org/10.1103/PhysRevD.62.114005}{\emph{Phys. Rev.}
  {\bfseries D62} (2000) 114005}
  [\href{https://arxiv.org/abs/hep-ph/0004014}{{\ttfamily hep-ph/0004014}}].

\bibitem{Weigert:2000gi}
H.~Weigert, \emph{{Unitarity at small Bjorken x}},
  \href{https://doi.org/10.1016/S0375-9474(01)01668-2}{\emph{Nucl. Phys.}
  {\bfseries A703} (2002) 823}
  [\href{https://arxiv.org/abs/hep-ph/0004044}{{\ttfamily hep-ph/0004044}}].

\bibitem{Iancu:2000hn}
E.~Iancu, A.~Leonidov and L.D.~McLerran, \emph{{Nonlinear gluon evolution in
  the color glass condensate. I}},
  \href{https://doi.org/10.1016/S0375-9474(01)00642-X}{\emph{Nucl. Phys.}
  {\bfseries A692} (2001) 583}
  [\href{https://arxiv.org/abs/hep-ph/0011241}{{\ttfamily hep-ph/0011241}}].

\bibitem{Iancu:2001ad}
E.~Iancu, A.~Leonidov and L.D.~McLerran, \emph{{The renormalization group
  equation for the color glass condensate}},
  \href{https://doi.org/10.1016/S0370-2693(01)00524-X}{\emph{Phys. Lett.}
  {\bfseries B510} (2001) 133}
  [\href{https://arxiv.org/abs/hep-ph/0102009}{{\ttfamily hep-ph/0102009}}].

\bibitem{Ferreiro:2001qy}
E.~Ferreiro, E.~Iancu, A.~Leonidov and L.~McLerran, \emph{{Nonlinear gluon
  evolution in the color glass condensate. II}},
  \href{https://doi.org/10.1016/S0375-9474(01)01329-X}{\emph{Nucl. Phys.}
  {\bfseries A703} (2002) 489}
  [\href{https://arxiv.org/abs/hep-ph/0109115}{{\ttfamily hep-ph/0109115}}].

\bibitem{Mueller:2012uf}
A.H.~Mueller, B.-W.~Xiao and F.~Yuan, \emph{{Sudakov Resummation in Small-$x$
  Saturation Formalism}},
  \href{https://doi.org/10.1103/PhysRevLett.110.082301}{\emph{Phys. Rev. Lett.}
  {\bfseries 110} (2013) 082301}
  [\href{https://arxiv.org/abs/1210.5792}{{\ttfamily 1210.5792}}].

\bibitem{Mueller:2013wwa}
A.~Mueller, B.-W.~Xiao and F.~Yuan, \emph{{Sudakov double logarithms
  resummation in hard processes in the small-x saturation formalism}},
  \href{https://doi.org/10.1103/PhysRevD.88.114010}{\emph{Phys. Rev. D}
  {\bfseries 88} (2013) 114010}
  [\href{https://arxiv.org/abs/1308.2993}{{\ttfamily 1308.2993}}].

\bibitem{Altinoluk:2024vgg}
T.~Altinoluk, J.~Jalilian-Marian and C.~Marquet, \emph{{Sudakov double logs in
  single-inclusive hadron production in DIS at small x from the color glass
  condensate formalism}},
  \href{https://doi.org/10.1103/PhysRevD.110.094056}{\emph{Phys. Rev. D}
  {\bfseries 110} (2024) 094056}
  [\href{https://arxiv.org/abs/2406.08277}{{\ttfamily 2406.08277}}].

\bibitem{Collins:1981uk}
J.C.~Collins and D.E.~Soper, \emph{{Back-To-Back Jets in QCD}},
  \href{https://doi.org/10.1016/0550-3213(81)90339-4}{\emph{Nucl. Phys. B}
  {\bfseries 193} (1981) 381}.

\bibitem{Collins:1981uw}
J.C.~Collins and D.E.~Soper, \emph{{Parton Distribution and Decay Functions}},
  \href{https://doi.org/10.1016/0550-3213(82)90021-9}{\emph{Nucl. Phys. B}
  {\bfseries 194} (1982) 445}.

\bibitem{Collins:1984kg}
J.C.~Collins, D.E.~Soper and G.F.~Sterman, \emph{{Transverse Momentum
  Distribution in Drell-Yan Pair and W and Z Boson Production}},
  \href{https://doi.org/10.1016/0550-3213(85)90479-1}{\emph{Nucl. Phys. B}
  {\bfseries 250} (1985) 199}.

\bibitem{Gribov:1972ri}
V.~Gribov and L.~Lipatov, \emph{{Deep inelastic e p scattering in perturbation
  theory}}, {\emph{Sov.J.Nucl.Phys.} {\bfseries 15} (1972) 438}.

\bibitem{Altarelli:1977zs}
G.~Altarelli and G.~Parisi, \emph{{Asymptotic Freedom in Parton Language}},
  \href{https://doi.org/10.1016/0550-3213(77)90384-4}{\emph{Nucl.Phys.}
  {\bfseries B126} (1977) 298}.

\bibitem{Dokshitzer:1977sg}
Y.L.~Dokshitzer, \emph{{Calculation of the Structure Functions for Deep
  Inelastic Scattering and e+ e- Annihilation by Perturbation Theory in Quantum
  Chromodynamics.}}, {\emph{Sov.Phys.JETP} {\bfseries 46} (1977) 641}.

\bibitem{Ellis:1997ii}
R.K.~Ellis and S.~Veseli, \emph{{$W$ and $Z$ transverse momentum distributions:
  Resummation in $q_{T}$ space}},
  \href{https://doi.org/10.1016/S0550-3213(97)00655-X}{\emph{Nucl. Phys. B}
  {\bfseries 511} (1998) 649}
  [\href{https://arxiv.org/abs/hep-ph/9706526}{{\ttfamily hep-ph/9706526}}].

\bibitem{Shao:2024nor}
D.Y.~Shao, Y.~Shi, C.~Zhang, J.~Zhou and Y.-j.~Zhou, \emph{{Revisiting
  azimuthal angular asymmetries in diffractive di-jet production}},
  \href{https://doi.org/10.1007/JHEP07(2024)189}{\emph{JHEP} {\bfseries 07}
  (2024) 189} [\href{https://arxiv.org/abs/2402.05465}{{\ttfamily
  2402.05465}}].

\bibitem{Shi:2023ejp}
Y.~Shi, S.-Y.~Wei and J.~Zhou, \emph{{Parton shower algorithm with the
  saturation effect}},
  \href{https://doi.org/10.1103/PhysRevD.108.096025}{\emph{Phys. Rev. D}
  {\bfseries 108} (2023) 096025}
  [\href{https://arxiv.org/abs/2307.04185}{{\ttfamily 2307.04185}}].

\bibitem{vanHameren:2025hyo}
A.~van Hameren and M.~Nefedov, \emph{{Hybrid high-energy factorization and
  evolution at NLO from the high-energy limit of collinear factorization}},
  \href{https://doi.org/10.1007/JHEP02(2025)160}{\emph{JHEP} {\bfseries 02}
  (2025) 160} [\href{https://arxiv.org/abs/2501.02619}{{\ttfamily
  2501.02619}}].

\bibitem{Caucal:2025mth}
P.~Caucal, E.~Iancu, F.~Salazar and F.~Yuan, \emph{{Gluon splitting at small
  $x$: a unified derivation for the JIMWLK, DGLAP and CSS equations}},
  \href{https://arxiv.org/abs/2510.08454}{{\ttfamily 2510.08454}}.

\bibitem{Ebert:2022cku}
M.A.~Ebert, J.K.L.~Michel, I.W.~Stewart and Z.~Sun, \emph{{Disentangling long
  and short distances in momentum-space TMDs}},
  \href{https://doi.org/10.1007/JHEP07(2022)129}{\emph{JHEP} {\bfseries 07}
  (2022) 129} [\href{https://arxiv.org/abs/2201.07237}{{\ttfamily
  2201.07237}}].

\bibitem{delRio:2024vvq}
O.~del Rio, A.~Prokudin, I.~Scimemi and A.~Vladimirov, \emph{{Transverse
  momentum moments}},
  \href{https://doi.org/10.1103/PhysRevD.110.016003}{\emph{Phys. Rev. D}
  {\bfseries 110} (2024) 016003}
  [\href{https://arxiv.org/abs/2402.01836}{{\ttfamily 2402.01836}}].

\bibitem{McLerran:1993ni}
L.D.~McLerran and R.~Venugopalan, \emph{{Computing quark and gluon distribution
  functions for very large nuclei}},
  \href{https://doi.org/10.1103/PhysRevD.49.2233}{\emph{Phys. Rev.} {\bfseries
  D49} (1994) 2233} [\href{https://arxiv.org/abs/hep-ph/9309289}{{\ttfamily
  hep-ph/9309289}}].

\bibitem{McLerran:1993ka}
L.D.~McLerran and R.~Venugopalan, \emph{{Gluon distribution functions for very
  large nuclei at small transverse momentum}},
  \href{https://doi.org/10.1103/PhysRevD.49.3352}{\emph{Phys. Rev.} {\bfseries
  D49} (1994) 3352} [\href{https://arxiv.org/abs/hep-ph/9311205}{{\ttfamily
  hep-ph/9311205}}].

\bibitem{Beuf:2014uia}
G.~Beuf, \emph{{Improving the kinematics for low-x QCD evolution equations in
  coordinate space}},
  \href{https://doi.org/10.1103/PhysRevD.89.074039}{\emph{Phys.Rev.} {\bfseries
  D89} (2014) 074039} [\href{https://arxiv.org/abs/1401.0313}{{\ttfamily
  1401.0313}}].

\bibitem{Iancu:2015vea}
E.~Iancu, J.~Madrigal, A.~Mueller, G.~Soyez and D.~Triantafyllopoulos,
  \emph{{Resumming double logarithms in the QCD evolution of color dipoles}},
  \href{https://doi.org/10.1016/j.physletb.2015.03.068}{\emph{Phys.Lett.}
  {\bfseries B744} (2015) 293}
  [\href{https://arxiv.org/abs/1502.05642}{{\ttfamily 1502.05642}}].

\bibitem{Ducloue:2019ezk}
B.~Duclou{\'e}, E.~Iancu, A.H.~Mueller, G.~Soyez and D.N.~Triantafyllopoulos,
  \emph{{Non-linear evolution in QCD at high-energy beyond leading order}},
  \href{https://doi.org/10.1007/JHEP04(2019)081}{\emph{JHEP} {\bfseries 04}
  (2019) 081} [\href{https://arxiv.org/abs/1902.06637}{{\ttfamily
  1902.06637}}].

\bibitem{Boussarie:2025mzh}
R.~Boussarie, P.~Caucal and Y.~Mehtar-Tani, \emph{{Collinear Structure of
  Nonlinear Small-$x$ Evolution}},
  \href{https://arxiv.org/abs/2509.20236}{{\ttfamily 2509.20236}}.

\bibitem{Caucal:2023nci}
P.~Caucal, F.~Salazar, B.~Schenke, T.~Stebel and R.~Venugopalan,
  \emph{{Back-to-back inclusive dijets in DIS at small x: gluon
  Weizs\"acker-Williams distribution at NLO}},
  \href{https://doi.org/10.1007/JHEP08(2023)062}{\emph{JHEP} {\bfseries 08}
  (2023) 062} [\href{https://arxiv.org/abs/2304.03304}{{\ttfamily
  2304.03304}}].

\bibitem{Caucal:2023fsf}
P.~Caucal, F.~Salazar, B.~Schenke, T.~Stebel and R.~Venugopalan,
  \emph{{Back-to-Back Inclusive Dijets in Deep Inelastic Scattering at Small x:
  Complete NLO Results and Predictions}},
  \href{https://doi.org/10.1103/PhysRevLett.132.081902}{\emph{Phys. Rev. Lett.}
  {\bfseries 132} (2024) 081902}
  [\href{https://arxiv.org/abs/2308.00022}{{\ttfamily 2308.00022}}].

\bibitem{Hentschinski:2021lsh}
M.~Hentschinski, \emph{{Transverse momentum dependent gluon distribution within
  high energy factorization at next-to-leading order}},
  \href{https://doi.org/10.1103/PhysRevD.104.054014}{\emph{Phys. Rev. D}
  {\bfseries 104} (2021) 054014}
  [\href{https://arxiv.org/abs/2107.06203}{{\ttfamily 2107.06203}}].

\bibitem{Ayala:1995hx}
A.~Ayala, J.~Jalilian-Marian, L.D.~McLerran and R.~Venugopalan, \emph{{Quantum
  corrections to the Weizsacker-Williams gluon distribution function at small
  x}}, \href{https://doi.org/10.1103/PhysRevD.53.458}{\emph{Phys. Rev. D}
  {\bfseries 53} (1996) 458}
  [\href{https://arxiv.org/abs/hep-ph/9508302}{{\ttfamily hep-ph/9508302}}].

\bibitem{Zhou:2018lfq}
J.~Zhou, \emph{{Scale dependence of the small x transverse momentum dependent
  gluon distribution}},
  \href{https://doi.org/10.1103/PhysRevD.99.054026}{\emph{Phys. Rev. D}
  {\bfseries 99} (2019) 054026}
  [\href{https://arxiv.org/abs/1807.00506}{{\ttfamily 1807.00506}}].

\bibitem{Mueller:2018llt}
A.H.~Mueller, \emph{{Conformal spacelike-timelike correspondence in QCD}},
  \href{https://doi.org/10.1007/JHEP08(2018)139}{\emph{JHEP} {\bfseries 08}
  (2018) 139} [\href{https://arxiv.org/abs/1804.07249}{{\ttfamily
  1804.07249}}].

\bibitem{ATLAS:2022cbd}
{\scshape ATLAS} collaboration, \emph{{Photo-nuclear jet production in
  ultra-peripheral Pb+Pb collisions at $\sqrt{s}_\text{NN} = 5.02$ TeV with the
  ATLAS detector}}, {\emph{ATLAS-CONF-2022-021} (2022) }.

\bibitem{CMS:2022lbi}
{\scshape CMS} collaboration, \emph{{Azimuthal Correlations within Exclusive
  Dijets with Large Momentum Transfer in Photon-Lead Collisions}},
  \href{https://doi.org/10.1103/PhysRevLett.131.051901}{\emph{Phys. Rev. Lett.}
  {\bfseries 131} (2023) 051901}
  [\href{https://arxiv.org/abs/2205.00045}{{\ttfamily 2205.00045}}].

\bibitem{ATLAS:2024mvt}
{\scshape ATLAS} collaboration, \emph{{Measurement of photonuclear jet
  production in ultraperipheral Pb+Pb collisions at sNN=5.02{\,}{\,}TeV with
  the ATLAS detector}},
  \href{https://doi.org/10.1103/PhysRevD.111.052006}{\emph{Phys. Rev. D}
  {\bfseries 111} (2025) 052006}
  [\href{https://arxiv.org/abs/2409.11060}{{\ttfamily 2409.11060}}].

\bibitem{Iancu:2023lel}
E.~Iancu, A.H.~Mueller, D.N.~Triantafyllopoulos and S.Y.~Wei, \emph{{Probing
  gluon saturation via diffractive jets in ultra-peripheral nucleus-nucleus
  collisions}},
  \href{https://doi.org/10.1140/epjc/s10052-023-12165-8}{\emph{Eur. Phys. J. C}
  {\bfseries 83} (2023) 1078}
  [\href{https://arxiv.org/abs/2304.12401}{{\ttfamily 2304.12401}}].

\bibitem{Zheng:2014vka}
L.~Zheng, E.~Aschenauer, J.~Lee and B.-W.~Xiao, \emph{{Probing Gluon Saturation
  through Dihadron Correlations at an Electron-Ion Collider}},
  \href{https://doi.org/10.1103/PhysRevD.89.074037}{\emph{Phys. Rev. D}
  {\bfseries 89} (2014) 074037}
  [\href{https://arxiv.org/abs/1403.2413}{{\ttfamily 1403.2413}}].

\bibitem{Albacete:2018ruq}
J.L.~Albacete, G.~Giacalone, C.~Marquet and M.~Matas, \emph{{Forward dihadron
  back-to-back correlations in $pA$ collisions}},
  \href{https://doi.org/10.1103/PhysRevD.99.014002}{\emph{Phys. Rev. D}
  {\bfseries 99} (2019) 014002}
  [\href{https://arxiv.org/abs/1805.05711}{{\ttfamily 1805.05711}}].

\bibitem{Stasto:2018rci}
A.~Stasto, S.-Y.~Wei, B.-W.~Xiao and F.~Yuan, \emph{{On the Dihadron Angular
  Correlations in Forward $pA$ collisions}},
  \href{https://doi.org/10.1016/j.physletb.2018.08.011}{\emph{Phys. Lett. B}
  {\bfseries 784} (2018) 301}
  [\href{https://arxiv.org/abs/1805.05712}{{\ttfamily 1805.05712}}].

\bibitem{vanHameren:2016ftb}
A.~van Hameren, P.~Kotko, K.~Kutak, C.~Marquet, E.~Petreska and S.~Sapeta,
  \emph{{Forward di-jet production in p+Pb collisions in the small-x improved
  TMD factorization framework}},
  \href{https://doi.org/10.1007/JHEP12(2016)034}{\emph{JHEP} {\bfseries 12}
  (2016) 034} [\href{https://arxiv.org/abs/1607.03121}{{\ttfamily
  1607.03121}}].

\bibitem{vanHameren:2023oiq}
A.~van Hameren, H.~Kakkad, P.~Kotko, K.~Kutak and S.~Sapeta, \emph{{Searching
  for saturation in forward dijet production at the LHC}},
  \href{https://doi.org/10.1140/epjc/s10052-023-12120-7}{\emph{Eur. Phys. J. C}
  {\bfseries 83} (2023) 947}
  [\href{https://arxiv.org/abs/2306.17513}{{\ttfamily 2306.17513}}].

\bibitem{Cassar:2025vdp}
K.~Cassar, Z.~Wang, X.~Chu and E.-C.~Aschenauer, \emph{{Investigating the
  broadening phenomenon in two-particle correlations induced by gluon
  saturation}},  \href{https://arxiv.org/abs/2503.08447}{{\ttfamily
  2503.08447}}.

\bibitem{Binosi:2003yf}
D.~Binosi and L.~Theussl, \emph{{JaxoDraw: A Graphical user interface for
  drawing Feynman diagrams}},
  \href{https://doi.org/10.1016/j.cpc.2004.05.001}{\emph{Comput.Phys.Commun.}
  {\bfseries 161} (2004) 76}
  [\href{https://arxiv.org/abs/hep-ph/0309015}{{\ttfamily hep-ph/0309015}}].

\end{thebibliography}
\end{document}